\documentclass[%
 aip,
 jmp,%
 amsmath,amssymb,
preprint,%
 reprint,onecolumn,%
nofootinbib]{revtex4-1}

\usepackage{dcolumn}
\usepackage{bm}
\usepackage{amsmath,amssymb,amsfonts}
\usepackage{graphicx} 
\usepackage{txfonts}
\usepackage{epstopdf}
\usepackage[T1]{fontenc}
\usepackage{color}
\usepackage{bbold}
\usepackage{verbatim}
\usepackage{enumerate}

\usepackage[unicode]{hyperref}
\hypersetup{
    unicode=true,                                           
    plainpages=false,
    pdffitwindow=true,                                      
    colorlinks=true,                                        
    linkcolor=blue,                                        
    citecolor=blue,                                         
    filecolor=blue,                                        
    urlcolor=blue                                          
}
\newcommand{\bra}[1]{\langle #1\vert}

\newcommand{\ket}[1]{\vert #1\rangle}

\newcommand{\bket}[2]{\langle #1\vert#2\rangle}

\newcommand{\refeq}[1]{Eq.~(\ref{#1})}

\newcommand{\reffig}[1]{Fig.~\ref{#1}}

\newcommand{\Gprop}[0]{\mathcal{G}}
\newcommand*\colvec[3][]{
\begin{pmatrix}\ifx\relax#1\relax\else#1\\\fi#2\\#3\end{pmatrix}
}    

 \newcommand{\polB}[0]{\mathbf{p}}
  \newcommand{\pol}[0]{p}
\newcommand{\muB}[0]{\boldsymbol{\mu}}

\begin{document}

\preprint{AIP/123-QED}

\title{Coherence specific signal detection via chiral pump-probe spectroscopy}
\author{David I. H. Holdaway}
\email{d.holdaway@ucl.ac.uk}
\affiliation{Department of Physics and Astronomy, University College London, Gower Street, WC1E 6BT, London, United Kingdom}
\author{Elisabetta Collini}
\email{elisabetta.collini@unipd.it}
\affiliation{Department of Chemical Sciences, University of Padova, I-35131 Padova, Italy}
\author{Alexandra Olaya-Castro}
\email{a.olaya@ucl.ac.uk}
\affiliation{Department of Physics and Astronomy, University College London, Gower Street, WC1E 6BT, London, United Kingdom}

\date {28/04/16}

 \begin{abstract}
We examine transient circular dichroism spectroscopy (TRCD) as a technique to investigate signatures of exciton coherence dynamics under the influence of structured vibrational environments. We consider a pump-probe configuration with a linearly polarized pump and a circularly polarized probe, with a variable angle $\theta$ between the two directions of propagation.  In our theoretical formalism the signal is decomposed in chiral and achiral doorway and window functions. Using this formalism, we show that the chiral doorway component, which beats during the population time, can be isolated by comparing signals with different values of $\theta$.  As in the majority of time-resolved pump-probe spectroscopy, the overall TRCD response shows signatures of both excited and ground state dynamics. However, we demonstrate that the chiral doorway function has only a weak ground state contribution, which can generally be neglected if an impulsive pump pulse is used. These findings suggest that the pump-probe configuration of optical TRCD in the impulsive limit has the potential to unambiguously probe quantum coherence beating in the excited state. We present numerical results for theoretical signals in an example dimer system.
%
 \end{abstract}

\maketitle

\section{Introduction}
\label{S:0}

Chirality (handedness) occurs in systems which are distinguishable from their mirror image, that is, systems which lack reflection symmetry. Both matter and light can display chirality, and the chirality of certain molecules leads to a difference in the absorption of left and right circularly polarized light and therefore exhibit circular dichroism (CD) and optical rotation dichroism (ORD). CD and ORD are not independent, as they are linked by Kramers--Kronig relations~\cite{Polavarapu2006}. 
Linear CD and ORD spectroscopy have been used extensively to determine more accurately the structure of  proteins~\cite{Kelly2000,Kelly2005,Greenfield2006} and the excitonic states in photosynthetic antennae including the Fenna-Matthews-Olson complex (FMO) from Green Sulfur bacteria~\cite{Furumaki2012}, and light harvesting complex I~\cite{Georgakopoulou2006} and II~\cite{Hemelrijk1992,Buchel1997}. Transient circular dichroism (TRCD), that is, the time-domain CD spectra of a system away from equilibrium, has been used within the infra-red~\cite{Bonmarin2008,Rhee2009} and ultra-violet frequency ranges~\cite{Hache2009,Abramavicius2005}. This technique has proved useful in measuring protein conformation and lipid structure~\cite{Qiu2003,Chen1998}. 

In the optical regime, however, TRCD has been developed very little and there is a less comprehensive understanding of what signals should be expected.  While some investigations of the relations between photoexcitation dynamics and chiral response have been carried out using pump-probe configurations~\cite{Niezborala2007} and, more recently, two-dimensional (2D) spectroscopy configurations~\cite{FidlerEngel2014}, the full potential of time-resolved CD is yet to be explored.  The reluctance to use this technique originates from the lower signal to noise ratios; typically CD signals are 3 to 4 orders of magnitude weaker compared to non-chiral origin signals~\cite{ParsonMOS}. Quantum transitions between electronic states of matter due to the interaction with light can be understood within a multipole expansion: electric dipoles give the most significant (lowest order) contribution, but give the same absorption for left and right circularly polarized light.  The next most significant terms are magnetic dipole and electric quadrapole terms, which are typically weaker by a factor proportional to the chromophore size divided by an optical wavelength~\cite{MukamelPoNLS}. The combination of these higher order moments and an electric dipole moment results in a difference in left-right absorption and hence CD. {
Systems of multiple chromophores such as photosynthetic pigment-protein complexes may show circular dichroism even when individual choromophores do not.  The origin of chirality here is both the  electronic interactions between chromophores and the lack of reflection symmetry in the transition dipole moments of each chromophore and the vectors describing the relative displacements and orientation between them~\cite{NobuyukiCDS}.  In the absence of electronic coupling there will be no CD and if all the relative displacement vectors lie in a plane, the system is no longer chiral and will not exhibit CD either. Therefore CD signals from such systems show a strong dependence on excited state delocalization and on the relative positions and orientations between individual dipole moments~\cite{Abramavicius2006-2}. This indicates that TRCD can probe dynamic exciton localization~\cite{FidlerEngel2014} as well as the sensitivity of electronic transitions to structural changes following photoexcitation~\cite{Abramavicius2006-2}.

Furthermore, transitions to states which have negligible electric dipole moments (and are thus "dark" to non-chiral spectroscopy), can potentially have magnetic dipole transitions which can be of comparable order to bright  states within the complex and therefore be directly probed via TRCD.  While it may be possible to infer the dynamics related to a dark state from the ground state dynamics or transitions to double excited states~\cite{Egorova2015} it is not possible to directly detect signals relating to coherences between the dark state and other "bright" states (as this term would vanish by definition) without observing the chiral signals.

In the context of photosynthetic excitons in light-harvesting antennae a current problem under scrutiny is to probe quantum coherences among exciton states \cite{Fassioli2013}. In these systems the comparable strength of interactions between site electronic excitations and the coupling to a structured vibrational background lead to a complex exciton and vibrational dynamics that has been incisively investigated in the last decade by  2D optical spectroscopy \cite{Engel2007, Calhoun2009, Collini2010, Panitchayangkoon2010}. These experiments have revealed pico-second coherence beating in a variety of photosynthetic complexes \cite{Engel2007, Calhoun2009, Collini2010, Panitchayangkoon2010} and chemical systems \cite{Collini2009, Halpin2014} generating an intense debate about the electronic and vibrational origin of such signals. There is  mounting theoretical \cite{Kolli2012, Christensson2012, Chin2013,  OReilly2014} and experimental \cite{Womick2011, Richards2012, Halpin2014, Novelli2015,Singh2015} evidence that coupling to some well resolved vibrational modes may indeed enable quantum coherent dynamics of excitons through vibronic coupling. Notwithstanding, coherence beating in 2D spectroscopy signals is influenced by both excited state and ground state vibrational coherences \cite{Tiwari2012, Plenio2013}.  

There is therefore a need for experimental approaches that can isolated coherence specific signals in the excited state with minimal or no contribution of the ground states vibrational coherences. This is a quite challenging problem as in the majority of the experimental conditions one would expect such contributions and therefore the experimental scenarios for achieving isolation of excited state coherences seem to be system specific \cite{Liebel2015, Lim2015}. In this paper we present a theoretical description of TRCD in excitonic systems and show that chiral time-resolved experiments in pump-probe configurations and in the impulsive limit could be potentially used to probe excited state coherences.  We develop a doorway window formalism to analyse the chiral and non-chiral contributions to the signal and discuss the conditions under which ground state coherences have negligible influence in the chiral doorway functions.  We illustrate the technique by presenting numerical results for a dimer exciton system, which shows signatures of vibronic dynamics.} The paper is organized as follows: Sec.~\ref{S:1} discusses exciton physics and our model system, along with origins of linear circular dichroism in excitonic systems and main assumptions we make.  Sec.~\ref{S:2} describes the extension to  TRCD in a pump-probe configuration in a doorway window formalism, along with discussions of how to obtain coherence specific signals.  Sec.~\ref{S:results} includes numerical results for our example system and Sec.~\ref{S:conclusion} outlines our main results and conclusions. 


\section{Origins of CD in isotropic ensembles of excitonic systems}
\label{S:1}

\subsection{Excitons and their structured vibrational environments}
\label{SS:1}

We are interested in systems consisting of interacting chromophores which have negligible overlap in their electronic wavefunctions. Each chromophore site is located at a position $\boldsymbol{R}_j$ and has one excited state with a transition dipole moment $\boldsymbol{\mu}_j$. The electronic degrees of freedom for each chromophore  are linearly coupled to independent baths of harmonic oscillators. The strength of such interaction is given by identical structured spectral densities of the form:
\begin{equation}
J(\omega) = \frac{2  \lambda_D \gamma_D \omega}{\gamma_D^2 + \omega^2} + \frac{2  \omega_0^2 \lambda_B \gamma_B \omega}{\gamma_B^2 \omega^2 + (\omega^2-\omega_0^2)^2} \;.
\label{eq:spec_den}
\end{equation}
The first term in Eq. (\ref{eq:spec_den}) is a smooth Drude component describing the interaction with a continuous distribution of modes with a characteristic cutoff frequency $\gamma_D$ and reorganization energy  $\lambda_D$. The second term describes the interaction with a well resolved, under-damped vibrational mode of frequency $\omega_0$, damping rate $\gamma_B$ and reorganization energy  $\lambda_B$. These expressions are associated with an overdamped and underdamped Brownian oscillator model, respectively.

Through out this work we use $\textrm{cm}^{-1}$ units, which involves setting $h = c = 1$cm$ = 1$. However, when we plot angular frequencies we scale these by a factor of $2 \pi c$ to keep consistency with the energy $\hbar \omega$.

The total Hamiltonian (consisting of the system and bath together) is split into three components
\begin{equation}
 \hat{H}_{S+B} = \hat{H}_{S} + \hat{H}_{B} + \hat{H}_{SB} \;,
\end{equation}
with $\hat{H}_{S}$ a the system Hamiltonian relating to the electronic degrees of freedom, $\hat{H}_{B}$ the bath of harmonic oscillators couple to each site excitation and  $\hat{H}_{SB}$ is the excitation-bath interaction.  The primary dynamics are determined by the system Hamiltonian~\cite{ValkunasMEDaR}
\begin{align}
H_{S} = \sum_{j=1}^N \omega_{j} \ket{j}\bra{j} + \sum_{j \ne j'} \left[ V_{k,k'}\ket{j}\bra{j'} +(\omega_{j} +\omega_{j'}+\kappa_{j,j'})  \ket{j;j'}  \bra{j;j'}\right]   \;,
\label{eq:Hsys}
\end{align}
where$\ket{j}$ is the excited state of chromophore $j$ (while all other chromophores are in the ground state) and $\ket{j;j'}$ is state where chromophores $j$ and $j'$ are both in their excited states (and all other in their ground state). The transitions frequencies $\omega_{j} = \omega_{j}^{(0)} + \lambda_D + \lambda_B$ are shifted from the "bare" excitation frequency $\omega_{j}^{(0)}$ by the reorganization energies due to the interaction with the structured thermal bath.  Transition frequencies to double excited states are furthermore shifted by $ \kappa_{j,j'}$, which for simplicity will be set to zero.

The electronic excitation dynamics under the influence of the vibrational environment will be exactly computed via a hierarchy of equations of motion (HEOM)~\cite{Chen2010}. This formalism, being non-perturbative, tracks down the influence of higher-order correlations of the vibrational environment via auxiliary density matrices, thereby capturing the information flow from system to environment and back and therefore capable of accounting for non-Markovian effects \cite{BreuerTOQS}

We consider the exciton basis that diagonalize $H_{s}$ into the form
\begin{align}
\hat{H}_{S} =\sum_{\xi=1}^N \tilde{\omega}_\xi \ket{\xi}\bra{\xi} + \sum_{f=1 }^{N(N-1)/2} \tilde{\omega}_{f}\ket{f,2}\bra{f,2}  \;,
\end{align}
where $\tilde{\omega}_\xi$ and $\tilde{\omega}_{f}$ are the energies of the  single and double exciton states, which are related to chromophore excitations via
\begin{equation}
\ket{\xi}  =  \sum_{j}  C_{\xi,j} \ket{j}\;, \quad \ket{f,2}  =  \sum_{j,j'}  C_{f,j;j'} \ket{j;j'} \;.
\end{equation}
Associated to this basis it is helpful to define effective dipole moments for transitions to different exciton states as follows:
\begin{equation}
\boldsymbol{\mu}_{\xi} = \sum_{j}  \bket{j}{\xi} \boldsymbol{\mu}_j \;, \quad \boldsymbol{\mu}_{\xi,f} = \sum_{j,j'}  \bket{\xi}{j'}\bket{j';j}{f,2} \boldsymbol{\mu}_j \;.
\label{eq:ex_dip_mom}
\end{equation}

\subsection{Effects from non-local terms in the polarization operator.}

Circular dichroism, the difference in absorption of left and right circularly polarized light, is due to the lack of reflection symmetry in the matter involved.  In excitonic systems such signals can have two origins.  The first is from the intrinsic chirality (due to the lack of reflection symmetry) in the chromophores themselves,  giving transitions a finite magnetic-dipole and/or electric quadrapole moment.  We do not consider this case explicitly in this work, although we briefly discuss how the effects can be included within this formalism without significantly changing our key results.  The second 
originates from the combination of the spatial separation and orientation of the chromophores and their dipole moments within a molecular complex, or multichromophore system, and the position dependence of the phase of the laser.  

Treating the light as a classical field, the time dependent interaction Hamiltonian between the light and the $m$th molecular complex in our sample reads~\cite{MukamelPoNLS}
\begin{equation}
\hat{H}_m(t) = -\int \hat{\mathbf{P}}_m \cdot \mathbf{E}^{\bot}(\mathbf{r},t) d \mathbf{r} \;,
 \label{eq:Int_op}
\end{equation}
with $\mathbf{E}^{\bot}(\mathbf{r},t)$ the transverse component of the electric field and 
\begin{equation}
\hat{\mathbf{P}}_m \approx  \sum_{j=1}^N \hat{\boldsymbol{\mu}}_{j;m}(\vec{q}) \delta(\mathbf{r}-\mathbf{R}_m - \mathbf{R}_{m;j})\; ,
 \label{eq:Pol_op}
\end{equation}
the polarization operator for the $m$ complex.  In \refeq{eq:Pol_op} we have assumed a dipole interaction for transitions to the single excitation  states of each chromophore, which are displaced by $\mathbf{R}_{m;j}$ from the center. The parameter $\vec{q}$ is the set of coordinates for the normal modes associated to the vibrational bath.  In our numerical calculations we will neglect this dependence (i.e.~make the Condon approximation~\cite{Condon1926}) and take $\muB_{j;m}(\vec{q}) \sim \muB_{j;m}(\vec{0})$. However, in a realistic system, linear order terms in a Taylor expansion of $\muB_{j;m}(\vec{q})$ will couple vibrational states to states with $\pm 1$ quanta of excitation (within the harmonic oscillator approximation), hence we discuss the impact of these couplings in Sec.~\ref{S:2}.

In order to simplify \refeq{eq:Pol_op}, we make the rigidity approximation $\mathbf{R}_{m;j}=T_m \mathbf{R}_{0;j}$ in which we have introduced the rotation operator $T_m = T(\theta_{1,m},\theta_{2,m},\theta_{3,m})$. This approximation assumes all chromophore positions relative to the center are identical up to a rotation.  Fluctuations in the positions and dipole moments could be accounted for as static disorder, assuming the distribution is known. The rotation operator describes the orientation relative to our reference molecule ($m=0$) in terms of the three Euler angles.  The transition dipole operator $\hat{\boldsymbol{\mu}}_{j;m}$ for chromophore $j$ in complex $m$ can now be written as:
\begin{equation}
\bra{g_{j,m}}\hat{\boldsymbol{\mu}}_{j;m}\ket{j'_{m'}}  = \delta_{m,m'} \delta_{j,j'} T_m\boldsymbol{\mu}_{j,m} \; , 
\label{eq:mu_op}
\end{equation}
where $\ket{g_{j,m}}$ is the ground state of chromophore $j$ of complex $m$. Hence we can express this dipole operator for a given chromophore within a complex with orientation operator $T_m$, as
\begin{equation}
\hat{\boldsymbol{\mu}}_{j;m}= T_m\boldsymbol{\mu}_{j,m} \ket{g_{j;m}}\bra{j_m} + (T\boldsymbol{\mu}_{j;m})^* \ket{j_m}\bra{g_{j;m}} \; .
\label{eq:mu_op2}
\end{equation}
Throughout this work we also make a single wavevector approximation for the pulses of light which will be relevant here, hence the electric field is a sum of terms of the form
\begin{equation}
 \mathbf{E}(\mathbf{r},t) = \left(\polB  E(t) e^{i  (\mathbf{k} \cdot \mathbf{r} - \omega_j t + \varphi)} +\polB^* E^*(t) e^{-i  (\mathbf{k} \cdot \mathbf{r} - \omega_j t+\varphi)} \right) .
 \label{eq:pump_Efield}
\end{equation}

We can calculate the expectation value of the polarization operator $\mathbf{P}(\mathbf{r},t) = \sum_m \mathrm{Tr}\{\hat{\mathbf{P}}_m \rho(t)\}$ via time-dependent perturbation theory in $\hat{H}_{int}=\sum_m \hat{H}_{m}(t)$. Assuming the initial state is the equilibrium density matrix for the system $\hat{\rho}_{eq}$, then to lowest order we have
\begin{align}
\mathbf{P}^{(1)}(\mathbf{r},t)= &i\int d \mathbf{r}_1 \int_0^{\infty} dt_1 \sum_m \sum_{j_2,j_1=1}^N  \delta(\mathbf{r}-\mathbf{R}_m - T_m\mathbf{R}_{j_2})  \delta(\mathbf{r}_1-\mathbf{R}_m -T_m\mathbf{R}_{j_1}) \nonumber \\ 
&\mathrm{Tr}\{ \boldsymbol{\mu}_{j_2;m} \Gprop(t_1)[\mathbf{E}(\mathbf{r}_1,t-t_1) \cdot  \hat{\boldsymbol{\mu}}_{j_1;m},\hat{\rho}_{eq}]\} \nonumber \\
 = &i\int_0^{\infty} dt_1 \sum_m \sum_{j_2,j_1=1}^N  \delta(\mathbf{r}-\mathbf{R}_m - T_m\mathbf{R}_{j_2})  \mathrm{Tr}\{  \boldsymbol{\mu}_{j_2;m} \Gprop(t_1) [\mathbf{E}(\mathbf{R}_m + T_m\mathbf{R}_{j_1},t-t_1) \cdot \hat{\boldsymbol{\mu}}_{j_1;m},\hat{\rho}_{eq}]\} \; ,
\label{eq:Pol_op2}
\end{align}
with $[,]$ a commutator bracket and $\Gprop(t_1)$ the time propagation operator for the system alone, i.e.~without coupling to the electric field.  As the ground, single and double excited state manifolds are uncoupled without light, we can propagate coherences between these levels separately and use the notation $\Gprop_{ab}(t_1)$ for the component acting on a particular manifold. For linear spectroscopy it is sufficient to consider only coherences between ground and excited state and therefore $ab \to eg$.  This equation describes a very general situation in which the complexes need to be uniformly distributed or randomly orientated (and can easily incorporate static disorder in terms of the variation in site transition energies and dipole moments).  We consider an isotropic system and take a continuum approximation for the molecular positions and for simplicity we do not average over static disorder.  This modifies the sum over $m$ to a integral over the positions $\mathbf{R}_m$ and the three Euler angles in $T \equiv T(\theta_1,\theta_2,\theta_3)$, we also make the substitution in \refeq{eq:mu_op2} to separate the vector components of the operators

\begin{align}
\mathbf{P}^{(1)}(\mathbf{r},t) \approx  &i\int_0^{\infty} dt_1 \int d \mathbf{R} \int_0^{2\pi} d\theta_1 \int_0^{\pi} d\theta_2 \int_0^{2\pi}d\theta_3  \frac{n'\sin(\theta_2)}{8 \pi^2} \sum_{j_2,j_1=1}^N \delta(\mathbf{r}-\mathbf{R} - T \mathbf{R}_{j_2}) \nonumber \\
&  \mathrm{Tr}\{  (T\boldsymbol{\mu}_{j_2} \ket{g}\bra{j_2} + h.c.) \Gprop(t_1)[\mathbf{E}(\mathbf{R} + T\mathbf{R}_{j_1},t-t_1) \cdot (T\boldsymbol{\mu}_{j_1} \ket{g}\bra{j_1} + h.c.),\hat{\rho}_{eq}]\}  \nonumber \\
\approx  &i \iint d\theta_1 d\theta_2 d\theta_3  \frac{n'\sin(\theta_2)}{8 \pi^2} \sum_{j_2,j_1=1}^N T\boldsymbol{\mu}_{j_2}  T\boldsymbol{\mu}_{j_1} \cdot \polB \; e^{i \mathbf{k} \cdot (\mathbf{r} + T\mathbf{R}_{j_1}-T\mathbf{R}_{j_2})} \nonumber \\ 
&\int_0^{\infty} dt_1 \mathrm{Tr}\{ e^{ -i \omega_1 (t-t_1)} E(t-t_1)   \ket{g}\bra{j_2}\Gprop_{eg}(t_1)\ket{j_1}\bra{g}\hat{\rho}_{eq} - h.c.\} 
\;. 
\label{eq:Pol_op3}
\end{align}
Here $\ket{g}$ is the state where all chromophores are in their ground state state and we have made the the rotating wave approximation to write the final line. This expression is valid for $\mathbf{r}$ inside the sample (edge effects which are assumed to be negligible) with $n'$ complexes per unit volume.  In this form we have separated off the parts which are affected by the average over the Euler angles. Assuming our system is significantly smaller than an optical wavelength, we can expand the exponential term to first order
\begin{align}
T\boldsymbol{\mu}_{j_2}  (T\boldsymbol{\mu}_{j_1} \cdot \polB ) e^{i  \mathbf{k} \cdot  T(\mathbf{R}_{j_1}-\mathbf{R}_{j_2})} \approx  T\boldsymbol{\mu}_{j_2}  (T\boldsymbol{\mu}_{j_1} \cdot \polB ) [1 + i  \mathbf{k} \cdot  T(\mathbf{R}_{j_1}-\mathbf{R}_{j_2}) + \ldots]
\; .
\label{eq:Ori_av_fo}
\end{align}
In the right hand side of \refeq{eq:Ori_av_fo}, isotropic averages can be performed analytically.  If we have a pulse with polarization $\polB_1$ and take averages for $\mathbf{P}^{(1)} \cdot \polB_2$ (the component of the polarization along direction $\polB_2$)
\begin{align}
\langle(\boldsymbol{\mu}_{j_1} \cdot \polB_1) (\boldsymbol{\mu}_{j_2} \cdot \polB_2)\rangle &\equiv \iint d\theta_1 d\theta_2 d\theta_3   \frac{\sin(\theta_2)}{8 \pi^2} (T\boldsymbol{\mu}_{j_2} \cdot \polB_2 )( T\boldsymbol{\mu}_{j_1} \cdot \polB_1) = (\polB_1 \cdot \polB_2) (\boldsymbol{\mu}_{j_1} \cdot  \boldsymbol{\mu}_{j_2})/3 \\
\langle(\boldsymbol{\mu}_{j_1} \cdot \polB_1) (\boldsymbol{\mu}_{j_2} \cdot \polB_2)(\mathbf{k} \cdot \Delta \mathbf{R}_{j_1,j_2})  \rangle  &= [ \polB_{2} \cdot (\polB_{1} \times \mathbf{k})][\muB_{j_2} \cdot (\muB_{j_1} \times \Delta \mathbf{R}_{j_1,j_2}) ]/6 \;.
\label{eq:2nd_3rd_av}
\end{align}
The first average therefore produces no polarization orthogonal to the input field (or $\polB_{2} \cdot \polB_{1} = 0$ and the second produces only polarization orthogonal to  $\polB_{1}$ (as $\mathbf{k}$ must be orthogonal to the polarization since light is a transverse wave), and vanishes for $\muB_{j_1} = \muB_{j_2}$. It is this latter term that is responsible for the circular dichroism (and also optical rotation) as we will see in the next section.

\subsection{Linear CD and the chiral interaction operator}

Because excitons are the relevant basis for considering CD effects in our system, we introduce a third order tensor relating to the chiral interaction involving excitons $\xi_1$ and $\xi_2$
\begin{equation}
\psi_{\xi_1,\xi_2}^{\nu, \nu_1 , \nu_2} = \sum_{j_1,j_2=1}^N  C_{\xi_1}^{j_1}C_{\xi_2}^{j_2} \mu_{j_1}^{\nu_1}\mu_{j_2}^{\nu_2} (R_{j_1}^{\nu} - R_{j_2}^{\nu}) \;.
\label{eq:psi_factor}
\end{equation}
This chiral factor will be useful for the extension to non-linear effects which are the primary consideration of this work. We have introduced tensor notation and assume Einstein summation notation in which repeated indices are summed over, i.e.~$A^{\nu} B ^{\nu} \equiv \sum_{\nu =x,y,z} A^{\nu} B ^{\nu} = \mathbf{A} \cdot \mathbf{B}$. Averages of this operator can be expressed as
\begin{align}
\langle k_j^{\nu} p_j^{\nu_j} p^{\nu'}  \psi_{\xi_j,\xi}^{\nu, \nu_j , \nu'}\rangle &= -\frac{[ \polB' \cdot \polB_j \times \mathbf{k}_j ]}{6}\sum_{\ell,\ell'} C^{\ell}_{\xi} C^{\ell'}_{\xi_j} [ \muB_{\ell} \cdot \muB_{\ell'} \times \mathbf{r}_{\ell'}+\muB_{\ell'} \cdot \muB_{\ell} \times \mathbf{r}_{\ell} ]  \nonumber \\
&=-|\mathbf{k}_j|\frac{[ \polB' \cdot \mathbf{b}_j ]}{3} [ \muB_{\xi} \cdot \tilde{\mathbf{m}}_{\xi_j} +  \muB_{\xi_j} \cdot \tilde{\mathbf{m}}_{\xi}]\;,
\label{eq:psi_factor_av}
\end{align}
where we have defined an effective magnetic dipole moment of  the one-exciton and double-exciton states as:
\begin{equation}
\tilde{\mathbf{m}}_{\xi} = \sum_{j}  \bket{j}{\xi} \boldsymbol{\mu}_j \times \mathbf{R}_j/2  \;, \quad \tilde{\mathbf{m}}_{\xi,f} = \sum_{j,j'}   \bket{\xi}{j'} \bket{j',j}{f,2} \boldsymbol{\mu}_j \times \mathbf{R}_j/2 \;.
\label{eq:ex_mag_dip}
\end{equation}
We are departing from the usual convention for the magnetic dipole moment here, which would have an additional factor of $i|k| = 2 \pi i / \lambda$ (such that it can be combined with a unit vector in the direction of the magnetic field). In our case we have extracted this complex factor so that the magnetic moments are real.

The frequency space absorption profile is related to the Fourier transform of the polarization in \refeq{eq:Pol_op2}, using the new notation and the convolution theorem is given by
\begin{align}
 \mathbf{P}^{\nu_2}(\mathbf{k},\omega_s) \propto  \sum_{\xi_2,\xi_1=1}^N  
 p_j^{\nu_1} \langle\mu_{\xi_1}^{\nu_1}\mu_{\xi_2}^{\nu_2} + k_j^{\nu}\psi_{\xi_1,\xi_2}^{\nu, \nu_1 , \nu_2} \rangle \mathrm{Tr}\{ \tilde{E}(\omega-\omega_s)  \hat{B}_{\xi_2}\tilde{\Gprop}_{eg}(\omega)\hat{B}^{\dagger}_{\xi_1}\hat{\rho}_{eq} - h.c.\} 
\; ,
\label{eq:Pol_op_ft}
\end{align}
Here we have introduced the lowering operator for an exciton $\hat{B}_{\xi} \equiv \ket{g}\bra{\xi}$, $\tilde{E}(\omega)$ the Fourier transform of $E(t)$ and $\tilde{\Gprop}_{ab}(\omega)$ the one sided Fourier transform of the propagation operator:
\begin{equation}
\tilde{\Gprop}_{ab}(\omega) = -i\int_{0}^{\infty} \Gprop_{ab}(t) e^{i \omega t}\;.
\label{eq:freq_prop_op}
\end{equation}
Assuming the excitons are a good basis for the electronic dynamics, the cross terms with $\xi_1 \neq \xi_2$ will be small in \refeq{eq:Pol_op_ft} and can often be neglected. This polarization of the medium produces a signal electric field $\mathbf{E}_s$ proportional to the imaginary component of the polarization.  Assuming the medium is optically thin, we can consider the total intensity to be the combination of this signal field and a local oscillator field $E_{LO}$ with wavevector $\mathbf{k}_s$. Typically in a linear or pump probe experiment $\mathbf{E}_{\rm LO}$ is just the input probe field, but we consider a general field as this allows us to consider frequency resolved detection. We therefore measure the intensity
\begin{equation}
I_{\rm LO+s}(t) = |\mathbf{E}_{\rm LO}|^2 + 2 \mathrm{Re}[\mathbf{E}_{\rm s} \cdot \mathbf{E}^*_{\rm LO} ] \quad \left(+|\mathbf{E}_{\rm s}|^2 \right)\;,
\end{equation}
the $|\mathbf{E}_{\rm s}|^2$ term is usually small enough to be neglected because $|E_{\rm LO}| \gg |E_{\rm s}|$ and, as we know $\mathbf{E}_{\rm LO}$, we can subtract the first term to leave $\mathrm{Re}[\mathbf{E}_{\rm s} \cdot \mathbf{E}^*_{\rm LO}] \propto -\omega_s \mathrm{Im}(\mathbf{P}^{(1)} \cdot \mathbf{E}^*_{\rm LO}) $ and therefore extract the signal.  

We assume the probe field, which is also our local oscillator, consists of a single frequency $\omega$ (therefore $\tilde{\mathbf{E}}(\omega) \to \delta(\omega)$).  We obtain the signal by subtracting the absorption signal with left circularly polarized (cp) light, with $\mathbf{p}_L = (\hat{x} + i \hat{y})/\sqrt{2}$ from right cp light ($\mathbf{p}_R = (\hat{x} - i \hat{y})/\sqrt{2} = \mathbf{p}_L^*$) giving
\begin{align}
 S_{cd}(\omega) \equiv S(\omega ; \polB_L)-S(\omega ; \polB_R) \propto  \omega \; \mathrm{Re} \left\{\sum_{\xi_2,\xi_1=1}^N  
 ( p_L^{\nu_1} (p_L^*)^{\nu_2} - p_R^{\nu_2} (p_R^*)^{\nu_1}) \langle\mu_{\xi_1}^{\nu_1}\mu_{\xi_2}^{\nu_2} + k_j^{\nu}\psi_{\xi_1,\xi_2}^{\nu, \nu_1 , \nu_2} \rangle \{\hat{B}_{\xi_2}\tilde{\Gprop}(\omega)\hat{B}^{\dagger}_{\xi_1}\hat{\rho}_{eq} - h.c.\} \right\}
\; .
\label{eq:CD_sig}
\end{align}
Using the results from \refeq{eq:2nd_3rd_av}, we can show the term proportional to $\mu_{\xi_1}^{\nu_1}\mu_{\xi_2}^{\nu_2}$ vanishes. Finally using \refeq{eq:psi_factor_av} and the assumption that only $\xi_1 = \xi_2$ terms are significant, we can show this average is equal to
\begin{align}
 S_{cd}(\omega) \sim \omega \sum_{\xi=1}^N  \muB_{\xi} \cdot \tilde{\mathbf{m}}_{\xi}
 \mathrm{Tr} \{\hat{B}_{\xi}\tilde{\Gprop}_{ge}(\omega)\hat{B}^{\dagger}_{\xi}\hat{\rho}_{eq} - h.c.\} 
\; .
\label{eq:CD_sig2}
\end{align}
The term $(\muB_{\xi} \cdot \tilde{\mathbf{m}}_{\xi})$ is proportional to the rotational strength $\mathcal{R}_{\xi}$ of the transition to exciton $\ket{\xi}$ \cite{Rosenfeld1928}. The total rotational strength is defined via
\begin{align}
\mathcal{R} = C \frac{n}{f^2} \int d \omega \; \frac{\alpha_{L}(\omega)-\alpha_{R}(\omega)}{\omega} = \sum_{\xi} \; \mathcal{R}_{\xi} \; ,
\label{eq:rot_strength}
\end{align}
with $\alpha_{L/R}(\omega)$ molar extinction coefficients for left- and right-circularly polarized light, $n$ the refractive index, $f$ the local field correction and $C$ a constant.  

\subsection{Information from linear CD and its limits}

We can use~\refeq{eq:CD_sig2} to extract useful information from a linear CD experiment.  Assuming the exciton lineshapes ($ \mathrm{Tr} \{\hat{B}_{\xi}\tilde{\Gprop}(\omega)\hat{B}^{\dagger}_{\xi}\hat{\rho}_{eq} - h.c.\}$) are known from non-chiral linear spectroscopy, we can fit these to the CD data and gain more information about the dipole orientation and exciton delocalization.  Taking the simple example of a dimer, we can rotate our coordinates such that $\boldsymbol{\mu}_1 = |\boldsymbol{\mu}_1| \hat{\mathbf{x}}$,  $\boldsymbol{\mu}_2 = |\boldsymbol{\mu}_2| (\cos(\theta_1) \hat{\mathbf{x}} + \sin(\theta_1) \hat{\mathbf{y}})$, and $\mathbf{R}_1 - \mathbf{R}_2 \equiv \delta \mathbf{R} = |\delta \mathbf{R}| (\cos(\theta_2)\sin(\phi_2) \hat{\mathbf{x}} + \sin(\theta_2)\sin(\phi_2) \hat{\mathbf{y}} +  \cos(\phi_2)\hat{\mathbf{z}})$; in this way the weights for the non-chiral and chiral exciton lineshapes are 
\begin{align}
|\boldsymbol{\mu}_{\xi}|^2 &= |C_{1\xi}|^2 |\boldsymbol{\mu}_1|^2 + |C_{2\xi}|^2 |\boldsymbol{\mu}_2|^2 +2 C_{1\xi} C_{2\xi} \cos(\theta_1) |\boldsymbol{\mu}_1||\boldsymbol{\mu}_2| \nonumber \\
\boldsymbol{\mu}_{\xi} \cdot \mathbf{m}_{\xi} &= |\delta \mathbf{R}||\boldsymbol{\mu}_1||\boldsymbol{\mu}_2| C_{1\xi} C_{2\xi}  \sin(\theta_1) \cos(\phi_2)  \;.
\end{align}
The weight terms for the CD signal have a maximum value of $|\delta \mathbf{R}||\boldsymbol{\mu}_1||\boldsymbol{\mu}_2|/2$.  This maximum occurs when $\theta_1 = \pi/2$ and $\phi_2=0$ (the two dipole moments are orthogonal and the displacement vector between them is orthogonal to the plane spanned by the two dipoles); and $C_{1\xi} = C_{2\xi}=1/\sqrt{2}$ and $C_{1\xi'} = -C_{2\xi'}=1/\sqrt{2}$ (maximum delocalization of eigenstates), with opposite sign for each exciton $\xi$ and $\xi'$.  Note that if there is no difference in the locations of the peaks or the lineshapes of the two excitons, the CD signals will cancel. More generally, the sum of a linear CD signal over all frequency space should be zero if it originates from excitonic delocalization~\cite{NobuyukiCDS,ParsonMOS}.  This is equivalent to stating that the rotational strength, given in~\refeq{eq:rot_strength}, is zero. 

This extra information can be used to improve estimates for dipole orientations and positions of chromophores (thereby electronic couplings) compared to ordinary linear absorption spectroscopy alone.  However, steady state spectroscopy cannot directly inform us of excited state dynamics, such as exciton migration, dynamic localization and coherent evolution. In the following section we extend this discussion to nonlinear CD in a pump-probe configuration, which can be understood as a linear probe of a density matrix which is not in equilibrium.  For this work we are primarily interested in observation of signatures of quantum superpositions as given by coherence oscillations. We therefore focus on strategies that allow us to extract coherence specific signals.

\section{Doorway window formulation for a TRCD experiment}

\label{S:2}

\subsection{Pump-probe spectroscopy in the doorway window formulation}

Pump-probe spectroscopy is a nonlinear technique in which a sample is excited by a strong "pump" pulse which is followed by a "probe" pulse, which is measured after leaving the sample.  The pump-probe signal is obtained by comparing the measurements with and without the pump being used.  For TRCD we assume the pump is linearly polarized in a direction orthogonal to the probe propagation and the probe is circularly polarized as illustrated in Fig. 1(a). We are interested in a situation in which ultra-short ($< 150$fs full-width at half maximum (FWHM)) pulses are used for the pump and probe, and will ultimately assume we have frequency resolved detection for the probe, which can be achieved by passing the probe through a monochrometer~\cite{Deflores2007}.

The pump pulse is assumed to be composed of a single wavevector $\mathbf{k}_u$ at a variable angle $\theta$ to the probe wavevector $\mathbf{k}_r$ (see Fig. 1(a)). We note that increasing the angle between the pump and probe pulses also increases the uncertainty in the time difference, $\tau$, between the interactions with the pump and probe. This uncertainty is unhelpful when we wish to observe signals from coherence beatings, and may need to be compensated for in the signal analysis. These effects are discussed further in App.~\ref{App:spat_overlap}. For practical considerations, the pump might need to be focused within the sample to improve time resolution, increasing the amount of wavevectors contribution to the signal.  

Generally the amplitude of the pump and probe are chosen to be small enough such that (third order) time-dependent perturbation theory can be used, and thus the polarization of the medium is linear in the probe electric field and quadratic (linear) with the electric field amplitude (intensity) of the pump and so we do not need to consider these explicitly. We assume the pump and probe pulses are of the form given in \refeq{eq:pump_Efield}, with $E(t)$ a Gaussian envelope.  Using the labels $r$ for the probe and $u$ for the pump, the above requires the definition of a carrier frequency $\omega_{r/u}$, polarization $\polB_{r/u}$ and FWHM $\sqrt{8 \ln(2)} \sigma_{r/u}$ for each pulse.  Additionally we consider a "local oscillator" (LO) at a relative phase of $\varphi$ to the probe, in order allow for generalization to both frequency and non-frequency resolved detection.

We can express the full pump-probe signal (chiral and non-chiral components) for a linearly polarized pump in terms of a third order response function $\tensor{S}(t_3,t_2,t_1;\mathbf{k}_3,\mathbf{k}_2,\mathbf{k}_1)$. This response function is a 4th rank tensor, however, as it is also explicitly contracted over the three wavevectors; it therefore has components which are averaged as 4th or 5th rank tensors respectively (higher orders terms are neglected).  The pump probe signal can therefore be expressed
\begin{align}
 \mathrm{S}_{PP} \propto & \omega_r {\rm Re} \left[ e^{i\varphi}\int_{-\infty}^{\infty}dt E_{LO}(t)\int_{0}^{\infty}dt_3\int_{0}^{\infty}dt_2 \int_{0}^{\infty}dt_1 \sum_{s=\pm} \pol^{\nu_1}_u \pol^{\nu_2}_u \pol^{\nu_3}_r \pol^{\nu_4}_{LO} e^{i[ (\omega_{LO}-\omega_r)t +  \omega_r t_3- s\omega_u t_1]} \right.
 \nonumber \\
&E_r(t-t_3)E_u(t-t_3-t_2)E_u(t-t_3-t_2-t_1) \left. S^{\nu_1,\nu_2,\nu_3\nu_4}(t_3,t_2,t_1;\mathbf{k}_3,s\mathbf{k}_u,-s\mathbf{k}_u) \right]\;,
\label{eq:full_response}
\end{align}
with $s$ taking values $\pm 1$. The TRCD signal obtained from the difference signal woth right/left circularly polarized probe pulses. Numerically we calculate a response function in this form and then obtain pump-probe signals from it, however this expression is not particularly enlightening in terms of understanding our results.

The doorway window formalism is useful to understand pump-probe experiments when the pump and probe pulses are well separated ($\tau \gg \sigma_{r}+\sigma_{u}$). The time variables related to non-linear response function within this formalism are illustrated in Fig. 1(b). After interacting with the pump pulse, the density matrix describing our system is out of equilibrium. At some time $\tau \gg \sigma_{u}$ after the first pulse, we can calculate this nonequilibirum density matrix to second order in time-dependent perturbation theory $\rho_{ne}(\tau) = \rho_{eq}  + \rho_{ne}^{(1)}(\tau) + \rho_{ne}^{(2)}(\tau) + h.o.t.$ where the second order correction is of the form
\begin{equation}
\hat{\rho}^{(2)}_{ne}(\tau) = \Gprop(\tau) \left( \begin{array}{cc}
D_g & 0  \\
0 & D_e  \end{array} \right) \equiv  \left( \begin{array}{cc}
\Gprop_{gg}(\tau) D_g & 0  \\
0 & \Gprop_{ee}(\tau) D_e  \end{array} \right)
\end{equation}
The two "doorway" functions $D_g$ and $D_e$ describe the changes to components in the ground and excited state manifolds respectively. This separation is sensible because the time propagator of the join system and bath $\Gprop(\tau)$ does not couple ground and excited states and hence $\Gprop(\tau) D_g \equiv \Gprop_{gg}(\tau) D_g$ and so on.  $D_g$ will have a negative trace because population is removed from the ground state after photoexcitation (bleaching) and so $-D_g$ represents a "hole" population. Notice also that there are also second order corrections related to coherences between the ground and the two-exciton states but these do not contribute to a third order signal in the direction of the probe so can be ignored.

The final signal is obtained by combining each doorway function with corresponding window functions $W_{g/e/f}$ representing the interaction with the probe and the detection which follows:  
\begin{equation}
 \mathrm{S}_{PP}= \omega_r \langle\mathrm{Tr}[\underbrace{W_{g} \Gprop_{gg}(\tau) D_{g}}_\text{GSB}  + \underbrace{W_{e} \Gprop_{ee}(\tau) D_{e}}_\text{SE}+ \underbrace{W_{f} \Gprop_{ee}(\tau) D_{e}}_\text{ESA}]\rangle  \;.
 \label{eq:PP_DW}
\end{equation}
The $\langle \ldots \rangle$ brackets denote an average over all molecular orientations, which are assumed to be effectively static during the wait time $\tau$.  
The three contributions from each window function are the Ground state bleaching (GSB, from $W_{g}$) stimulated emission (SE, from $W_{e}$) and excited state absorption (ESA, from $W_{f}$) to the two-exciton  states. 

The three terms in \refeq{eq:PP_DW} will, in general, have both chiral and non-chiral contributions. In the next subsections we will derive mathematical expressions for these terms in different limits and focus on identifying the contribution to TRCD. Finally in Sec. \ref{ssec:Iso_av} we show how coherence specific components can be isolated.

\begin{figure}[t]
\centering\includegraphics[width=0.5\linewidth]{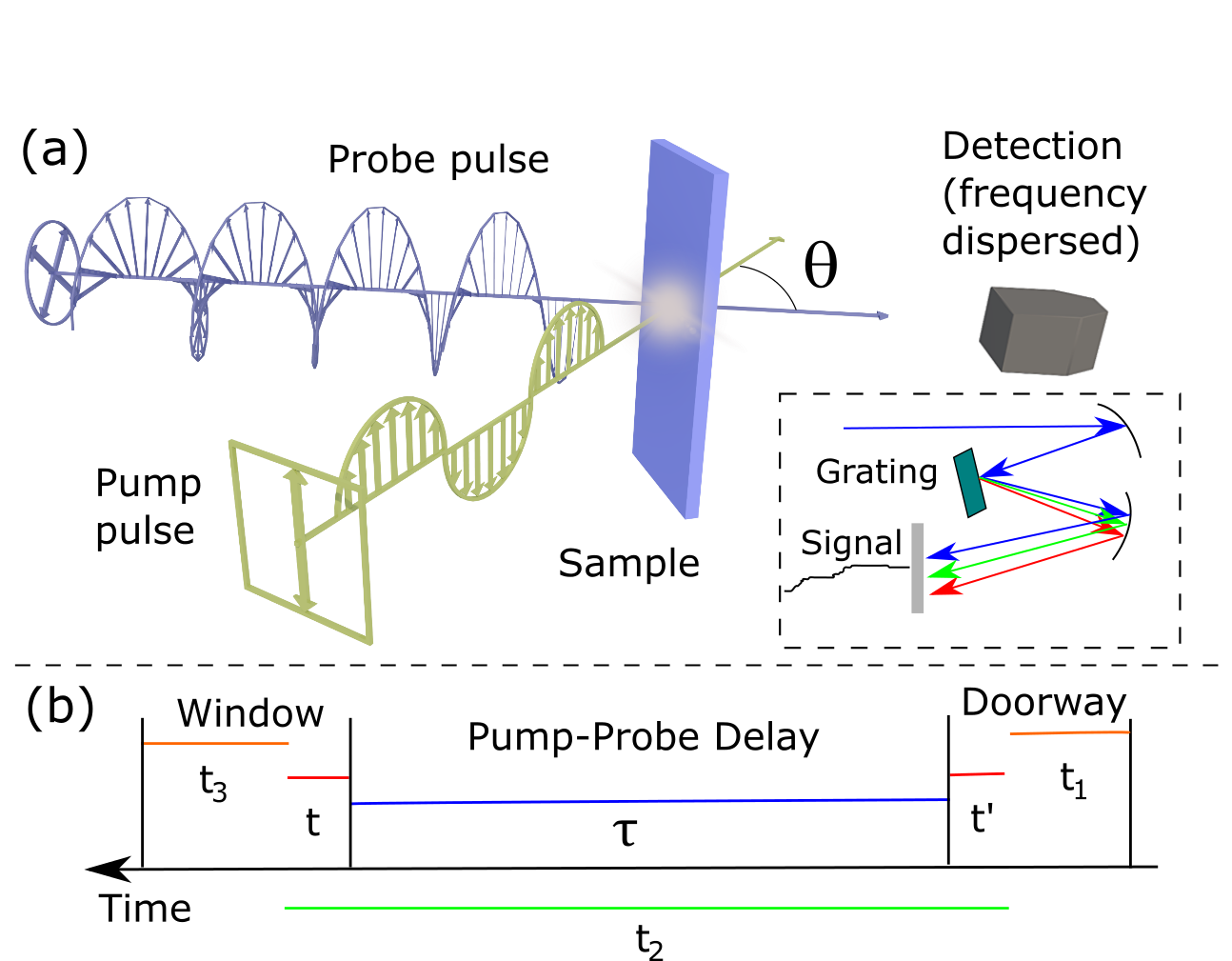}
\caption{(a) Experimental configuration with linearly polarized pump and circularly polarized probe; inset shows frequency resolved detection via monochromatic. (b) Doorway window time variables related to non-linear response function time variables.  The time $t_2 = \tau +t - t'$ is split into the delay time and two additional time variables. These extra time variables include the uncertainty in the population time due to the finite width of the two pulses.}
\label{Fig:door_win}
\end{figure}

\subsection{Doorway functions in different limits}

\subsubsection{Doorway functions with chiral and non-chiral components}
Expressions for the non-chiral (NC) doorway functions can be found regularly in the literature (see for example~\cite{MukamelPoNLS}). Within second order time-dependent perturbation theory, we have two interactions with the pump; assuming the rotating wave approximation interaction is with the forward component (wavevector $\mathbf{k}_u$) and the other with the backward component (wavevector $-\mathbf{k}_u$).  As we must include both chiral interactions, the additional effects can be expressed in the form $k^{\nu} \psi^{\nu,\nu_1,\nu_2}_{\xi_1,\xi_2}$ as defined in \refeq{eq:psi_factor}.  
Recalling that $E_{\rm u}$ is electric field envelope of the pump pulses with a (linear) polarization $\polB_u$ (in tensor notation $\pol_u^{\nu}$) we have 
\begin{subequations}
\begin{align}
D_{g} &\propto -\int_{-\infty}^{t+\tau} dt'\int_{0}^{\infty} dt_1  e^{i\omega_u t_1} E_u^*(t') E_u(t'-t_1) \pol_u^{\nu_1} \pol_u^{\nu_2} \sum_{\xi_1,\xi_2}  (\mu^{\nu_1}_{\xi_1}  \mu^{\nu_2}_{\xi_2} + i k_u^{\nu}  \psi^{\nu,\nu_1,\nu_2}_{\xi_1,\xi_2}) \Gprop_{gg}^{\dagger}(t') \hat{V}_{g,\xi_2} \Gprop_{eg}(t_1) \hat{B}^{\dagger}_{\xi_1} \hat{\rho}_{eq} + h.c. \\
D_{e} &\propto \int_{-\infty}^{t+\tau} dt'\int_{0}^{\infty} dt_1 e^{i\omega_u t_1} E_u^*(t') E_u(t'-t_1) \pol_u^{\nu_1} \pol_u^{\nu_2} \sum_{\xi_1,\xi_2}  (\mu^{\nu_1}_{\xi_1}  \mu^{\nu_2}_{\xi_2} + i k_u^{\nu}  \psi^{\nu,\nu_1,\nu_2}_{\xi_1,\xi_2})  \Gprop_{ee}^{\dagger}(t')  [\Gprop_{eg}(t_1) \hat{B}^{\dagger}_{\xi_1} \hat{\rho}_{eq}]\hat{B}_{\xi_2} + h.c.
\;.
\label{eq:Doorway_fun}
\end{align}
\end{subequations}
Here $h.c.$ denotes the Hermitian conjugate,  $\hat{B}^{\dagger}_{\xi_j}\ket{g} =\ket{\xi_j}$, $\hat{B}_{\xi_j}\ket{\xi_k} =\delta_{\xi_j , \xi_k}$ and $\Gprop_{ab}(t)$ the time propagation operator for the sub-block $ab$ of the system (for example $ee$ is the first excited state manifold).  The time variable $t'$ is a result of the transformation displayed in \reffig{Fig:door_win}(b). As we only consider a linearly polarized pump, the chiral contribution will vanish for terms in the sum with $\xi_1 = \xi_2$ because $\psi^{\nu,\nu',\nu'}_{\xi,\xi} = 0$. The contributions for a circularly polarized pump would not vanish and we would need to take the conjugate of one of the polarizations in \refeq{eq:Doorway_fun} as has been considered by Cho~\cite{Cho2003}.  Moreover, we assume the pulses are well separated and hence we can take the limit of the integral over $t'$ to $+\infty$.

\subsubsection{Limit of a low frequency-bandwidth pump}

We first investigate the limit in which the frequency bandwidth of the pump is much narrower than the energy gaps between any exciton transition, but not so wide in time space that it overlaps with the probe (hence time ordering still applies).  In this regime the time-width of the pump is much greater than electronic dephasing times~\cite{MukamelPoNLS} and we can make the approximation $E_u^*(t')E_u(t'-t_1) \approx |E_u(t')|^2$. Assuming our exciton states $\ket{\xi}$ are good approximations to the true electronic eigenstates we have 
\begin{align}
D_{e} &\sim \int_{-\infty}^{\infty} d t'  \vert E_u(t') \vert ^2 \pol_u^{\nu_1} \pol_u^{\nu_2} \sum_{\xi}  \mu^{\nu_1}_{\xi}  \mu^{\nu_2}_{\xi}    \Gprop^{\dagger}_{ee}(t')  [\tilde{\Gprop}_{eg}(\omega_1) \hat{B}^{\dagger}_{\xi} \hat{\rho}_{eq}]\hat{B}_{\xi} + h.c.
\;.
\label{eq:Doorway_fun_long}
\end{align}
Here $\tilde{\Gprop}_{ge}$ is as defined in \refeq{eq:freq_prop_op}. When vibrational modes are strongly coupled to exciton states, leading to splitting of energy levels as shown in \reffig{Fig:system_states}, we would instead need to take $\ket{\xi}$ as the hybrid "vibronic" states which we discuss later.


 As the pump is frequency resolved in this limit, $\vert E(t') \vert^2$ varies quite slowly and hence $\Gprop^{\dagger}_{ee}(t')$ propagates the system for a significant time interval.   
The signal we receive is therefore effectively a weighted average (with a slowly varying weight function) of many different values of $\tau$. This has two important consequences. Firstly, all terms in the sum must have $\xi_1 = \xi_2$ because coherences between states of different energies will evolve in phase during the population time, leading to cancellation when time averaged. We therefore excite only a statistical mixture of the system eigenstates. A similar effect 
occurs due to the wavevector mismatch between the pump and probe as discussed in App.~E (cf. \refeq{eq:coh_sup}).  The difference here is that exciton coherences are always excited in individual complexes but the spatial variation causes cancellation in the total signal. Secondly, the chiral contribution in~\refeq{eq:Doorway_fun_long} has vanished. Note that the chiral contribution will still vanish when the chromophores have intrinsic magnetic dipole or electric quadrupole moments as $\psi^{\nu,\nu',\nu'}_{\xi,\xi} = 0$ (cf App.~\ref{App:trans_current}). Furthermore, in App.~\ref{App:beyond_condon} we show that $\psi^{\nu,\nu',\nu'}_{\xi,\xi} = 0$ remains true when we have transitions to different vibrational states after relaxing the Condon approximation.



\subsubsection{Limit of a short time-width (impulse) pump}

We now consider the \textit{impulsive limit}, opposite to the frequency resolved situation, the pump pulse is extremely short compared to electronic dephasing times (and therefore very broad in frequency) such that $E_u^*(t') E_u(t'-t_1) \approx |E_0|^2\delta(t')\delta(t_1)$. Within the Condon approximation, \refeq{eq:Doorway_fun} reduces to
\begin{subequations}
\begin{align}
D_{g} &\sim  -|E_0|^2 \pol_u^{\nu_1} \pol_u^{\nu_2} \sum_{\xi_1,\xi_2}  (\mu^{\nu_1}_{\xi_1}  \mu^{\nu_2}_{\xi_2} + i k_u^{\nu}  \psi^{\nu,\nu_1,\nu_2}_{\xi_1,\xi_2})  \hat{B}_{\xi_2} \hat{B}^{\dagger}_{\xi_1} \hat{\rho}_{eq} + h.c. \\
D_{e} &\sim |E_0|^2 \pol_u^{\nu_1} \pol_u^{\nu_2} \sum_{\xi_1,\xi_2}  (\mu^{\nu_1}_{\xi_1}  \mu^{\nu_2}_{\xi_2} + i k_u^{\nu}  \psi^{\nu,\nu_1,\nu_2}_{\xi_1,\xi_2})    [\hat{B}^{\dagger}_{\xi_1} \hat{\rho}_{eq}]\hat{B}_{\xi_2} + h.c.
\;.
\label{eq:Doorway_fun_short}
\end{align}
\end{subequations}
This form inevitably excites coherences between all states to which transitions are possible in the excited state manifold such that $D_{e}$ will evolve during the wait time $\tau$. The decomposition of $D_{e}$ into vibronic states is discussed in App.\ref{sec:imp_pump_vibronic}. On the other hand, the semiclassical Condon approximation we have made means that the nuclear (vibrational) degrees of freedom remain static during the short pump-length. This implies that the doorway function for the ground state hole $D_{g}$ will be proportional to the equilibrium state as $\Gprop_{gg}(\tau) \hat{B}_{\xi_2} \hat{B}^{\dagger}_{\xi_1} \hat{\rho}_{eq} = \delta_{\xi_1,\xi_2} \hat{\rho}_{eq}$. Then, in this view, $D_{g}$ is independent of time and has no chiral contribution.


In general, the Franck-Condon approximation is not completely adequate for describing the response of a system, as the vibrational modes will slightly affect dipole moments as indicated in~\refeq{eq:Pol_op}.  
In this situation, transitions from $\ket{g}_{el} \ket{n_1,n_2,\ldots}_{vib}$ to $\ket{e}_{el} \ket{\ldots,n_{\ell}+ \ell,\ldots}_{vib}$ will have a finite electric dipole moment $\muB_{g,n_{\ell};e,n_{\ell}+\ell}$.  
This leads to correction terms to $D_{g}$, which we denote $D'_{g}$, involving vibrational coherences that will beat during in the wait time.  These terms are related to resonant impulsive Raman scattering~\cite{Liebel2015}.  Our analysis presented App.~\ref{App:beyond_condon} show that even in this more general situation the chiral contribution to $D'_{g} = \sum_{n,n'} \ket{n}\bra{n'} \rho_D(n,n')$ will cancel completely in the impulsive limit. 
The key physical rationale underlying this result is the conservative nature of excited state CD which can be seen as follows. In the impulsive limit $\tilde{E}_u(\omega)$ is very broad and thus has roughly equal amplitudes over the entire range of excited state transition frequencies. The vanishing rotational strength \refeq{eq:rot_strength}, implies $\sum_{\xi}\mathbf{m}_{\xi} \cdot \muB_{\xi} =0$. Hence, we expect the Liouville pathways of the form $\ket{g,n}\bra{g,n} \to \ket{\xi,n}\bra{g,n} \to \ket{g,n+1}\bra{g,n}$ (similar to that shown in \reffig{Fig:Fman_diag}, but with both interactions occurring on the same side) to give a net contribution of zero.   More generally, cancellation of the ground state coherence contribution to the chiral signal in the impulsive limit still occurs due to the opposite signs for the forward and backward propagating terms in \reffig{Fig:Fman_diag}.  
For example, the pathways $\ket{g,n}\bra{g,n} \to \ket{e,n}\bra{g,n} \to \ket{g,n+1}\bra{g,n} $ and $\ket{g,n}\bra{g,n}  \to \ket{e,n+1}\bra{g,n}  \to \ket{g,n+1}\bra{g,n} $ both contribute to the same ground state coherence; the pathways have prefactors of $\muB_{e,1} \cdot \mathbf{m}_{e} -   \mathbf{m}_{e,1} \cdot \muB_{e}$ and $\muB_{e} \cdot \mathbf{m}_{e,1} - \mathbf{m}_{e} \cdot \muB_{e,1}$, which cancel exactly. We discuss this further in App.~\ref{App:beyond_condon}.

The above discussion for the chiral doorway in the impulsive limit shows the potential of time-resolved, impulsive chiral spectroscopy to probe excited state dynamics free of ground-state vibrational contributions that are difficult to prevent in other set-ups. In a more general experimental situation, the pump pulse will have a finite time width and lie between these two extremes.   
This will mean some dependence on the carrier frequency is present (allowing us to address particular transitions), but vibrational coherences can be formed in $D_{g}$ and also in the chiral component of $D'_{g}$. We however expect this effect to be small. Therefore in what follows we will neglect any effects due to the breakdown of the Condon approximation. 

\subsection{Chiral and non-chiral doorway functions}
\label{sec:Chiral doorway} 

\begin{figure}[t]
\centering\includegraphics[width=0.4\linewidth]{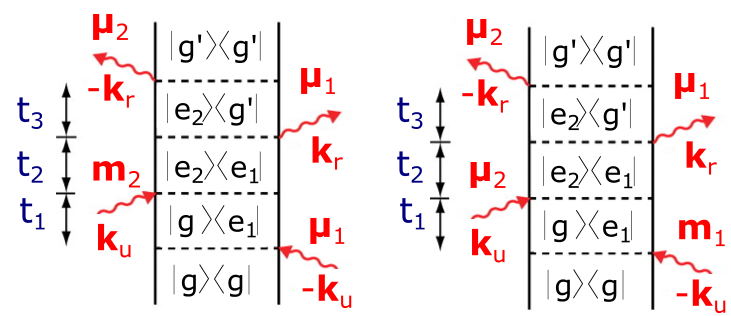}
\caption{Two Feynman diagrams contributing to a stimulated emission pathway in $S_{ChD}$.  Here $\ket{e_k}$ represent arbitrary electronically excited eigenstates and $\ket{g}$ and $\ket{g'}$ the ground electronic state with different vibrational quanta present. For large $t_2$, it is possible $\ket{\xi_1}\bra{\xi_2}$ mixes to a different coherence before the probe interaction as a result of bath interactions.}
\label{Fig:Fman_diag}
\end{figure}

We now proceed to show that chiral doorway functions in the excited state are in fact \textit{coherence specific} signals. We proceed by separating the doorway functions into chiral (C) and non-chiral (NC) components.  The chiral terms give a contribution to the output signal of the form 
\begin{equation}
S_{ChD}(\tau) =  \omega_r\langle\mathrm{Tr}[W^{(NC)}_{g} \Gprop_{gg}(\tau) D_{g}^{(C)}]\rangle + \langle\mathrm{Tr}[(W'^{(NC)}_{f}+W^{(NC)}_{e}) \Gprop_{ee}(\tau) D_{e}^{(C)}]\rangle \;,
\label{eq:chiraldoorsig}
\end{equation}
with $W^{(NC)}_{\alpha}$ the non-chiral window functions, which can be found, for example, in Mukamel (2006)~\cite{MukamelPoNLS}. In \reffig{Fig:Fman_diag} we show a typical stimulated emission pathway contributing to $S_{ChD}$. The effective magnetic dipole interaction can occur on either the forward or backward interaction; with a linearly polarized pump, this results in terms with opposite sign and hence full cancellation for $\xi_1 = \xi_2$. 
For a general profile $E_u(t)$ real we have
\begin{subequations}
\begin{align}
D_{g}^{(C)} &\propto -i\int_{-\infty}^{\infty} dt'\int_{0}^{\infty} dt_1  E_u(t') E_u(t'-t_1) \pol_u^{\nu_1} \pol_u^{\nu_2} k_u^{\nu} \sum_{\xi_1,\xi_2}    \psi^{\nu,\nu_1,\nu_2}_{\xi_1,\xi_2} \Gprop^{\dagger}_{gg}(t') ( e^{i\omega_u t_1} \hat{B}_{\xi_2} \Gprop_{eg}(t_1) \hat{B}^{\dagger}_{\xi_1} \hat{\rho}_{eq} - h.c.) \\
D_{e}^{(C)} &\propto i\int_{-\infty}^{\infty} dt'\int_{0}^{\infty} dt_1 E_u(t') E_u(t'-t_1) \pol_u^{\nu_1} \pol_u^{\nu_2}k_u^{\nu}\sum_{\xi_1 , \xi_2}    \psi^{\nu,\nu_1,\nu_2}_{\xi_1,\xi_2}  \Gprop^{\dagger}_{ee}(t') (e^{i\omega_u t_1} [\Gprop_{eg}(t_1) \hat{B}^{\dagger}_{\xi_1} \hat{\rho}_{eq}]\hat{B}_{\xi_2} - h.c.)
\;.
\label{eq:chiral_doorway}
\end{align}
\end{subequations}
Notice that in $D_{e}^{(C)}$ the term $e^{i\omega_u t_1} [\Gprop_{eg}(t_1) \hat{B}^{\dagger}_{\xi_1} \hat{\rho}_{eq}]\hat{B}_{\xi_2}$  is subtracted from its Hermitian conjugate (and similarly for $D_{g}^{(C)}$); if these matrices have real numbers on the diagonal these will cancel out exactly, leaving only off diagonal elements i.e.~coherences. This chiral density matrix can be traceless (it is only a component of $\hat{\rho}_{e}$) but the global imaginary prefactor of $i$ means it is Hermitian. To illustrate these points consider the simplest system with two electronic excited states $\ket{1}$ and $\ket{2}$ coupled to a vibrational bath described by $\hat{\rho}_{\rm vib}$, initially at thermal equilibrium. For an impulsive pump pulse we have via \refeq{eq:Doorway_fun_short}
\begin{equation}
D_{e} \sim 2\{ \underbrace{|A|^2\ket{1}\bra{1}+|B|^2\ket{2}\bra{2} + AB(\ket{1}\bra{2} + \ket{2}\bra{1})}_{D_e^{(NC)}} + i \underbrace{C(\ket{1}\bra{2} - \ket{2}\bra{1})}_{D_e^{(C)}} \} \otimes \hat{\rho}_{\rm vib} \;,
\end{equation}
with $A = E_0 (\polB_u \cdot \muB_1)$,  $B = E_0 (\polB_u \cdot \muB_2)$ coming from the non-chiral terms, and $C = |E_0|^2\pol_u^{\nu_1} \pol_u^{\nu_2} k_u^{\nu} \psi^{\nu,\nu_1,\nu_2}_{1,2}$ from the chiral interactions. Moreover we have $D_{g}\sim  -2(|A|^2+|B|^2) \hat{\rho}_{eq}$ with no chiral contribution. The key issue here is that the chiral contribution in the excited state $D_e^{(C)}$ is traceless and therefore the signal component $S_{ChD}$ in \refeq{eq:chiraldoorsig} is \textit{coherence specific} and has no contribution from ground state hole as discussed in the previous section.  Together with the arguments presented for the impulsive regime of the chiral doorway, our analysis shows the potential advantages of this technique to probe excited state coherences.

\subsection{Chiral window functions}
{Besides the the chiral doorway function, we also have the contribution from the chiral window function
\begin{equation}
S_{ChW}(\tau) = \omega_r \langle \mathrm{Tr}[W^{(C)}_{g} \Gprop_{gg}(\tau) D_{g}^{(NC)}] \rangle + \langle\mathrm{Tr}[(W^{(C)}_{f}+W^{(C)}_{e})\Gprop_{ee}(\tau) D_{e}^{(NC)}] \rangle \;.
\label{eq:chiralwindowsig}
\end{equation}
Here we explicitly show the chiral component of the GSB window function $W^{(C)}_{g}$ while the remaining components  $W^{(C)}_{e}$ and$W^{(C)}_{f}$ are presented in Appendix~\ref{App:window}.
\begin{equation}
W_{g}^{(C)} \propto \mathrm{Re} \int_{-\infty}^{\infty} dt \int_{0}^{\infty} dt_3 \; e^{i \varphi + i\omega_{\rm LO} t_3 + i(\omega_{\rm LO}-\omega_{r}) t} E_{\rm LO}^*(t+t_3) E_r(t) \sum_{\xi_3,\xi_4}   k_r^{\nu} p_{\rm LO}^{\nu_4} p_r^{\nu_3} \psi^{\nu,\nu_3,\nu_4}_{\xi_3,\xi_4}  \hat{B}_{\xi_4} \Gprop_{ge}(t_3) \hat{B}^{\dagger}_{\xi_3} \Gprop_{gg}(t)  \;,
\label{eq:Chiral_window_gg}
\end{equation}
with the $\mathrm{Re}$ indicating the real part, which should be taken after combining this expression with the doorway function.  
In a CD setup, $\polB_r =\polB_{L/R} = (\hat{x} \pm i \hat{y})/\sqrt{2}$ and $\polB_{\rm LO} = (\polB_r)^*$, hence the contraction over $\psi^{\nu,\nu_3,\nu_4}_{\xi_3,\xi_4}$ will be non-zero when $\xi_3 = \xi_4$. We note that if the first two interactions came from different pulses with different wavevectors (as would be the case in 2DS), we would have more complicated expressions which cannot be expressed just using the factor $\psi$.  

We consider frequency resolved detection with an ultrafast probe pulse $E_{r}(t) \sim  E_{r0}\delta(t)$; the signal with a frequency component $\omega_{LO}$ can be obtained by setting the amplitude of that frequency component in the probe pulse i.e. $E_{\rm LO}(t) =  \tilde{E}_{r}(\omega_{LO}-\omega_r)$ and the relative  phase between LO and probe to be zero i.e. $(\varphi=0)$ in \refeq{eq:Chiral_window_gg}.  The modified window function after considering difference between left $(\polB_L)$ and right $(\polB_R)$ circularly polarized light becomes:
\begin{equation}
\tilde{W}_{g}^{(C)} \sim \mathrm{Re} \; \tilde{E}^*_{r}(\omega_{LO}-\omega_r) E_{r0}   \sum_{\xi_3,\xi_4}   k_r^{\nu} (p_{L}^{\nu_4} (p_L^*)^{\nu_3} - p_{R}^{\nu_4} (p_R^*)^{\nu_3})\psi^{\nu,\nu_3,\nu_4}_{\xi_3,\xi_4}  \hat{B}_{\xi_4} \tilde{\Gprop}_{ge}(\omega_{\rm LO}) \hat{B}^{\dagger}_{\xi_3}   \;.
\label{eq:Chiral_window_gg_short}
\end{equation}
The factor of $\tilde{E}^*_{r}(\omega_{LO}-\omega_r)$ is included here for completeness, but we will ignore this term in the numerical calculations as it can simply be scaled out.  \refeq{eq:Chiral_window_gg_short} represent a configuration that achieves the best possible time and frequency resolution and is also the most simple theoretically.

Other experimental geometries for TRCD can still be described by \refeq{eq:Chiral_window_gg}. A setup with non-frequency resolved detection would be described by $\mathbf{E}_{\rm LO}(t) \to \mathbf{E}_{r}(t)$ and $\omega_{LO} \to \omega_{r}$.  Alternative configurations like that of Niezborala (2007)~\cite{Niezborala2007} involve manipulating the output light by using polarizing beam splitters to select output light. This is equivalent to a taking a linearly polarized probe and a local oscillator polarized orthogonal to this at relative phase of $\varphi = \pi/2$ (taking $\varphi = 0$ would measure the transient optical rotation instead). }

\subsection{Isotropic averaging and separation of chiral doorway and window components}
\label{ssec:Iso_av}
\subsubsection{Linearly independent contributions to the signal}
As the system is isotropic we must average the signals over all possible sample orientations. There are six linearly independent chiral contributions to the third order response tensor~\cite{Abramavicius2006} (although additional degrees of freedom are associated with the wavevectors); from these, only three are ever required for pump probe with a linearly polarized pump.  As the system is isotropic, we can simply fix the probe direction along the $z$ axis without any loss of generality. Denoting the polarizations of the pump, the probe and the local oscillator in a bracket as $[\polB_{\rm pump},\polB_{\rm probe},\polB_{\rm LO}]$, we can access the three independent configurations with polarizations $[x,y,x]$, $[y,y,x]$ and $[z,y,x]$.  The TRCD signal is an average of the $[x,y,x]$ and $[x,x,y] \equiv [y,y,x]$ signals. There is also an additional degree of freedom relating to angle between the wavevectors of the pump and probe $\theta$ (see Fig. 1(a)), hence we must also consider colinear and non-colinear contributions (where possible) to cover all possibilities.  Note that the pump-probe angle $\theta$ lies in the plane orthogonal to $\polB_{\rm pump}$.

\subsubsection{Relevant averages for TRCD}
In order to calculate the TRCD signals with our method, we have to compute the isotropic average of the three electric dipole transitions and one effective magnetic transition dipole, for every possible pathway in Liouville space.  These can be calculated via fourth rank averages (see for example \cite{Wagniere1982}).  We derive the averages in appendix \ref{sec:ori_av} and summarize the key results here. 

For compactness we denote 
$\mu_{jk} = \muB_{\xi_{j}} \cdot \muB_{\xi_{k}}$ and $m_{jk} = \mathbf{m}_{\xi_{j}} \cdot \muB_{\xi_{k}}$.  The average of a pathway contributing to the chiral doorway contribution $S_{ChD}$ is of the form 
\begin{equation}
\mathrm{Av}(\xi_1,\xi_2,\xi_3,\xi_4)_{ChD} = \frac{\mathbf{k}_u \cdot \hat{z}}{12} [ (m_{13}\mu_{42}  -m_{24}\mu_{13} ) - (m_{23}\mu_{41}  -m_{14}\mu_{23} )] \;,
\label{eq:CP_diff}
\end{equation}
with $\hat{z}$ a unit vector in the direction of the probe. This average vanishes when $\xi_1 = \xi_2$ as expected, indicating that this is indeed a coherence specific pathway.  
The equivalent average for the chiral window $S_{ChW}$ is calculated to be
\begin{equation}
\mathrm{Av}(\xi_1,\xi_2,\xi_3,\xi_4)_{cW} = \frac{|k_r|}{60} [6 \mu_{12} (m_{34} + m_{43}) - (\mu_{24}m_{31} +\mu_{14}m_{32} +\mu_{23}m_{41} +\mu_{32}m_{42} ) ] \;.
\label{eq:CP_diff2}
\end{equation}
Notably, \refeq{eq:CP_diff2} is independent of the angle between the pump and probe but \refeq{eq:CP_diff} is not and, in fact, vanishes if the pump and probe are orthogonal. It is precisely this dependence that will allow us to separate the chiral-window and chiral-doorway contributions. The numerical factor of $1/60$ is a combination of the $1/30$ factor in fourth rank averages and $1/2$ from the circular $x$ and $y$ components of the circular polarization, the components in \refeq{eq:CP_diff} sum together with a factor of 5 leading to cancellation.

\subsubsection{Isolating chiral-doorway and chiral-window contributions by manipulating the angle between the pump and probe}

One of the most important consequences of the relations presented in equations~\ref{eq:CP_diff} and \ref{eq:CP_diff2} is that manipulation of the angle $\theta$ between the pump and the probe allows to obtain the chiral doorway signal. Specifically, the chiral doorway function ca be obtained by computing the difference between the TRCD signals obtained via a colinear and a orthogonal pump configurations.

More generally, we can consider taking two otherwise identical measurements with the probe traveling along the $z$ axis, but with the pump (linearly polarized in the $x-y$ plane) at angles $\theta_1$ and $\theta_2$ to the z-axis. Neglecting any differences in the overlap of the paths, these two signals can be broken down as $S_1 = \cos(\theta_1)S_{ChD} +  S_{ChW}$ and $S_2 = \cos(\theta_2)S_{ChD} +  S_{ChW}$.  To extract the two chiral contributions we can therefore combine these signal as
\begin{subequations}
\begin{align}
S_{ChD} &= \frac{S_1- S_2}{\cos(\theta_1)-\cos(\theta_2)}  \\
S_{ChW} &= \frac{S_1 \cos(\theta_2) - S_2 \cos(\theta_1)}{\cos(\theta_2)-\cos(\theta_1)} \;.
\label{eq:chiral_cmps}
\end{align}
\end{subequations}
Both the non-chiral and $S_{ChW}$ contributions are assumed independent of the pump angle for this direct subtraction to work.  Within this approximation, the difference in signals with a forward/backward propagating pump pulse would provide ideal contrast. 
In reality, the wavevector mismatch also affects the signal by turning it into a convolution over a range of $\tau$ values, as we discuss in App.~\ref{App:spat_overlap}.  In practical terms it may therefore be better to compare two (or more) signals with smaller angle differences.     
Fortunately this convolution effect will be the same for the chiral and non-chiral components, hence the change in the non-chiral signals (which have much better signal to noise) could be used as a calibration measure.  We can therefore re-scale / numerically compensate for this effect before making the subtraction.  

\section{Results}
\label{S:results}
\subsection{Example system}

In this section we examine theoretical TRCD signals for our example system, consisting of an electronic coupled dimer subject to the influence of a thermal  bath with spectral density of fluctuations  that includes an overdamped and a well resolved, underdamped vibrational mode as discussed in Sec. \ref{SS:1}.  Systems of this kind are expected to show signatures of vibronic states leading to long-lived excitonic coherences \cite{Kolli2012,Christensson2012, Chin2013}. We assume the probe pulse is circularly polarized and is of an extremely short time width. The detection is assumed to be frequency resolved. We start by discussing the results when the pump is short enough to be considered impulsive. In particular, we examine the differences between the non-chiral (standard pump probe signal) and the chiral doorway and chiral window contributions discussed Sec. \ref{S:2}. We then move on to analyzing the situation where the pump is of finite time and we can achieve frequency resolution.

\begin{figure}[t]
\centering\includegraphics[width=0.9\linewidth]{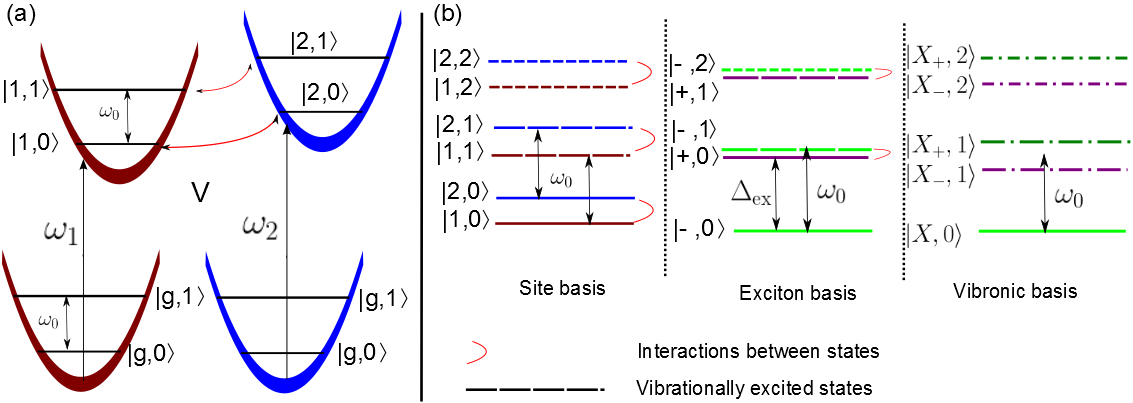}
\caption{(a) Energy landscape of the two chromophores in the ground and excited states (b) Energy levels (vertical scale) of the system in the site, exciton and vibronic basis when the anti-correlated component of the vibrational mode ($\hat{b}_R$ in \refeq{eq:norm_modes}) is included in the Hamiltonian.  Red arrows indicate strong coupling between states in the Hamiltonian.}
\label{Fig:system_states}
\end{figure}

The electronic parameters for the dimer with Hamiltonian as given in\refeq{eq:Hsys} are $\hbar\omega_1 = 12328$cm$^{-1}$, $\hbar\omega_2  = 12472$cm$^{-1}$ and $V_{1,2} = 70.7$cm$^{-1}$.  The energy gap between the upper ($\ket{+}$) and lower ($\ket{-}$) excitons states, $\Delta_{\rm ex}\sim 202$cm$^{-1}$, is close to the energy quanta of the underdamped mode in \refeq{eq:spec_den} i.e. $\hbar\omega_0\sim 222$cm$^{-1}\sim\Delta_{\rm ex}$. The quantum interaction between the electronic excitations and this well resolved vibration leads to hybridized exciton-vibration states that enables non-exponential population transfer between exciton states i.e. coherent energy transfer \cite{OReilly2014}. In our numerical calculations we compute non-perturbative electronic dynamics via the HEOM as outlined in Appendix~\ref{sec:num_meth}. To gain a better understanding of the relations between the reduced dynamics for the excitonic system and the transitions between exciton-vibration states, in Appendix~\ref{sec:Vibronic} we reformulate the problem by explicitly considering the exciton-vibration Hamiltonian. Within this approach, we use vibronic eigenstates $\ket{X_{\pm},n}$ (defined in \refeq{eq:vibronic_def}) which are a mixture of $\ket{+}$ with $n-1$ quanta in the vibrational mode and $\ket{-}$ with $n$ quanta in the mode (note $\ket{X,0}$ is just the lower exciton state $\ket{-}$).  The states $\ket{X_{\pm},1}$ have energies $230$cm$^{-1}$ and $193$cm$^{-1}$ larger than $\ket{X,0}$.  This is shown schematically in \reffig{Fig:system_states}.

The remaining parameters relating to the spectral density are $\lambda_D = 20$cm$^{-1}$ and $\gamma_D = 630$cm$^{-1}$ for reorganization energy and cutoff frequency for the Drude mode and $\lambda_B = 4.4$cm$^{-1}$ and damping $\gamma_B = 19$cm$^{-1}$ for the near resonant underdamped mode. The sample temperature is assumed to be cryogenic $T=77$K and no static disorder in energy levels or dipole moments is included.  These parameters are chosen to be similar to a dimer system of the  FMO complex as investigated in~\cite{Plenio2013}. For simplicity we have considered a larger value for the cutoff frequency $\gamma_D$ such that the overall decay of exciton-vibration coherences is Markovian thereby simplifying the analysis. The choice of a cryogenic temperature reduces broadening but is not essential for the techniques we propose.

\subsection{Chiral signals with an impulsive pump}
As we discussed in section section IIIB, in the limit of an impulsive pump the chiral doorway component of the ground state bleaching vanishes. Furthermore, we ignore the time-dependence of GSB in the chiral window component of the signal $S_{ChW}(\tau)$. This means that in our case the chiral  doorway and window functions given in equations \refeq{eq:chiraldoorsig} and \refeq{eq:chiralwindowsig} become:
\begin{subequations}
\begin{align}
S_{ChD}(\tau) &\simeq \omega_r \langle\mathrm{Tr}[(W'^{(NC)}_{f}+W^{(NC)}_{e}) \Gprop_{ee}(\tau) D_{e}^{(C)}]\rangle \\
S_{ChW}(\tau) & \simeq \omega_r \left(\langle \mathrm{Tr}[W^{(C)}_{g} \Gprop_{gg}(0) D_{g}^{(NC)}] \rangle +\langle\mathrm{Tr}[(W^{(C)}_{f}+W^{(C)}_{e})\Gprop_{ee}(\tau) D_{e}^{(NC)}]\rangle\right)
\end{align}
\end{subequations}
where chiral component $D_{e}^{(C)}$ can be extracted from  \refeq{eq:Doorway_fun_short}.  We now proceed to describe our numerical results for the $S_{ChW}(\tau)$ and $S_{ChD}(\tau)$ in our example system.

\begin{figure}[ht]
\centering\includegraphics[width=0.7\linewidth]{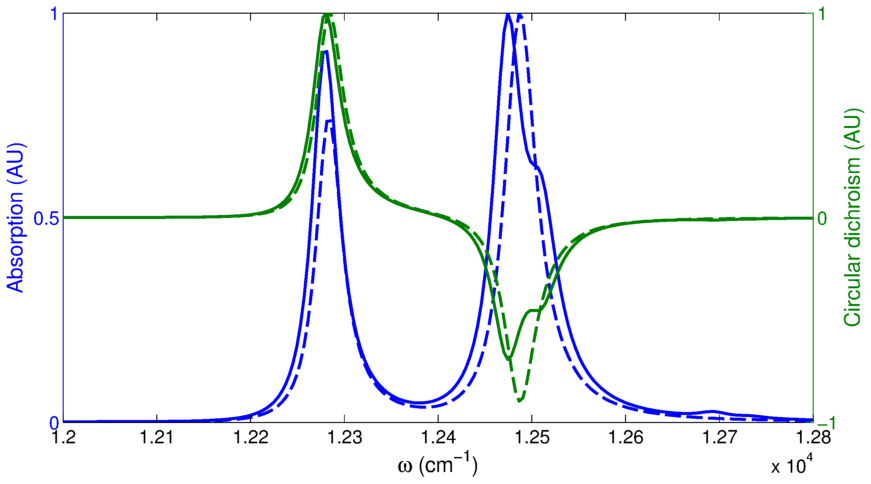}
\caption{Linear absorption (left axis) and CD signal (right axis), both normalized to a magnitude of unity, with/without the underdamped mode present (solid/dotted lines).  Angular frequency is scaled by a factor of $2 \pi c$. The mode red shifts the lower exciton peak and the coupling splits the upper exciton peak into two components. An extra peak appears in the absorption relating to higher vibronic states, but contributes only weakly to the CD signal.}
\label{Fig:linear_CD}
\end{figure}

\begin{figure}[ht]
\centering\includegraphics[width=0.9\linewidth]{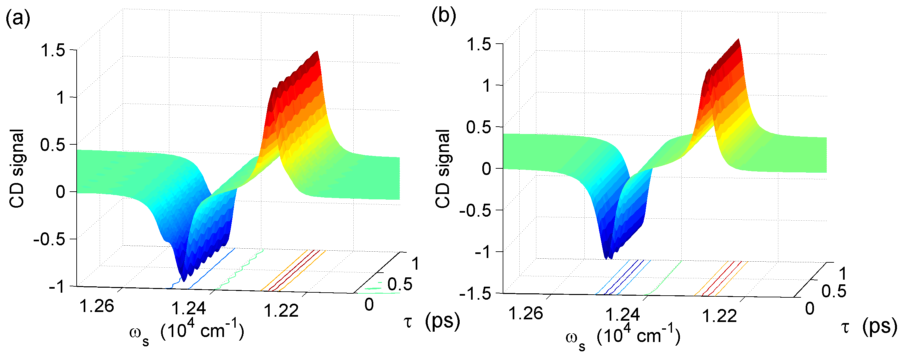}
\caption{TRCD signal from a noncolinear configuration (equal to the chiral window signal $S_{ChW}$) with an impulsive pump and (a) the full spectral density and (b) just a Drude spectral density (no underdamped mode).  Some oscillation is visible in (a) where as (b) appears relatively constant indicating the mode is responsible for prolonging coherent beating in this system.  Angular frequency is scaled by a factor of $2 \pi c$. }
\label{Fig:IP_cW}
\end{figure}

\subsubsection{Chiral window contributions}
The total signals from the non-colinear configuration are shown in \reffig{Fig:IP_cW} with the underdamped mode included (excluded) in a (b).  This contributes only to $S_{ChW}$ as defined in~\refeq{eq:chiral_cmps} because $S_{ChD}$ vanishes due to isotropic averaging in this geometry, as can be seen in \refeq{eq:CP_diff2}.  Note that this CD signal is scaled relative to maximum non-chiral signal (given by the average left and right absorption) multiplied by $\lambda_{typ}/(2 \pi R_{12})$ to allow for arbitrary separation between our chromophores. We have chosen  $\lambda_{typ} = 806$nm as the mean signal wavelength considered. If we take $R_{12} \sim 2.6$nm, the maximum amplitude in \reffig{Fig:IP_cW} would be about 1/100 of the non-chiral signal. 

The signals in both \reffig{Fig:IP_cW}(a) and \reffig{Fig:IP_cW}(b) are somewhat masked by the GSB component. The GSB is constant due to the fast pump approximation and is very similar in shape to the linear CD spectra shown in \reffig{Fig:linear_CD}, with a small relative difference ($<10^{-3}$ of the relative signal) due to changes in the dipole averaging. In \reffig{Fig:IP_cW}(b) the signal is almost constant after $~250$fs, whereas in \reffig{Fig:IP_cW}(a), where the underdamped mode is included, oscillations are visible even at long times.  While oscillations must arise purely from the excited state contributions due to the impulsive pump limit in our model, it is still possible for coherences between states which differ only in vibrational degrees of freedom to contribute to this chiral window signal. 

\subsubsection{Chiral doorway contributions}

We next examine the chiral doorway contribution $S_{ChD}$ by subtracting the signal of the non-colinear configuration (pump along $y$ probe along $z$) from the signal from the colinear configuration.  Or more generally via the relationship in \refeq{eq:chiral_cmps}. In an experiment, noise and changes to the spatial overlap of the pump and probe pulse may make a direct subtraction unfeasible. As such extracting $S_{ChD}$ may require looking at the emergence of new or modified amplitude beating peaks when the pump angle is changed.

Within the fast (impulsive) pump limit, the only terms which contribute to $S_{ChD}$ are proportional to coherences of the type $(\ket{\xi_1}\bra{\xi_2}-\ket{\xi_2}\bra{\xi_1})$, which we show can be decomposed into different vibronic states in App.~\ref{App:vibronic_derivation}. Notably we have no contribution from purely vibrational coherences, which have been shown to dominate 2DS signals in FMO~\cite{Tempelaar2014}. 

As contributions only come from terms which undergo a quantum beating during the population time $\tau$, this is referred to as coherence specific contribution.  This is a novel feature of the chiral spectroscopy, as it is not possible to extract coherence specific features in ordinary pump-probe setups.  

\begin{figure}[ht]
\centering\includegraphics[width=0.9\linewidth]{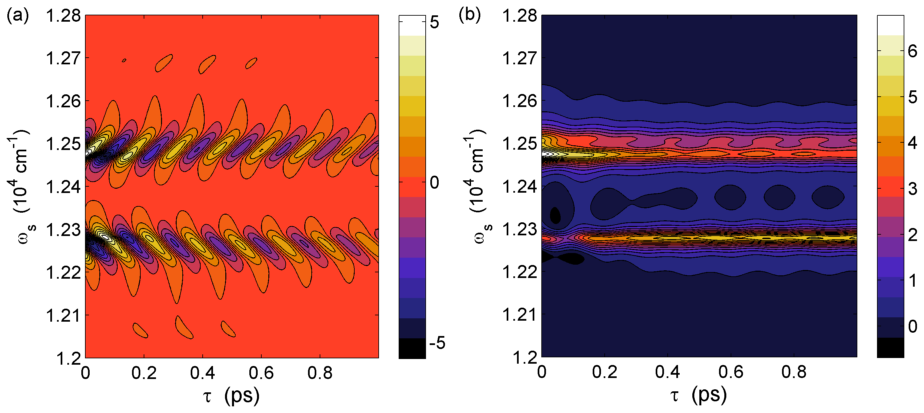}
\caption{(a) Excited state contribution to the $S_{ChD}$ (doorway) signal component, defined in \refeq{eq:chiral_cmps}, multiplied by a factor of 100.  (b) Ordinary pump-probe signal with the pump and probe linearly polarized along the x-axis.  The non-chiral pump-probe signal included dynamics from population evolution which complicates the interpretation. Note that both signals can be obtained simultaneously as the difference/sum of the left and right absorption.}
\label{Fig:SP_EScD}
\end{figure}

The doorway contribution $S_{ChD}$ is plotted in \reffig{Fig:SP_EScD} (a) along with the non-chiral pump-probe signal in \reffig{Fig:SP_EScD} (b). The lines of oscillating contours come from the beating coherences. Two distinct branches of oscillating peaks are visible at $\omega_s \sim 12500$cm$^{-1}$ and $\omega_s \sim 12250$cm$^{-1}$ Feynman diagram analysis shows each branch will have contributions from both SE and ESA pathways due to the fact that both rephasing and non-rephasing contributions are present in a pump-probe signal.  

The sustained oscillations in \reffig{Fig:SP_EScD} (a) can be explained as originated from a coherence between the lower exciton state and a vibronic state~\cite{Halpin2014} because the coherence time is longer than would be excepted for pure exciton coherences and contributions from pure vibrational coherences are not here. In \reffig{Fig:SP_EScD} (b) the oscillations originate from a coherence between the lower and upper exciton states, which decay much faster.  The side bands visible at the end of the frequency range are associated to transitions between different harmonic oscillator energy levels of the center-of-mass mode (c.f. App.~\ref{App:rel_mode_co}), which will differ by around $\omega_0$ or can also correspond to transitions to higher energy vibronic states.

In order to better understand these oscillations, we consider particular frequency slices of these plots and perform a Prony decomposition~\cite{Hauer1991} as shown in App.~\ref{App:Prony}.  The simplicity of the coherence specific signal allows for an easier decomposition and a more accurate determination of the beat frequencies than in a non-chiral impulsive pump-probe signal.

\subsection{Frequency resolution with finite width Gaussian pulses}

The opposite limit to the ultra-fast time resolution setup is the frequency resolved configuration, in which the carrier frequency of the pump is well defined.  Within this limit, a Gaussian pulse cannot excite coherences between the states with different energies and only population type transfers would be possible, completely eliminating the chiral doorway contribution $S_{ChD}$ and hence giving no difference in the chiral signals from the colinear and non-colinear geometries. 

Since we are interested in measuring signals from coherences, we consider an intermediate and more experimentally relevant limit in which the pump pulse has a finite duration and frequency bandwidth. The probe is still assumed to be short with frequency resolved detection employed.  Such a signal is calculated from our data using \refeq{eq:PP_sig}.  Our pump pulse is therefore described by
\begin{equation}
E_u(t) = E_u(0) \exp\left(-\frac{t^2}{4\sigma_{u}^2}\right) \;,
\end{equation}
note that $\sigma_u$ is the standard deviation of the intensity profile $|E_u(t)|^2.$

\begin{figure}[ht]
\centering\includegraphics[width=0.95\linewidth]{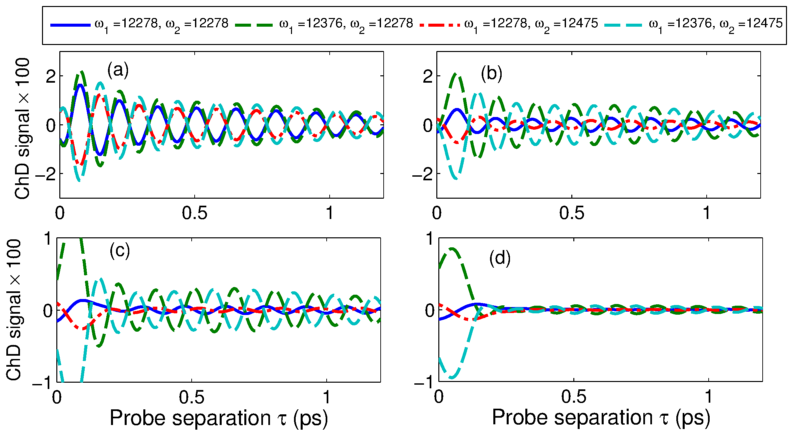}
\caption{Full contribution to the $S_{ChD}$ (chiral doorway) signal component defined in \refeq{eq:chiral_cmps} with pulse widths $\sigma$ of a) $25$fs, b) $50$fs, c) $75$fs and d) $100$fs.  Oscillations become less visible as the pulse time-width increases because insufficient bandwidth is available to excite coherent superposition of states and the lines with different pump carrier frequencies become distinct.  Note these labels denote $\omega / 2 \pi c$ in inverse cm units. The signal in c) and d) also decay with time faster than typical vibronic coherences, indicating they have a different origin.}
\label{Fig:SP_cD_finitew}
\end{figure}

As the pump pulse is of a finite duration, the coherences will give a suppressed contribution to the total signal, but must still be accounted for.  The chiral doorway signal $S_{ChD}$ is plotted in~\reffig{Fig:SP_cD_finitew} for a range of Gaussian pulse standard deviations $\sigma_{pulse}$ and a carrier frequencies.  
 The time widths $\sigma_{u}=25,50,75,100$fs correspond to frequency FWHM of $500, 250, 167,  125$cm$^{-1}$ (or standard deviations of $\tilde{\sigma}_u= 212,106, 70.8, 53$cm$^{-1}$) respectively. 
The initial rise in~\reffig{Fig:SP_cD_finitew} is due to the assumption of strict time ordering. In reality, there are additional contributions with the probe giving rise to the 1st or 2nd interaction instead of the 3rd, and therefore also a coherence component, sometimes unhelpfully referred to as the coherence artifact~\cite{MukamelPoNLS}.  Therefore these graphs are only valid at times larger than around $2\sigma_u$.

When the pump carrier frequency is set at $\hbar \omega_1 = 12278$cm$^{-1}$, resonant with the lower exciton state $\ket{-}\equiv \ket{X,0}$, the transitions to the two vibronic states $\ket{X_{\pm},1}$ (predicted to be around $193$cm$^{-1}$ and $230$cm$^{-1}$ higher) lie in the frequency tail of this pulse. It is therefore possible to excite the coherences between the vibronic states and the ground exciton state, denoted $\ket{X_{\pm},1}\bra{X,0}$, but with decreasing amplitudes as the pulses get longer. The $50$fs pulse has a FWHM of $250$cm$^{-1}$ and so can non-negligibly excite such a coherence. However, the contribution to our signal is reduced by a factor of $\exp(\Delta E^2/4\hbar^2 \tilde{\sigma}_u^2)$, the relative amplitude of $\tilde{E}(\omega)$ at the frequency of the upper exciton transition. With our parameters, the signal reduces to around $\exp(-230^2/[4 \times 106^2]) \sim 0.3$ of the impulsive pump limit. When the carrier frequency is set at $\hbar\omega_1 = 12376 $cm$^{-1}$, which lies at the mid point between transitions to $\ket{-}$ and $\ket{X_{-},1}$, both transitions will be slightly off resonant. In this case the signal is reduced to around $\sim \exp(-(132^2+98^2)/[4 \times 106^2]) \sim 0.55$ of the impulsive pump limit, thus giving more signal as shown in~\reffig{Fig:SP_cD_finitew}(b).  This analysis is only approximate since effects like transition broadening are not taken into account.

This effect is even more pronounced for the longer pulses in~\reffig{Fig:SP_cD_finitew}(c) and (d), with significant drops in signal and more pronounced differences between the two carrier frequencies. We also note the coherence  $\ket{X_{-},1}\bra{X,0}$ will be preferentially excited (as $\ket{X_{-},1}$ is lower in energy than $\ket{X_{+},1}$). This coherence decays faster than $\ket{X_{+},1}\bra{X,0}$ [visible in \reffig{Fig:prony_cD}(a) in App.~\ref{App:Prony}], hence the apparently more rapid signal decay. We also note that a small thermal population is present in vibrational excited electronic ground state $\ket{g,1}$ can potentially allow the $\ket{X_{+},1}\bra{X_{-},1}$ coherence to contribute when the lower carrier frequency is used and the pulses are narrow band.

\begin{figure}[ht]
\centering\includegraphics[width=0.95\linewidth]{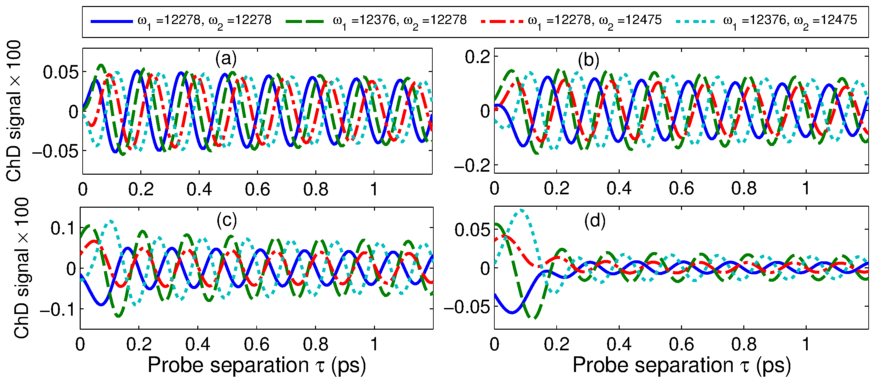}
\caption{GSB contribution to the $S_{ChD}$ (doorway) signal component, defined in \refeq{eq:chiral_cmps}, for pulse widths $\sigma$ of a) $25$fs, b) $50$fs, c) $75$fs and d) $100$fs.  The contribution is much less than that from the excited state, with the minimum difference being an order of magnitude at $\sigma = 50$fs.}
\label{Fig:GScD_finitew}
\end{figure}

The chiral doorway function will now have a finite contribution from the ground state hole, which consists of purely vibrational coherences. This contribution is plotted in \reffig{Fig:GScD_finitew}. Initially the contribution grows as the pulse width increases since the coherence time is longer and the exciton-vibration coupling allows vibrational coherences to form.  However, in the range we are showing, the contribution decreases as the pulse width grows because the frequency range is too narrow to excite coherences (the same reason as the excited state contribution).  This is still a fairly minor contribution to the overall signal (difference).

By manipulating carrier frequency and pulse width it should be possible to identify the energies of states which are responsible for the beatings observed in the signals. In order to select coherences between particular states one can extend the technique to include chirped pulses that effectively have two carrier frequencies.

\subsection{Comparison with other techniques}

Coherence specific polarization configurations in 2D optical spectroscopy are possible~\cite{SchlauCohen2011}, but present additional experimental difficulties due to the number of independent pulse polarizations which must be controlled.  Pump-probe experiments are also generally easier to perform than 2D spectroscopy as there is no need to control relative phase (no phasing  problem) and no Fourier artifact from measurement.  Additionally, the ability to use a short pump pulse allows us to limit the chiral contribution of ground state vibrational coherences, which may otherwise mask quantum beats from vibronic states.

Pump-Probe Polarization Anisotropy spectroscopy has also been used to study the evolution of coherences within multichromophore systems~\cite{Savikhin1997,Edington1995,Smith2011}.  However this requires specific geometrical properties of the chromophores, namely degenerate perpendicular transition dipole moments and thus it is not fully general~\cite{Collini2009}. Excitonic TRCD requires chirality, which for a dimer system of chromophores means the electric transition dipole moments and the displacement vector between the centers of each chromophore are not parallel and do not lie in the same plane. 

There are two primary challenges associated with TRCD bases spectroscopy.  Firstly we expect a 3 - 4 fold reduction in signal compared with non-chiral techniques~\cite{ParsonMOS}.  Secondly the positions of chromophores may also drift relative to one another, on timescales much longer than measurements, and in ways that are not well understood. This can lead to cancellation of certain signals from an ensemble measurement and generally complicate analysis. Moreover, circular polarization can introduce experimental complexity, though alternative methods exist to overcome this~\cite{Niezborala2007}.

\section{Conclusions}
\label{S:conclusion}
We investigated the different contributions to the time-resolved circular dichroism signal from a system of electronically coupled chromophores. Our formalism separates out doorway ($S_{ChD}$) and window ($S_{ChW}$) components to this chiral signal. This distinction is useful as the chiral doorway component is dependent on the presence of coherences within the non-equilibrium electronic density matrix formed after the interaction with the pump pulse. Comparing signals from two experimental geometries enables us to isolate $S_{ChD}$ thereby directly probing excited state coherences that beat sinusoidally in the pump-probe delay time. Our numerical results focus on the example of a dimer system, similar to a subsystem found within the FMO complex.  We have shown that for this system the oscillations due to the vibronic coherences excited by a pump pulse are visible in the time-domain (short pump pulse) transient CD signal, free from any ground state contribution, and  therefore provide evidence of exciton coherence being mediated by hybrid exciton-vibration states. We also show that vibrational coherences with no coupling to excitons do not contribute to this signal.

In the frequency domain (well resolved pump carrier frequency) spectroscopy, $S_{ChD}$ is found to vanish if the pump beam is linearly polarized.  Using a pump pulse with a finite width in time, $S_{ChD}$ is non zero and has the ability excite coherences between specific states and have an additional contribution originating from vibrational coherences in the ground state. This contribution is however found to be at least an order of magnitude lower than that of the excited state contribution for our system, allowing for unambiguous signatures of excited state dynamics. This is different to techniques such as 2D Fourier transform optical spectroscopy and ordinary pump-probe, in which the ground state contribution can conceal excited state coherences.

By systematically changing the pulse width and carrier frequency, it would be possible to eliminate the participation of coherences between states with energy differences larger than the pulses frequency width, or which are out of resonance.  This spectroscopy technique can therefore help to isolate and characterize coherences which have been predicted from theory or other experiments using different techniques. Additional this technique could be used to identify coherences between exciton states with a negligible electric dipole moment, too small to be identified with non-chiral techniques.

\begin{acknowledgments}
The authors would like to thank Camilla Ferrante, University of Padova, for helpful discussions.  This work was supported by the EU FP7 project PAPETS -- Phonon-Assisted Processes for Energy Transfer and Sensing (GA 323901); and ERC Starting Grant QUENTRHEL (GA 278560).
\end{acknowledgments}

\begin{appendix}

\section{General derivation of the chiral interaction operator including intrinsic supra molecular chirality}
\label{App:trans_current}
Starting with the minimum coupling Hamiltonian for light and matter in the semi classical approximation\cite{MukamelPoNLS}
\begin{equation}
 \hat{H}'(t) = -\int d \mathbf{r} \left[ \hat{\mathbf{J}}(\mathbf{r},t) \cdot \mathbf{A}(\mathbf{r},t) + \hat{Q}(\mathbf{r},t) : \mathbf{A}(\mathbf{r},t) \mathbf{A}(\mathbf{r},t)\right] \; .
 \label{eq:min_coup_Ham_full}
\end{equation}
We then neglect the term proportional to the charge density $\hat{Q}(\mathbf{r},t)$ times square of the (classical) electromagnetic vector potential $\mathbf{A}(\mathbf{r},t)^2$. Notice that this term is typically much smaller in experiments with visible light probing matter with strong dipole transitions.
We can then express the effective semi-classical Hamiltonian in $\mathbf{k}$ space as
\begin{equation}
 \hat{H}'(t) \approx -\int d \mathbf{k}  \hat{\mathbf{J}}(\mathbf{k},t) \cdot \mathbf{A}(-\mathbf{k},t) \; .
 \label{eq:min_coup_Ham_FT}
\end{equation}
Denoting the creation and annihilation operators for the $a$th excited state of the $\ell$th chromophore $\hat{B}^{\dagger}_{\ell a}$ and $\hat{B}_{\ell a}$, 
we can express current density operator in momentum space as
\begin{equation}
 \hat{\mathbf{J}}(\mathbf{k},t) = \sum_{\ell,a} \left( \overline{j}^*_{\ell a}(-\mathbf{k}) \hat{B}^{\dagger}_{\ell a} + \overline{j}_{\ell a}(-\mathbf{k}) \hat{B}_{\ell a} \right) \;.
\end{equation}
The terms $\overline{j}_{\ell a}(\mathbf{k})$ can in principle be calculated from the many-body wavefunctions of the ground and excited states via a multipole expansion in the displacement of charges $q_{\alpha}$ from the chromophore center~\cite{Abramavicius2006}:
 \begin{align}
 \overline{j}_{\ell a}(-\mathbf{k}) =  -i e^{i \mathbf{k} \cdot \mathbf{r}_j} \sum_{\alpha} &q_{\alpha} \bra{\phi_{\ell a}}   \omega \left[(\mathbf{r}_{\alpha}-\mathbf{r}_{j}) -i \mathbf{k} \cdot (\mathbf{r}_{\alpha}-\mathbf{r}_{j}) \otimes (\mathbf{r}_{\alpha}-\mathbf{r}_{j})/2 + \ldots\right]  \nonumber \\ 
 + &\mathbf{k} \times \left[ (\mathbf{r}_{\alpha}-\mathbf{r}_{j}) \times \mathbf{p}_{\alpha}/2j_{\alpha} + \ldots \right]  \ket{\phi_{jg}}  \;.
\label{eq:full_trans_current}  
\end{align}
Note that $c=\hbar =1$ in the above expression and the "$\ldots$" denote higher order magnetic / electric multipole moments~\cite{MukamelPoNLS}. Here $q$ denotes the charge of the $\alpha$th particle in the system.  
When the sum over all charges is performed, the first two terms are the electric transition dipole moment $\boldsymbol\mu_{\ell a}$ and quadrapole $Q^{\nu_1,\nu_2}_{\ell a}$ moment (contracted over $\mathbf{k}$). The only term explicitly written term in the second bracket is the magnetic dipole moment $\mathbf{m}_{\ell a}$.  

Naturally \refeq{eq:full_trans_current} is very complicated. To simplify it, we Taylor expand the exponential prefactor (assuming that all $\mathbf{r}_m$ are much smaller than an optical wavelength) and truncate terms of order $\mathbf{k}^2$ or higher: 
  \begin{equation}
 \overline{j}_{\ell a}(-\mathbf{k}) = -i\omega \boldsymbol\mu_{\ell a} - \omega \mathbf{k} \cdot (Q_{\ell a} + \boldsymbol\mu_{\ell a} \mathbf{r}_j) + i\mathbf{k} \times \mathbf{m}_{\ell a}   \;.
\label{eq:full_trans_current2}  
\end{equation}
Here $\omega= |\mathbf{k}|/c$, $Q_{\ell a}$ is the electric quadrapole tensor and $\mathbf{m}_{\ell a}$ is the magnetic dipole for the transition from the ground state to excited state $a$ of chromophore $\ell$. The term $\boldsymbol\mu_{\ell a} \mathbf{r}_m$ is the familiar super-molecule coupling term considered in the bulk of this paper. 

We now have the added complication that our coupling Hamiltonian is defined in terms of a vector potential.  However we can still make a single mode approximation with a gauge choice $\nabla \phi = 0$ and the slowly varying envelope approximate to use the expression
\begin{equation}
 \mathbf{A}_{\ell}(\mathbf{r},t) \approx -\frac{i}{\omega_\ell}\left(\polB_\ell  E_\ell(t) e^{i  (\mathbf{k}_\ell \cdot \mathbf{r} - \omega_\ell t + \varphi_\ell)} -\polB_\ell^* E_\ell^*(t) e^{-i  (\mathbf{k}_\ell \cdot \mathbf{r} - \omega_\ell t+\varphi_\ell)} \right) \;,
 \label{eq:vector_pot}
\end{equation}
for the vector potential for each of our pulses. The approximation \refeq{eq:vector_pot} is less valid when pulses excite a wide range of frequencies. In this case it is also preferable to scale the effective transition moments by the wavelength of light resonant with that transition and contract over a unit vector $\hat{k}$.

The factors of $1/\omega_j$ in \refeq{eq:vector_pot} will cancel those in \refeq{eq:full_trans_current2} and we can derive similar expression to those in the body of the paper.  However, we must now include the extra terms relating to supra-molecular chirality within \refeq{eq:psi_factor}. Hence we have a more general expression
\begin{equation}
\psi_{\xi_1,\xi_2}^{\nu, \nu_1 , \nu_2} = \mu_{\xi_2}^{\nu_2} ( Q^{\nu \nu_1}_{\xi_1} + \epsilon_{\nu_1\nu \nu'} \tilde{m}^{\nu'}_{\xi_1} ) - \mu_{\xi_1}^{\nu_1} ( Q^{\nu \nu_2}_{\xi_2} + \epsilon_{\nu_2\nu \nu'} \tilde{m}^{\nu'}_{\xi_2} ) \;,
\label{eq:psi_factor_general}
\end{equation}
with $\epsilon_{ijk}$ the Levi-civita tensor and 
\begin{equation}
Q^{\nu \nu'}_{\xi} = \sum_{\ell,a} \bra{g} \hat{B}_{\ell a} \ket{\xi} Q^{\nu \nu'}_{\ell a} \;, \qquad \tilde{m}^{\nu}_{\xi} = \sum_{\ell,a} \bra{g} \hat{B}_{\ell a} \ket{\xi}  (i\mathbf{m}^{\nu}_{\ell a} + \epsilon_{\nu \nu_1 \nu_2} R_m^{\nu_1}  \mu_{\ell a}^{\nu_2}/2) \;,
\end{equation}
the exciton basis electric quadrapole moment and magnetic dipole moments. We note \refeq{eq:psi_factor_general} will still vanish if $\xi_j = \xi_4$ and $\nu_j = \nu_4$ as before.  The intrinsic magnetic dipoles of each transition can easily be included into the expressing derived in this paper using effective magnetic dipoles but the electric quadrupole moments would require additional calculation for averaging.

\section{Derivation of vibronic states with an explicit vibrational mode}
\label{App:vibronic_derivation}
\subsection{Reformulation as Hamiltonian dynamics}
In our numerical calculations we compute the exact dynamics of electronic degrees of freedom using the HEOM. This approach will give exactly the same dynamics as that in which the underdamped mode is included in the system Hamiltonian and then traced out. However, accounting explicitly for quantum interaction between excitons and underdamped vibration allows to relate this dynamics to the transitions and coherences in the Hilbert space of the vibronic (exciton-vibration) states. The full system Hamiltonian now consists of the electronic Hamiltonian $\hat{H}_{\rm Elec}$ and two addition components:
\begin{equation}
\hat{H}_{\rm sys} = \hat{H}_{\rm Elec} + \hat{H}_{\rm Osc} + \hat{H}_{\rm El-Osc} \;;
\label{eq:fullHam}
\end{equation}
we denote the exact $k$th energy eigenstate in the single $(e)$ or double $(f)$ electronically excited manifold of this Hamiltonian as $\ket{\psi_{k,e/f}}$.  
Labelling the creation and annihilation operators for the mode on site $j$ as $\hat{b}^{\dagger}_j$ and $\hat{b}_j$, the individual components are 
\begin{subequations}
\begin{align}
\hat{H}_{\rm Elec} &=  E_1 \ket{1}\bra{1} + E_2 \ket{2}\bra{2} + V (\ket{1}\bra{2} + \ket{2}\bra{1}) \\
\hat{H}_{\rm Osc} &=  (\omega_0+1/2) (\hat{b}^{\dagger}_1\hat{b}_1 + \hat{b}^{\dagger}_2\hat{b}_2) \\
\hat{H}_{\rm El-Osc} &=  \sum_{j=1}^2 \sqrt{S_{\rm HR}} \omega_0 \ket{j}\bra{j} (\hat{b}^{\dagger}_j + \hat{b}_j) \;,
\end{align}
\end{subequations}
with the Huang-Reiss factor $\sqrt{S_{\rm HR}}$ related to the reorganization of the underdamped mode via $\lambda_{B} = S_{\rm HR} \omega_0$.  In our case the electronic coupling $V = 71 $cm$^{-1}$ is also comparable to the energy gap $E_2-E_1 = 144 $cm$^{-1}$ and larger than the effective coupling to Drude component of the bath i.e. $V > \sqrt{\lambda_D \omega_D}$. Then excitons are well defined. The upper and lower exciton states are denoted by $\ket{+} = \cos(\theta)\ket{2}+\sin(\theta)\ket{1} $ and $\ket{-} = \cos(\theta)\ket{1}-\sin(\theta)\ket{2}$ respectively, with energies $E_{\pm} = (E_2+E_1)/2 \pm\sqrt{V^2-(E_2-E_1)^2/4}$. The mixing angle $\theta$ is given by $\theta = \tan^{-1} \left( \frac{1}{\sqrt{1+\epsilon^2}-\epsilon} \right)$ for $\epsilon = (E_2-E_1)/2V$ . For our parameters the energy difference between the upper and lower exciton is $\Delta_{ex} \sim 202$~cm$^{-1}$ and  the mixing angle related to the degree of delocalization of an exciton is $\theta \sim 0.376 \pi$.  Note this angle is not related to the pump-probe angle mentioned in the main text.  

\subsection{Relative mode coordinates}
\label{App:rel_mode_co}
The assumption of identical modes for the two sites leads to consider two collective nuclear motions: a center-of-mass (correlated) oscillation and relative (anti-correlated) oscillation, with creation and annihilation operators
\begin{align}
\hat{b}_C &= (\hat{b}_1 + \hat{b}_2)/\sqrt{2} \\
\hat{b}_R &= (\hat{b}_1 - \hat{b}_2)/\sqrt{2} \;,
\label{eq:norm_modes}
\end{align}
with the oscillator Hamiltonian now given by
\begin{equation}
\hat{H}_{\rm Osc} =  (\omega_0+1/2) (\hat{b}^{\dagger}_C\hat{b}_C + \hat{b}^{\dagger}_R\hat{b}_R)
\end{equation}
The key feature of these collective motions is that the center-of-mass mode dynamics decouple from the exciton dynamics
\begin{align}
\hat{H}_{\rm El-Osc} = &g \left\{ 2\cos(\theta)\sin(\theta)^{\phantom{2}}[ \ket{-}\bra{+} + \ket{+}\bra{-}] + [  \cos^2(\theta)-\sin^2(\theta)][ \ket{+}\bra{+} - \ket{-}\bra{-}]   \right\}(\hat{b}_R + \hat{b}_R^{\dagger}) \nonumber \\
&+ [ \ket{+}\bra{+} + \ket{-}\bra{-}](\hat{b}_C + \hat{b}_C^{\dagger})  \;.
\label{eq:rel_mode_coords}
\end{align}
The second line of \refeq{eq:rel_mode_coords} is independent of population differences and coherences between the exciton states, only requiring that the system is in the excited state.  This center-of-mass therefore evolves as a damped quantum harmonic oscillator, initially displaced from equilibrium. We use the notation $\ket{+,n} \equiv \ket{+} \otimes \ket{n}$ for the joint exciton-vibration states, with the second index denoting the number of vibrational quanta in the relative mode.  Note that the vibrational states used are eigenstates within the ground electronic state.

\subsection{Vibronic states}
\label{sec:Vibronic}
For further analysis it is useful to split \refeq{eq:rel_mode_coords} into three different terms:
\begin{subequations}
 \begin{align}
 H_{\rm JC} =&  \sqrt{2}g_{k} \cos(\theta)\sin(\theta)[ \ket{-}\bra{+} \hat{b}_{R} + \ket{+}\bra{-} \hat{b}^{\dagger}_{R}] \;,  \\
H_{\rm NRW} =&  \sqrt{2}g_{k} \cos(\theta)\sin(\theta)[ \ket{-}\bra{+} \hat{b}^{\dagger}_{R} + \ket{+}\bra{-} \hat{b}_{R}] \;,  \\
H_{\rm NHD} =&  \sqrt{2} g_{k} [ \cos^2(\theta)-\sin^2(\theta)][ \ket{+}\bra{+} - \ket{-}\bra{-}]  (\hat{b}_{R} + \hat{b}_{R}^{\dagger})   \;. 
 \end{align}
\end{subequations}
Here JC stands for Jaynes Cummings, as the Hamiltonian including this form of interaction to the relative mode maps to the well-known Jaynes-Cummings model~\cite{Shore1993}. The subscripts NRW and NHD stand for ``non rotating wave'' and ``not Homodimer''. The former vanishes when we ignore the terms that couple subspaces with different number of excitations via the rotating wave approximation (although the system is still solvable without this approximation~\cite{Braak2011}) and the latter can be neglected if the onsite excitation energies are identical.

The coupling term $H_{\rm JC}$ strongly couples the upper and lower exciton states, particularly in the resonant case where $\omega_0$ matches the exciton splitting and the resulting states are degenerate.  This means the effective eigenstates are superpositions of exciton-vibration states referred from now on as vibronic states.  Denoting the states with population in the upper/lower state and $n$ quanta in the relative component of the vibrational mode by $\ket{\pm,n}$, a better basis for $n \ge 1$ are the states: 
\begin{align}
\ket{X_{+},n} &= \cos(\phi_n) \ket{+,n-1} +\sin(\phi_n) \ket{-,n} \nonumber \\
\ket{X_{-},n} &= \sin(\phi_n) \ket{+,n-1} - \cos(\phi_n) \ket{-,n}\;,
\label{eq:vibronic_def}
\end{align}
With our parameters these new states have energies $230$cm$^{-1}$ and $193$cm$^{-1}$ relative to the lower exciton state and $\phi_1 = 0.3410 \pi$ is the mixing angle. The higher energy state $\ket{X_{+},m}$ has more character from the lower exciton state.  Because our vibrational dephasing time is much longer than our electronic one, we therefore expect a longer coherence time associated with $\ket{X,0}\bra{X_{-},1}$ compared to $\ket{X,0}\bra{X_{+},1}$. 
  
When $\theta \approx \pi/2$, $\omega_0 \approx \Delta_{ex}$ and $|g/\Delta_{ex}| \ll 1$, these states are good approximations to the system eigenstates.  The new mixing angles $\phi_m$ are chosen to diagonalize $\hat{H}_{\rm Elec}+\hat{H}_{\rm Osc}+H_{\rm JC}$.
The lowest eigenstate in the excited manifold, denoted $\ket{X,0}$ is still the lower exciton state with zero quanta in the relative vibrational mode $\ket{-,0}$ as the weak mixing to the state $\ket{+,1}$ from $H_{\rm NRW}$ is neglected.

When $\theta \neq \pi/2$, which is the case in our system, the states $\ket{\pm,n}$ are coupled to $\ket{\pm,n\pm 1}$ from $H_{\rm NHD}$.  Combined with the damping this will lead to a Stokes shift on the relative mode.  
  Transitions dipole moments to all the exciton vibrational states (except $\ket{X,0}$) are a mixture of $\boldsymbol{\mu}_{-}$ and $\boldsymbol{\mu}_{+}$ with the equivalents for the beyond-dipole approximation moments.

\subsection{Impulsive pump doorway evolution in a vibronic state basis}
\label{sec:imp_pump_vibronic}
For the electronic excited state of our dimer, we noted that the initial chiral doorway function is given by
$D^{(C)}_e = iC(\ket{\xi_1} \bra{\xi_2}-\ket{\xi_2} \bra{\xi_1}) \otimes \hat{\rho}_{v,eq}$ where $\hat{\rho}_{v,eq}$ is the equilibrium density matrix for the vibrational degrees of freedom in the electronic ground state. In terms of  the vibronic eigenstates of the Hamiltonian \ref{eq:fullHam} with a single excitation, i.e.$\ket{\psi_{k,e}}$, we can express this initial condition as
\begin{equation}
D^{(C)}_e = i C \left\{\sum_{k \langle k'} \ket{\psi_k,e}\bra{\psi_{k',e}} [\bra{\psi_{k,e}}(\ket{\xi_1}\hat{\rho}_{v,eq} \bra{\xi_2})\ket{\psi_{k',e}} -\bra{\psi_{k,e}}(\ket{\xi_2}\hat{\rho}_{v,eq} \bra{\xi_1})\ket{\psi_{k',e}}]  - h.c. \right\} \;.
\label{eq:vibronic_impulsive}
\end{equation}
This operator still lacks any terms on the diagonal and so consists only of terms which beat during the population time (unless degenerate states are present). These decay due to the interaction with the remaining degrees of freedom of the thermal bath.  Since detection is performed in frequency space we expect to probe resonances related to transitions from vibronic states.  In the case of our dimer, the center-of-mass mode $\hat{b}_C$ is uncoupled from the electronic dynamics and evolves as a quantum harmonic oscillator with decoherence, effectively undergoing simple harmonic motion as the coherence time is much larger for vibrational modes than for the electronic degrees of freedom.  This lead to undressed oscillations with a period of $\omega_0$ on top of the vibronic dynamics, but no pure vibrational coherences can contribute because $(\bra{\psi_k}(\ket{\xi_1}\hat{\rho}_{v,eq} \bra{\xi_2})\ket{\psi_{k',e}} -\bra{\psi_{k,e}}(\ket{\xi_2}\hat{\rho}_{v,eq} \bra{\xi_1})\ket{\psi_{k',e}})$ would evaluate to zero and hence the mode $\hat{b}_C$ can only contribute overtones.  

\section{Chiral window functions}
\label{App:window} 

Additionally to the window function associated with ground state bleaching (GSB) we have the two excited state window functions $W_{e}$ and $W_{f}$ related to stimulated emission and excited state absorption.  These stimulated emission window function is given by
\begin{equation}
W_{e} \propto -\mathrm{Re} \int_{-\infty}^{\infty} dt \int_{0}^{\infty} dt_3 \; e^{i \varphi + i\omega_{\rm LO} t_3 + i(\omega_{\rm LO}-\omega_{r}) t} E_{\rm LO}^*(t+t_3) E_r(t) \sum_{\xi_3,\xi_4}    p_{\rm LO}^{\nu_4} p_r^{\nu_3} (\mu^{\nu_3}_{\xi_3} \mu^{\nu_4}_{\xi_4} + i k_r^{\nu} \psi^{\nu,\nu_3,\nu_4}_{\xi_3,\xi_4} ) [\hat{B}^{\dagger}_{\xi_3}\Gprop_{eg}(t_3)]\hat{B}_{\xi_4} \Gprop_{ee}(t)   \;.
\label{eq:window_ee}
\end{equation}
As before, the chiral part $W_{e}^{(C)}$ is given by the terms in the sum featuring $\psi^{\nu,\nu_3,\nu_4}_{\xi_3,\xi_4}$ and the non-chiral part by those with only dipole transitions.  For the ESA part we have
\begin{equation}
W_{f} \propto \mathrm{Re} \int_{-\infty}^{\infty} dt \int_{0}^{\infty} dt_3 \; e^{i \varphi + i\omega_{\rm LO} t_3 + i(\omega_{\rm LO}-\omega_{r}) t} E_{\rm LO}^*(t+t_3) E_r(t) \sum_{\xi_3,\xi_4}  \sum_{f_3,f_4}  p_{\rm LO}^{\nu_4} p_r^{\nu_3} (\mu^{\nu_3}_{\xi_3,f_3} \mu^{\nu_4}_{\xi_4,f_4} + i k_r^{\nu} \psi^{\nu,\nu_3,\nu_4}_{[\xi_3,f_3],[\xi_4,f_4]} ) \hat{V}_{\xi_4,f_4} \Gprop_{ef}(t_3) \hat{B}^{\dagger}_{\xi_3,f_3} \Gprop_{ee}(t)  \;,
\label{eq:window_ff}
\end{equation}
where the additional sum over the $N(N-1)/2$ double excited states has now been included and $\hat{B}_{\xi,f}=\ket{\xi} \bra{f,2}$.  For this work, $N=2$ and hence only one double excited state is possible; for larger systems the number of double excited state becomes unfavorable and models such the coherent exciton scattering model become favorable. 

\section{Impact of dipole coupling beyond the Condon approximation}
\label{App:beyond_condon}
All of our numerics feature dipole moments only for transitions in which the vibrational degrees of freedom are unaffected, known as the Condon approximation.  To go beyond this approximation, we can expand the molecular polarizability to first order in normal mode coordinates~\cite{Dhar1994}
\begin{equation}
\alpha(\vec{q}) = \alpha_0 + \sum_{j} \left( \frac{\partial q}{\partial q_j}\right)_{0} q_j + \ldots  \; ,
\end{equation}
with the $\ldots$ denoting higher order terms. The linear terms lead to dipole moments which couple vibrational states on different chromophores with $\pm 1$ quanta.  Our dipole moment operator can now be written as 
\begin{align}
\hat{\muB} =  \sum_{j}[\hat{B}_{j}+\hat{B}^{\dagger}_{j}] [\muB_{g j} + \muB_{g, j;1}(\hat{b}_{j}+\hat{b}^{\dagger}_{j}) ]  \;.
\label{eq:mu_gen}
\end{align}
More generally, anharmonicity with the mode coordinate allows some weak coupling to energy levels with a difference of multiple quanta.  As we mentioned in the main text, these additional couplings do not effect the chiral doorway component of our signal in the impulsive pump limit. To see this, we consider a generalized chiral interaction tensor which includes transitions between different levels of a particular mode:  
\begin{equation}
 \psi_{\xi_1,\xi_2;n,\ell,n'}^{\nu, \nu_1 , \nu_2} =\sum_{j_1,j_2=1}^N  C_{\xi_1}^{j_1}C_{\xi_2}^{j_2} \mu_{g,n;j_1,\ell}^{\nu_1}\mu_{j_2}^{g,n';j_2,\ell} (R_{j_1}^{\nu} - R_{j_2}^{\nu}) \;.
\end{equation}
In the impulsive limit, we excite all excitons with equal weight and hence we take the sum over all $\xi_1 = \xi_2$. The sum  $\sum_{\xi} C_{\xi}^{j_1}C_{\xi}^{j_2} = \delta_{j_1,j_2}$ as the states are orthonormal and $(R_{j_1}^{\nu} - R_{j_2}^{\nu})$ is clearly zero when $j_1 = j_2$, hence we have no chiral contribution.  Outside of the impulsive limit this logic no longer holds and the detuning between the carrier frequency of our pump pulse and the exciton energies becomes increasingly important, hence all excitons are not excited equally. 

More generally, when we have a transition with both an intrinsic magnetic dipole moment and an electric dipole moment the situation is now more complicated. We look at the chiral contribution to the projector $\ket{n}\bra{n'}$ in our doorway function correction $D_g'=\sum_{n,n'} \rho_D(n,n')\ket{n}\bra{n'}$, still in the impulsive limit:
\begin{align}
\rho_D(n,n') = \sum_{\xi,\tilde{n}} \{ &P(n)\left[ (\mathbf{b} \cdot \mathbf{m}_{g n;\xi \tilde{n}})(\polB \cdot \muB_{\xi \tilde{n};g, n'}) - (\polB \cdot \muB_{g, n;\xi \tilde{n}})(\mathbf{b} \cdot \mathbf{m}_{\xi \tilde{n};g,n'})\right]  \nonumber \\
- &P(n')  \left[ (\mathbf{b} \cdot \mathbf{m}_{g n';\xi \tilde{n}})(\polB \cdot \muB_{\xi \tilde{n};g, n}) - (\polB \cdot \muB_{g, n';\xi \tilde{n}})(\mathbf{b} \cdot \mathbf{m}_{\xi \tilde{n};g,n})\right] \} \;.
\end{align}
Here $\mathbf{b} = \mathbf{k} \times \polB$ is a vector in the direction of the magnetic field; the first term comes from the left acting Feynman diagram and the latter from the right side. Both terms will vanish for $n=n'$,  however in general  cancellation is less obvious.  The sum over all $\xi$ will not lead to full cancellation as the CD signal is no longer conservative;  however, the sum over $\tilde{n}$ {\it will} lead to cancellation. 

As all possible paths are summed over with equal weight, the terms to the left of each bracket will be repeated on the right. For example, taking $n=0$, $n'=1$ the $\tilde{n}=1$ term will give $(\mathbf{b} \cdot \mathbf{m}_{g 0;\xi 1})(\polB \cdot \muB_{\xi 1;g 1})$ for the left term in the square brackets and $\tilde{n}=0$ will give $(\polB \cdot \muB_{g 0;\xi 0})(\mathbf{b} \cdot \mathbf{m}_{\xi 0,g 1})$ on the right side.  As $\muB_{\xi 0;g, 0}=\muB_{\xi 1;g, 1}$ (this is simply the zero order term in $\vec{q}$ expansion) and $\mathbf{m}_{g 0;\xi 1} \approx \mathbf{m}_{\xi 0,g 1}$ these two terms approximately cancel.  The difference in resonant excitation frequency of the two transitions may mean cancellation is not complete; but this difference is small compared to optical frequencies, leaving only a negligible difference term. Since all terms can be paired off in this way, we have nearly full cancellation.

It therefore seems quite general that the chiral doorway signal remains free of significant ground state contributions in the impulsive limit and therefore only excited state electronic coherences will contribute to this signal component.  Outside of the impulsive limit, the pulse time width becomes comparable to electronic transition frequencies and dephasing rates and this is no longer true.  However in the frequency resolved limit we also expect the chiral doorway contribution to vanish anyway. Pinched between these two extremes, the ground state contribution to the chiral doorway function is expected to be weak in all cases.

\section{Including the effects of spatial overlap of the pump and probe}
\label{App:spat_overlap}
When the pump and probe beams are not exactly colinear, we need to consider a formalism that explicitly considers the locations where the pump and probe both interact with molecules.  This situation is barely covered in the literature, as typically the pump can be focused to a point much smaller than the probe width.  However, as we wish to make a precise subtraction of measurements with different angles, it is worth considering explicitly. Additionally, strong focusing is undesirable for us here, as it will mean more wavevectors-polarization combinations will contribute to the final signal, complicating our understanding. 

We assume cylindrical symmetry for our pulses, with Gaussian envelopes $E_r(\mathbf{R},t) = \exp(-(z-z_0(t))^2/2\tilde{\sigma}_r^2-(x^2+y^2)^2/2\sigma_r^2)$ for the probe and the pump at an angle of $\theta$ to this, lying in the $z-y$ plane.  

Within the doorway window formalism, the doorway and window functions will remain the same, but acquire an additional factor for the pulse intensity relative to the maximum. This factor is $\exp\left(-R_{\bot}^2/\sigma_r^2\right)$ for the window and
$\exp\left(-R_{\bot'}^2/\sigma_u^2\right)$ for the doorway; we have defined $R_{\bot} = \sqrt{x^2+y^2}$ and $R_{\bot'} = \sqrt{x^2+(y \cos(\theta) -z \sin(\theta))^2}$ the displacements from the primary maxima of each pulse, in the plane orthogonal to propagation.  More significantly, the time delay $\tau$ will now be a function of position within the sample;  $\tau$ will decrease along the positive $z$ and $y$ axes.

For the ground state contribution, we therefore have to evaluate a term of the form
\begin{equation}
\tilde{S}_{GSB}(\omega_s, \tau;E_u;E_r) = \omega_s \int \; d^3\textbf{R} \; \exp\left(-\frac{R_{\bot'}^2}{ \sigma_u^2}-\frac{R_{\bot}^2}{\sigma_r^2}\right) \langle \mathrm{Tr}[W_g(\omega_s;E_r) \mathcal{G}\left(\tau-\frac{n}{c}\left[z(1-\cos(\theta))-z_0+y \sin(\theta) \right]\right) D_g(E_u)]  \rangle   \;,
\end{equation}
where $n$ is the refractive index.  The most significant change here is that the time delay is now dependent on $z$ and $y$. We note the $y$ dependence will vary across the probe pulse, which means this could theoretically be separated with spatially resolved detection. The $z$ dependence is in the direction of propagation and therefore could not be removed within this experimental geometry.

In the case of a fully non-colinear experiment ($\theta = \pi/2$), we essentially have to integrate our "ideal" signal over a range of $\tau$ with a weight function equal to the radial width of our pump beam.  If our "ideal" phase matched signal is oscillating about some constant $A$ with a frequency $\omega$ (i.e.~$S(\tau) \approx A + \sin(\omega \tau)\exp(-\Gamma \tau)$) our actual signal is
\begin{equation}
\tilde{S}(\tau) \propto \int dz \int dy \; e^{-\frac{z^2}{ \sigma_u^2}-\frac{y^2}{\sigma_r^2}} \{ A + e^{-\Gamma [\tau-n(z+y)/c]} \sin(\omega [\tau-n(z+y)/c]) \} \propto A+\sin ( \omega t+\phi) e^{-\frac{n^2 (\omega^2-\Gamma^2) \left(\sigma_1^2+\sigma_2^2\right)}{4c^2} } \;,
\label{eq:coh_sup}
\end{equation}
with $\phi = \left(\sigma_1^2+\sigma_2^2\right)\Gamma \omega n^2/c^2$ a phase shift.  Unless $n \omega \sqrt{\sigma_1^2+\sigma_2^2} /c \lesssim 1 $, any weakly damped, oscillating component will be much weaker in this configuration. Additionally the overall amplitude will be reduced as the beam angle increases because less molecules are in the path of both pulses. For our signals we have $\omega/2 \pi c \approx 200$cm$^{-1}$ with $n \sim 1.3$ in water, hence we require $\mathrm{max}(\sigma_1,\sigma_2) \lesssim 6\mu$m. This is a fairly narrow waist and so some focusing (at least along the $z$ direction) is likely required to take measurements in this geometry.  

To directly compare signals taken at different angles we may have to numerically "undo" the effect of this convolution over $\tau$.  This would probably be easiest to achieve in Fourier space (over $\tau$), where it essentially involves a rescaling of peaks depending on their frequency and width.  Experimentalists would have to look at the change to these peaks when $\theta$ is varied, compared with that expected from the convolution effect alone, in order to extract the doorway contribution. 


\section{Isotropic averages for polarization configurations within TRCD}
\label{sec:ori_av} 

\subsection{Relevant orientation averages}

We consider the averaging by considering effective magnetic dipoles $\tilde{\mathbf{m}}_{\xi}$ rather than explicitly considering fifth rank tensors.  To indicate why this is possible we note that the result $(\mathbf{A} \cdot \hat{y})(\mathbf{B} \cdot \hat{z}) - (\mathbf{B} \cdot \hat{y})(\mathbf{A} \cdot \hat{z}) = (\mathbf{A} \times \mathbf{B}) \cdot \hat{x}$ combined with the fact odd rank tensors change sign~\cite{Wagniere1982} when two indices are permuted means
\begin{equation}
\langle\mathbf{A}_1 \cdot \hat{x}_1 \ldots (\mathbf{A}_n \cdot \hat{y})(\mathbf{A}_{n+1} \cdot \hat{z})\rangle = \langle\mathbf{A}_1 \cdot \hat{x}_1 \ldots (\mathbf{A}_n \times \mathbf{A}_{n+1}) \cdot \hat{x})/2\rangle \;.
\label{eq:odd_rank_av}
\end{equation}
As $\mathbf{r}_j \times \boldsymbol{\mu}_j$ transforms as a vector, these order $n+1$ averages can be calculated from $n$th rank tensors.
 
Applying this to the calculation of our chiral window averages, we have for $\polB_u = \hat{x}$ and $\polB_{r} =(\hat{x} \pm i \hat{y})/\sqrt{2}$
\begin{align}
k_r^{\nu} \pol_u^{\nu_1}  \pol_u^{\nu_2}  \pol_{L/R}^{\nu_3} \pol_{R/L}^{\nu_4} \langle\mu_{\xi_1}^{\nu_1} \mu_{\xi_2}^{\nu_2} \psi_{\xi_3,\xi_4}^{\nu, \nu_3 , \nu_4}\rangle &= \mp \frac{i|k_r|}{2} \langle\mu_{\xi_1}^x \mu_{\xi_2}^x  \sum_{j_3,j_4}  C^{j_3}_{\xi_3} C^{j_4}_{\xi_4} (\mu_{j_3}^x \mu_{j_4}^y - \mu_{j_3}^y \mu_{j_4}^x) \Delta \mathbf{R}^z_{j_3,j_4}  \rangle\;, \nonumber \\ 
&= \pm \frac{i|k_r|}{2} \langle\mu_{\xi_1}^x \mu_{\xi_2}^x  (\mu_{\xi_3}^x m_{\xi_4}^x + \mu_{\xi_4}^x m_{\xi_3}^x +   \mu_{\xi_3}^y m_{\xi_4}^y + \mu_{\xi_4}^y m_{\xi_3}^y )\rangle\;.
\label{eq:chiral_window_av}
\end{align}
Note that we use the notation $\mu_{\xi_1}^x \equiv \muB_{\xi_1} \cdot \hat{x}$ and hence the indices $x,y,z$ should not be subject the Einstein summation notation! The second line uses the results \refeq{eq:odd_rank_av} and \refeq{eq:ex_mag_dip}.  Equivalently for the chiral doorway average we have 
\begin{align}
k_u^{\nu} \pol_u^{\nu_1}  \pol_u^{\nu_2}  \pol_{L/R}^{\nu_3} \pol_{R/L}^{\nu_4} \langle\mu_{\xi_3}^{\nu_3} \mu_{\xi_4}^{\nu_4} \psi_{\xi_1,\xi_2}^{\nu, \nu_3 , \nu_4}\rangle &= \frac{k_u^z}{2} \langle[\mu_{\xi_3}^x \mu_{\xi_4}^x+\mu_{\xi_3}^y \mu_{\xi_4}^y \mp i(\mu_{\xi_3}^x \mu_{\xi_4}^y - \mu_{\xi_3}^y \mu_{\xi_4}^x ) ]\sum_{j_1,j_2}   C^{j_1}_{\xi_1} C^{j_2}_{\xi_2} (\mu_{j_1}^x \mu_{j_2}^x [\mathbf{R}^z_{j_1} - \mathbf{R}^z_{j_2}])\rangle\;, \nonumber \\ 
&=  \pm i \frac{k_u^z}{2} \langle (\mu_{\xi_3}^y \mu_{\xi_4}^x - \mu_{\xi_3}^x \mu_{\xi_4}^y )  ( m_{\xi_1}^y \mu_{\xi_2}^x-m_{\xi_2}^y \mu_{\xi_1}^x) \rangle\;.
\label{eq:chiral_doorway_av}
\end{align}

\subsection{Isotropic average formalism}

\label{sec:ori_av_CDF} 

In order to evaluate the averages \refeq{eq:chiral_doorway_av} and \refeq{eq:chiral_window_av}, we introduce the isotropic average for a 4th rank tensor:~\cite{Cho2003}
\begin{equation}
\pol^{v_1} \pol_2^{v_2} \pol_3^{v_3} \pol_4^{v_4} \langle \mu_1^{v_1} \mu_2^{v_2} \mu_3^{v_3} \mu_4^{v_4} \rangle_{\rm iso} =\colvec[(\polB_1 \cdot \polB_2)(\polB_3 \cdot \polB_4)]{(\polB_1 \cdot \polB_3)(\polB_2 \cdot \polB_4)}{(\polB_3 \cdot \polB_2)(\polB_1 \cdot \polB_4)}^{\dagger} M^{(4)} \colvec[(\muB_1 \cdot \muB_2)(\muB_3 \cdot \muB_4)]{(\muB_1 \cdot \muB_3)(\muB_2 \cdot \muB_4)}{(\muB_3 \cdot \muB_2)(\muB_1 \cdot \muB_4)} \;.
\label{eq:4th_rank_av}
\end{equation}
with $\polB_k$ the polarization of the field responsible for the $k$th interaction and
\begin{equation}
M^{(4)} = \frac{1}{30}\left( \begin{array}{ccc}
4 & -1 & -1 \\
-1 & 4 & -1 \\
-1 & -1 & 4 \end{array} \right) \;.
\end{equation}

Using \refeq{eq:4th_rank_av} we can show that the averages for the chiral window function evaluate to
\begin{align}
i k_r^{\nu} \pol_u^{\nu_1}  \pol_u^{\nu_2}  \pol_{L/R}^{\nu_3} \pol_{R/L}^{\nu_4} \langle\mu_{\xi_1}^{\nu_1} \mu_{\xi_2}^{\nu_2} \psi_{\xi_3,\xi_4}^{\nu, \nu_3 , \nu_4}\rangle 
= \mp \frac{|k_r|}{60} [6 \mu_{12} (m_{34} + m_{43}) - (\mu_{24}m_{31} +\mu_{14}m_{32} +\mu_{23}m_{41} +\mu_{32}m_{42} ) ] \;,
\label{eq:chiral_window_av2}
\end{align}
where we have introduced the compact notation $\mu_{jk} = \muB_{\xi_{j}} \cdot \muB_{\xi_{k}}$ and $m_{jk} = \mathbf{m}_{\xi_{j}} \cdot \muB_{\xi_{k}}$.  This result does not vanish for $\xi_{1} = \xi_{2}$ or $\xi_{3} = \xi_{4}$ and is therefore not coherence specific as expected.  The $\mp$ terms at the front relate to whether the probe is left or right circularly polarized and will vanish when we take one signal from the other.

The averages for the chiral doorway evaluate to
\begin{align}
i k_u^{\nu} \pol_u^{\nu_1}  \pol_u^{\nu_2}  \pol_{L/R}^{\nu_3} \pol_{R/L}^{\nu_4} \langle\mu_{\xi_1}^{\nu_1} \mu_{\xi_2}^{\nu_2} \psi_{\xi_1,\xi_2}^{\nu, \nu_1 , \nu_2}\rangle 
= \mp \frac{k_u^z}{12} [ (m_{13}\mu_{42}  -m_{24}\mu_{13} ) - (m_{23}\mu_{41}  -m_{14}\mu_{23} )] \;,
\label{eq:chiral_doorway_av2}
\end{align}
this average does vanish for $\xi_{1} = \xi_{2}$ as expected.  This average remains finite for $\xi_{3} = \xi_{4}$ (reducing to a term $\propto m_{13}\mu_{23}  -m_{23}\mu_{13}$), however for a pump probe configuration we always have the conjugate term with $\xi_1$ and $\xi_2$ reversed which leads to cancellation of these terms, and is therefore coherence specific.  Note that this latter effect is not down to orientation averaging.

\section{Methods for numerical calculation}
\label{sec:num_meth}
\subsection{Evaluation of third order response functions using the HEOM}
Calculating the pump-probe signals for a specific configuration of pulses with known shapes, carrier frequencies and delays is possible as a direct calculation, for example by using the doorway-window formalism outlined earlier.  However in this work we calculate the signals directly from the components of the response function relating to the rephasing and non-rephasing signals in two-dimensional spectroscopy. The pump probe signal is then obtained via the method outlined in App.~\ref{App:pp_from_resp}.

We calculate the rephasing and non-rephasing response functions, Fourier transformed over the first and last variables~\cite{SchlauCohen2011}
\begin{equation}
\mathrm{Sig}_{R/NR} \approx \int_0^{\infty} dt_3 \int_0^{\infty} dt_1 e^{i\omega_3 t_3 \mp i\omega_1 t_1} S (t_3,\tau,t_1;\mp k_u,\pm k_u,k_r) \equiv \tilde{S}_{R/NR}  (\omega_3,\tau, \omega_1;k_1,k_2,k_3) \;.   
\end{equation}
Here we have taken the contraction over polarizations and wavevectors as implicit.  
Each response function is further broken down into three components, the ground state bleaching (GSB) component, the stimulated emission (SE) component and the excited state absorption (ESA) component.  All subsequent quantities can be calculate from these response functions. We can also calculate these quantities in the time domain $t_1$ instead of $\omega_1$ when we wish to consider very short pump pulses, or simply at $t_1=0$ for the impulsive pump.

In order to perform this calculation we roughly follow the method of \cite{Chen2010}. We first solve for the Heisenberg picture interaction operators in Liouville space.  In terms of computation, our Liouville space terms $\langle \langle O \vert$ are the operator $\hat{O}$ (described by a matrix) flattened into a vector and $\langle \langle O \vert \rho \rangle \rangle \equiv \mathrm{Tr}\{ \hat{O} \hat{\rho}\}$. We compute $\langle \langle V_{\xi_4} (\omega_3) \vert$ with $V_{\xi_4} (0) = \ket{g}\bra{\xi_4}$ or $\ket{\xi_4}\bra{f_4,2}$ for transitions to the double excited states, for all $\xi_4$ and $f_4$ in order to make analytically calculating orientation averages easier.  We perform this calculation directly in frequency space via the following equation, valid for matrix $\mathcal{L}$ constant in time and $\mathbb{1}$ an identity matrix the size of $\mathcal{L}$
\begin{equation}
 \int_0^\infty dt e^{-i\omega t} e^{\mathcal{L} t} = \frac{\mathbb{1}}{i\omega \mathbb{1} - \mathcal{L}} \;.
 \label{eq:freq_space_tev}
\end{equation}
This method is generally more accurate than propagating in time and then taking a fast Fourier transform due to the periodicity implicit in a discrete Fourier transform. 
If the dynamics are Markovian, the Liouvillian can always be expressed as a matrix and our system as a vector obtained by flattening the density matrix. Within the Hierarchical equations of motion formalism~\cite{Tanimura1989,Ishizaki2005} additional tiers of "auxiliary" density matrices are included. The auxiliary matrices are coupled to those in the tier above and below and have additional decay terms; these matrices contain information about the displacement (and higher moments) of the bath\cite{Zhu2012}.  Assuming we truncate the hierarchy at a finite level (with what is essentially a Markovian assumption) we can combine all these dynamics into a single matrix operator $\Lambda$ which acts on the entire Hierarchy and \refeq{eq:freq_space_tev} can again be used. We use 3 Matsubara frequencies and four tiers in the hierarchy to achieve satisfactory numerical convergence. 

We also solve for $\vert  \rho_{e,g}(\omega_1;\xi_1) \rangle \rangle = G(\omega_1) \hat{B}_{\xi_1} \vert \rho_{eq} \rangle \rangle$, with $\xi_1$ a state in the single exciton manifold.  This quantity then relates to two different classes of second order density matrix elements, the ground state hole $\vert\rho_{g,g}(t_1,0;\xi_1,\xi_2)\rangle \rangle$ or excited state elements $\vert\rho_{e,e}(t_1,0;\xi_1,\xi_2)\rangle \rangle$. 
Each of these second order elements are then propagated in time by solving the set of coupled ODEs $\vert\dot{\rho}(t)\rangle \rangle = \mathcal{L} \vert\rho(t)\rangle \rangle$, the rephasing contributions can be calculated from the Hermitian conjugate of these matrices.  These terms can then be combined to calculate any of the response functions.

The pump probe signal can be calculated from both of these frequency domain response functions as we outline in \ref{App:pp_from_resp}. We consider the limit in which the pump pulse has a finite duration, but the probe is assumed to be short, and frequency resolved detection is employed via a heterodyne detection sequence.  Using these approximations our signal can be calculated from the rephasing and nonrephasing components as
\begin{equation}
S_{PP}(\omega_s,\omega_1) \approx \omega_s \mathrm{Re}\left\{ \int_{-\infty}^{\infty} d\omega' \int_{-\infty}^{\tau} dt' E_1(t') E_1(\omega')[ e^{i \omega' t'} S_{R}(\omega_s,\tau-t',\omega_1+\omega') +  e^{-i \omega' t'} S_{NR}(\omega_s,\tau-t',\omega_1-\omega') ]  \right\} \;.
\label{eq:PP_sig}
\end{equation}
We outline the derivation of this expression in the next subsection.  

\subsection{Derivation of the pump probe signal from the response functions}
\label{App:pp_from_resp}

For a pump-probe configuration with two pulses separated by a delay $\tau$ with resonant frequencies $\omega_1$, and envelope $E_1(t)$ for the pump and $\omega_2$ and $E_2(t)$ for the probe, we can express the signal within the rotating wave approximation as~\cite{MukamelPoNLS}:
\begin{align}
 S_{PP}(\omega_1,\omega_2,\omega_s;\tau) = &2 \omega_s {\rm Re} \int_{-\infty}^{\infty}dt  \int_{0}^{\infty}dt_3\int_{-\infty}^{t+\tau}dt'\int_{0}^{\infty}dt_1 \nonumber \\
&e^{i [(\omega_2-\omega_s) t + \omega_s t_3] }\left[ E_{\rm LO}^*(t+t_3) E_2(t) E_1^*(t')E_1(t'-t_1) e^{+ i \omega_1 t_1} S_{NR}(t_3,\tau+t-t',t_1) \right. \nonumber \\
&+ \left.E_{LO}^*(t+t_3) E_2(t) E_1(t')E_1^*(t'-t_1) e^{ - i \omega_1 t_1} S_{R}(t_3,\tau+t-t',t_1)\right] + S_{\rm coh} + S_{\rm nto}\;.
\label{eq:pp_from_RPNR}
\end{align}
Here $\omega_s$ is the signal frequency and LO stands for the local oscillator; in our case a particular frequency component of $E_2$.  We have also included the coherent (coh) and non-time ordered (nto) contributions, which only occurs if the pump and probe pulses overlap (i.e.~the pulse widths are not much smaller than $\tau$).   For this work we will ignore these terms and hence set $S_{\rm coh} = S_{\rm nto}=0$.  Using the convolution theorem, we can express the integrals over $t_1$ and $t_3$ as an inverse Fourier transform of the Fourier transforms of the convoluted quantities and obtain (up to constant prefactors)
\begin{align}
 S_{PP}(\omega_1,\omega_2,\omega_s;\tau) \approx & \omega_s {\rm Re} \int_{-\infty}^{\infty}dt  \int_{-\infty}^{\infty}d\omega   \int_{-\infty}^{t+\tau}dt'   \int_{-\infty}^{\infty}d\omega'  \nonumber \\
&\tilde{E}_{\rm LO}^*(\omega) E_2(t) e^{i (\omega_2-\omega_s) t  } e^{i(\omega t + \omega' t')} \left[  E_1^*(t')\tilde{E}_1(\omega')  \tilde{S}_{NR}(\omega_s-\omega,\tau+t-t',\omega_1-\omega') \right. \nonumber \\
&+ \left. E_1(t') \tilde{E}_1^*(\omega')  \tilde{S}_{R}(\omega_s-\omega,\tau+t-t',\omega_1+\omega')\right] \;.
\end{align}
For simplicity we assume that probe is short in time and the local oscillator (which represents frequency resolved detection of the probe, with some manipulation of the polarization) is well resolved in frequency with a carrier frequency $\omega_2$ not too different from the signal frequency. Hence, we can make the approximation $\exp[i (\omega_2-\omega_s) t] \tilde{E}_{\rm LO}^*(\omega) E_2(t) \sim \delta(t)\delta(\omega)$. In principle, one would need to account for the finite length of the probe pulse with a factor of $\tilde{E}^*_r(\omega_s-\omega_r)$, but this dependence is assumed to be scaled out of the final signal.  With this approximation we have 
\begin{align}
 S_{PP}(\omega_1,\omega_s;\tau) = & \omega_s {\rm Re}  \int_{-\infty}^{t+\tau}dt'   \int_{-\infty}^{\infty}d\omega' \; e^{i \omega' t'} \nonumber \\
& \left[  E_1^*(t')\tilde{E}_1(\omega')  \tilde{S}_{NR}(\omega_s,\tau-t',\omega_1-\omega') \right. \nonumber \\
&+ \left. E_1(t') \tilde{E}_1^*(\omega')  \tilde{S}_{R}(\omega_s,\tau-t',\omega_1+\omega')\right] \;,
\end{align}
noting that the dependence on the probe carrier frequency $\omega_2$ has now dropped out.  This expression can be used to calculate pump probe signals with finite during pump pulses.  When we also have a very short pump pulse we can ignore the $t' $ dependence in the response function during the population time and we have
\begin{align}
 S_{PP}(\omega_1,\omega_s;\tau) \approx & \omega_s {\rm Re}   \int_{-\infty}^{\infty}d\omega' |\tilde{E}_1(\omega')|^2  \nonumber \\
& \left[   \tilde{S}_{NR}(\omega_s,\tau,\omega_1-\omega') +  \tilde{S}_{R}(\omega_s,\tau,\omega_1+\omega')\right] \;.
\end{align}
Finally in the extreme short time limit we can ignore the frequency dependence in the Fourier transform of the pulse envelope and Fourier transform the response function back into time space in $t_1$ and set this to zero.  

\section{Prony Analysis}
\label{App:Prony}

In order to better understand the oscillations present in our results, we consider particular frequency slices through figure \reffig{Fig:SP_EScD} and perform a Prony decomposition. Prony decomposition takes an impulsive signal and decomposes it into oscillating and decaying exponential components. This technique has been used for analysis of impulse responses~\cite{Hauer1991}, notably in NMR signals~\cite{Piero1997,Luthon1989}. More noise tolerant methods are generally preferred for experimental data~\cite{Meunier1998} such as time-frequency and wavelet methods~\cite{Volpato2015}. Even so it remains a powerful tool in low noise systems (such as simulated data) as it can estimate both frequency, damping and relative phase of beating components.  
 
\begin{figure}[ht]
\centering\includegraphics[width=0.9\linewidth]{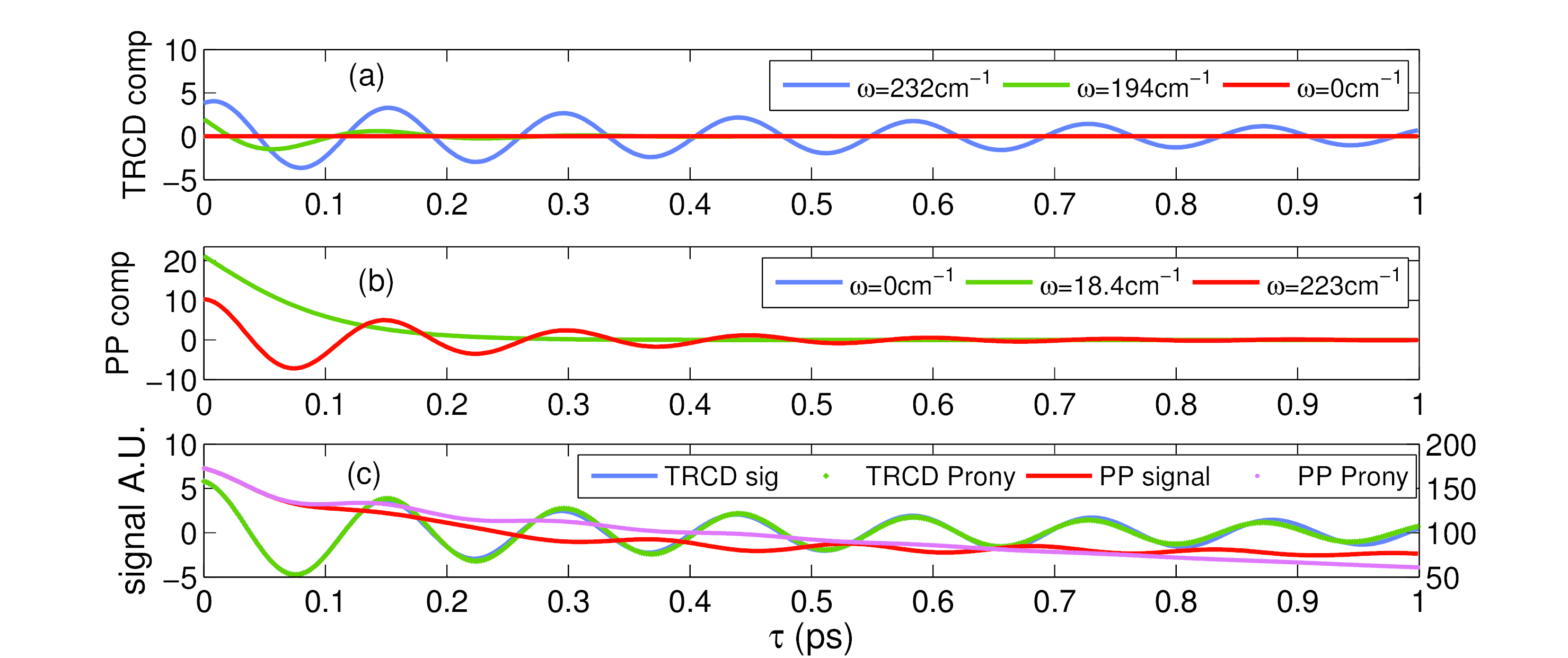}
\caption{Three largest components in the Prony decomposition of a slice at $\omega_s = 12495$cm$^{-1}$ (all angular frequencies quoted are scaled by $2 \pi c$) of $S_{ChD}$ ((a) top) and the non-chiral component ((b) middle) in the impulsive pump regime. The bottom graph (c) shows the raw signal for both $S_{ChD}$ (left axis), the non-chiral pump-probe (right axis) and the sum of all components in the $N =13$ Prony decomposition.  The pump probe signal is dominated by a decay due to population relaxation (not shown in b) and is not reproduced well by the Prony decomposition.}
\label{Fig:prony_cD}
\end{figure}

In \reffig{Fig:prony_cD} we show slices through $\hbar \omega_s = 12495 $cm$^{-1}$ and take a Prony decomposition (into 13 complex components, the three largest amplitude are shown) of the first picosecond of signal. The decomposition in (a) of the $S_{ChD}$ is mainly comprised of two oscillating signals with frequencies $232$cm$^{-1}$ and $194$cm$^{-1}$ close to the frequencies of the effective vibronic states which we introduced in App.~\ref{App:vibronic_derivation}. The ordinary (non-chiral) pump probe signal (b) is dominated by two exponential decays (the larger amplitude is not shown on this scale) and a single oscillation at $223$cm$^{-1}$, close to $\hbar\omega_0$. Beatings from both the vibronic states and pure vibrational coherences are present in the non-chiral pump probe (and cannot be in $S_{ChD}$), along with effects from population transfers, hence we are unable to resolve all contributions at this level.  This analysis shows the simplicity in extracting beating components from our coherence specific signal. 

The presence of multiple oscillating signals lying on top of one another is partly a weakness of the impulsive pump configuration, as it is not possible to excite coherences between two particular states.  The Prony analysis fails to reproduce the ordinary pump-probe signal at late times which may be due to the complicated coherent population transition dynamics, which cannot be represented by exponential decays and hence Prony analysis.  Slow bath induced processes during the population time $\tau$ can also mix coherences pathways and hence cause signals to deviate from oscillating exponentials, however this effect is much less significant. 

\end{appendix}


\begin{thebibliography}{65}%
\makeatletter
\providecommand \@ifxundefined [1]{%
 \@ifx{#1\undefined}
}%
\providecommand \@ifnum [1]{%
 \ifnum #1\expandafter \@firstoftwo
 \else \expandafter \@secondoftwo
 \fi
}%
\providecommand \@ifx [1]{%
 \ifx #1\expandafter \@firstoftwo
 \else \expandafter \@secondoftwo
 \fi
}%
\providecommand \natexlab [1]{#1}%
\providecommand \enquote  [1]{``#1''}%
\providecommand \bibnamefont  [1]{#1}%
\providecommand \bibfnamefont [1]{#1}%
\providecommand \citenamefont [1]{#1}%
\providecommand \href@noop [0]{\@secondoftwo}%
\providecommand \href [0]{\begingroup \@sanitize@url \@href}%
\providecommand \@href[1]{\@@startlink{#1}\@@href}%
\providecommand \@@href[1]{\endgroup#1\@@endlink}%
\providecommand \@sanitize@url [0]{\catcode `\\12\catcode `\$12\catcode
  `\&12\catcode `\#12\catcode `\^12\catcode `\_12\catcode `\%12\relax}%
\providecommand \@@startlink[1]{}%
\providecommand \@@endlink[0]{}%
\providecommand \url  [0]{\begingroup\@sanitize@url \@url }%
\providecommand \@url [1]{\endgroup\@href {#1}{\urlprefix }}%
\providecommand \urlprefix  [0]{URL }%
\providecommand \Eprint [0]{\href }%
\providecommand \doibase [0]{http://dx.doi.org/}%
\providecommand \selectlanguage [0]{\@gobble}%
\providecommand \bibinfo  [0]{\@secondoftwo}%
\providecommand \bibfield  [0]{\@secondoftwo}%
\providecommand \translation [1]{[#1]}%
\providecommand \BibitemOpen [0]{}%
\providecommand \bibitemStop [0]{}%
\providecommand \bibitemNoStop [0]{.\EOS\space}%
\providecommand \EOS [0]{\spacefactor3000\relax}%
\providecommand \BibitemShut  [1]{\csname bibitem#1\endcsname}%
\let\auto@bib@innerbib\@empty
\bibitem [{\citenamefont {Polavarapu}, \citenamefont {Petrovic},\ and\
  \citenamefont {Zhang}(2006)}]{Polavarapu2006}%
  \BibitemOpen
  \bibfield  {author} {\bibinfo {author} {\bibfnamefont {P.~L.}\ \bibnamefont
  {Polavarapu}}, \bibinfo {author} {\bibfnamefont {A.~G.}\ \bibnamefont
  {Petrovic}}, \ and\ \bibinfo {author} {\bibfnamefont {P.}~\bibnamefont
  {Zhang}},\ }\bibfield  {title} {\enquote {\bibinfo {title} {Kramers kronig
  transformation of experimental electronic circular dichroism: Application to
  the analysis of optical rotatory dispersion in dimethyl-l-tartrate},}\ }\href
  {\doibase 10.1002/chir.20310} {\bibfield  {journal} {\bibinfo  {journal}
  {Chirality}\ }\textbf {\bibinfo {volume} {18}},\ \bibinfo {pages} {723--732}
  (\bibinfo {year} {2006})}\BibitemShut {NoStop}%
\bibitem [{\citenamefont {Kelly}\ and\ \citenamefont
  {Price}(2000)}]{Kelly2000}%
  \BibitemOpen
  \bibfield  {author} {\bibinfo {author} {\bibfnamefont {S.~M.}\ \bibnamefont
  {Kelly}}\ and\ \bibinfo {author} {\bibfnamefont {N.~C.}\ \bibnamefont
  {Price}},\ }\bibfield  {title} {\enquote {\bibinfo {title} {{The use of
  circular dichroism in the investigation of protein structure and
  function.}}}\ }\href {\doibase 10.2174/1389203003381315} {\bibfield
  {journal} {\bibinfo  {journal} {Curr. Protein. Pept. Sci.}\ }\textbf
  {\bibinfo {volume} {1}},\ \bibinfo {pages} {349--384} (\bibinfo {year}
  {2000})}\BibitemShut {NoStop}%
\bibitem [{Kel(2005)}]{Kelly2005}%
  \BibitemOpen
  \bibfield  {title} {\enquote {\bibinfo {title} {How to study proteins by
  circular dichroism},}\ }\href {\doibase
  http://dx.doi.org/10.1016/j.bbapap.2005.06.005} {\bibfield  {journal}
  {\bibinfo  {journal} {Biochim. Biophys. Acta}\ }\textbf {\bibinfo {volume}
  {1751}},\ \bibinfo {pages} {119 -- 139} (\bibinfo {year} {2005})}\BibitemShut
  {NoStop}%
\bibitem [{\citenamefont {Greenfield}(2006)}]{Greenfield2006}%
  \BibitemOpen
  \bibfield  {author} {\bibinfo {author} {\bibfnamefont {N.~J.}\ \bibnamefont
  {Greenfield}},\ }\bibfield  {title} {\enquote {\bibinfo {title} {Using
  circular dichroism spectra to estimate protein secondary structure},}\ }\href
  {\doibase 10.1038/nprot.2006.202} {\bibfield  {journal} {\bibinfo  {journal}
  {Nat. Protoc.}\ }\textbf {\bibinfo {volume} {1}},\ \bibinfo {pages} {2876}
  (\bibinfo {year} {2006})}\BibitemShut {NoStop}%
\bibitem [{\citenamefont {Furumaki}\ \emph {et~al.}(2012)\citenamefont
  {Furumaki}, \citenamefont {Yabiku}, \citenamefont {Habuchi}, \citenamefont
  {Tsukatani}, \citenamefont {Bryant},\ and\ \citenamefont
  {Vacha}}]{Furumaki2012}%
  \BibitemOpen
  \bibfield  {author} {\bibinfo {author} {\bibfnamefont {S.}~\bibnamefont
  {Furumaki}}, \bibinfo {author} {\bibfnamefont {Y.}~\bibnamefont {Yabiku}},
  \bibinfo {author} {\bibfnamefont {S.}~\bibnamefont {Habuchi}}, \bibinfo
  {author} {\bibfnamefont {Y.}~\bibnamefont {Tsukatani}}, \bibinfo {author}
  {\bibfnamefont {D.~A.}\ \bibnamefont {Bryant}}, \ and\ \bibinfo {author}
  {\bibfnamefont {M.}~\bibnamefont {Vacha}},\ }\bibfield  {title} {\enquote
  {\bibinfo {title} {{Circular dichroism measured on single chlorosomal
  light-harvesting complexes of green photosynthetic bacteria}},}\ }\href
  {\doibase 10.1021/jz301671p} {\bibfield  {journal} {\bibinfo  {journal} {J.
  Phys. Chem. Lett.}\ }\textbf {\bibinfo {volume} {3}},\ \bibinfo {pages}
  {3545--3549} (\bibinfo {year} {2012})}\BibitemShut {NoStop}%
\bibitem [{\citenamefont {Georgakopoulou}, \citenamefont {{Van Grondelle}},\
  and\ \citenamefont {{Van Der Zwan}}(2006)}]{Georgakopoulou2006}%
  \BibitemOpen
  \bibfield  {author} {\bibinfo {author} {\bibfnamefont {S.}~\bibnamefont
  {Georgakopoulou}}, \bibinfo {author} {\bibfnamefont {R.}~\bibnamefont {{Van
  Grondelle}}}, \ and\ \bibinfo {author} {\bibfnamefont {G.}~\bibnamefont {{Van
  Der Zwan}}},\ }\bibfield  {title} {\enquote {\bibinfo {title} {{Explaining
  the visible and near-infrared circular dichroism spectra of light-harvesting
  1 complexes from purple bacteria: A modeling study}},}\ }\href {\doibase
  10.1021/jp051794c} {\bibfield  {journal} {\bibinfo  {journal} {J. Phys. Chem.
  B}\ }\textbf {\bibinfo {volume} {110}},\ \bibinfo {pages} {3344--3353}
  (\bibinfo {year} {2006})}\BibitemShut {NoStop}%
\bibitem [{\citenamefont {Hemelrijk}\ \emph {et~al.}(1992)\citenamefont
  {Hemelrijk}, \citenamefont {Kwa}, \citenamefont {van Grondelle},\ and\
  \citenamefont {Dekker}}]{Hemelrijk1992}%
  \BibitemOpen
  \bibfield  {author} {\bibinfo {author} {\bibfnamefont {P.~W.}\ \bibnamefont
  {Hemelrijk}}, \bibinfo {author} {\bibfnamefont {S.~L.}\ \bibnamefont {Kwa}},
  \bibinfo {author} {\bibfnamefont {R.}~\bibnamefont {van Grondelle}}, \ and\
  \bibinfo {author} {\bibfnamefont {J.~P.}\ \bibnamefont {Dekker}},\ }\bibfield
   {title} {\enquote {\bibinfo {title} {{Spectroscopic properties of LHC-II,
  the main light-harvesting chlorophyll a/b protein complex from chloroplast
  membranes}},}\ }\bibfield  {booktitle} {\emph {\bibinfo {booktitle}
  {Biochimica et Biophysica Acta (BBA) - Bioenergetics}},\ }\href {\doibase
  10.1016/S0005-2728(05)80331-7} {\ \textbf {\bibinfo {volume} {1098}},\
  \bibinfo {pages} {159--166} (\bibinfo {year} {1992})}\BibitemShut {NoStop}%
\bibitem [{\citenamefont {{C. B{\"{u}}chel}}(1997)}]{Buchel1997}%
  \BibitemOpen
  \bibfield  {author} {\bibinfo {author} {\bibfnamefont {G.~G.}\ \bibnamefont
  {{C. B{\"{u}}chel}}},\ }\bibfield  {title} {\enquote {\bibinfo {title}
  {{Orgzation of the pigment molecules in the chlorophyll a/c light-harvesting
  complex of Pleurochloris meiringensis (xanthophyceae). Characterization with
  circular dichroism and absorbance spectroscopy}},}\ }\href@noop {} {\bibfield
   {journal} {\bibinfo  {journal} {J. Photochem. Photobiol.}\ }\textbf
  {\bibinfo {volume} {37}},\ \bibinfo {pages} {118--124} (\bibinfo {year}
  {1997})}\BibitemShut {NoStop}%
\bibitem [{\citenamefont {Bonmarin}\ and\ \citenamefont
  {Helbing}(2008)}]{Bonmarin2008}%
  \BibitemOpen
  \bibfield  {author} {\bibinfo {author} {\bibfnamefont {M.}~\bibnamefont
  {Bonmarin}}\ and\ \bibinfo {author} {\bibfnamefont {J.}~\bibnamefont
  {Helbing}},\ }\bibfield  {title} {\enquote {\bibinfo {title} {{A picosecond
  time-resolved vibrational circular dichroism spectrometer.}}}\ }\href
  {\doibase 10.1364/OL.33.002086} {\bibfield  {journal} {\bibinfo  {journal}
  {Opt. Lett.}\ }\textbf {\bibinfo {volume} {33}},\ \bibinfo {pages}
  {2086--2088} (\bibinfo {year} {2008})}\BibitemShut {NoStop}%
\bibitem [{\citenamefont {{Rhee}}\ \emph {et~al.}(2009)\citenamefont {{Rhee}},
  \citenamefont {{June}}, \citenamefont {{Lee}}, \citenamefont {{Lee}},
  \citenamefont {{Ha}}, \citenamefont {{Kim}}, \citenamefont {{Jeon}},\ and\
  \citenamefont {{Cho}}}]{Rhee2009}%
  \BibitemOpen
  \bibfield  {author} {\bibinfo {author} {\bibfnamefont {H.}~\bibnamefont
  {{Rhee}}}, \bibinfo {author} {\bibfnamefont {Y.-G.}\ \bibnamefont {{June}}},
  \bibinfo {author} {\bibfnamefont {J.-S.}\ \bibnamefont {{Lee}}}, \bibinfo
  {author} {\bibfnamefont {K.-K.}\ \bibnamefont {{Lee}}}, \bibinfo {author}
  {\bibfnamefont {J.-H.}\ \bibnamefont {{Ha}}}, \bibinfo {author}
  {\bibfnamefont {Z.~H.}\ \bibnamefont {{Kim}}}, \bibinfo {author}
  {\bibfnamefont {S.-J.}\ \bibnamefont {{Jeon}}}, \ and\ \bibinfo {author}
  {\bibfnamefont {M.}~\bibnamefont {{Cho}}},\ }\bibfield  {title} {\enquote
  {\bibinfo {title} {{Femtosecond characterization of vibrational optical
  activity of chiral molecules}},}\ }\href {\doibase 10.1038/nature07846}
  {\bibfield  {journal} {\bibinfo  {journal} {Nature}\ }\textbf {\bibinfo
  {volume} {458}},\ \bibinfo {pages} {310--313} (\bibinfo {year}
  {2009})}\BibitemShut {NoStop}%
\bibitem [{Hac(2009)}]{Hache2009}%
  \BibitemOpen
  \bibfield  {title} {\enquote {\bibinfo {title} {{Picosecond transient
  circular dichroism of the photoreceptor protein of the light-adapted form of
  Blepharisma japonicum}},}\ }\href {\doibase 10.1016/j.cplett.2009.10.059}
  {\bibfield  {journal} {\bibinfo  {journal} {Chem. Phys. Lett.}\ }\textbf
  {\bibinfo {volume} {483}},\ \bibinfo {pages} {133--137} (\bibinfo {year}
  {2009})}\BibitemShut {NoStop}%
\bibitem [{\citenamefont {Abramavicius}\ and\ \citenamefont
  {Mukamel}(2005)}]{Abramavicius2005}%
  \BibitemOpen
  \bibfield  {author} {\bibinfo {author} {\bibfnamefont {D.}~\bibnamefont
  {Abramavicius}}\ and\ \bibinfo {author} {\bibfnamefont {S.}~\bibnamefont
  {Mukamel}},\ }\bibfield  {title} {\enquote {\bibinfo {title} {{Time-domain
  chirally-sensitive three-pulse coherent probes of vibrational excitons in
  proteins}},}\ }\href {\doibase 10.1016/j.chemphys.2005.06.046} {\bibfield
  {journal} {\bibinfo  {journal} {Chem. Phys.}\ }\textbf {\bibinfo {volume}
  {318}},\ \bibinfo {pages} {50--70} (\bibinfo {year} {2005})},\  \BibitemShut {NoStop}%
\bibitem [{\citenamefont {{Qiu}}, \citenamefont {{Zachariah}},\ and\
  \citenamefont {{Hagen}}(2003)}]{Qiu2003}%
  \BibitemOpen
  \bibfield  {author} {\bibinfo {author} {\bibfnamefont {L.}~\bibnamefont
  {{Qiu}}}, \bibinfo {author} {\bibfnamefont {C.}~\bibnamefont {{Zachariah}}},
  \ and\ \bibinfo {author} {\bibfnamefont {S.~J.}\ \bibnamefont {{Hagen}}},\
  }\bibfield  {title} {\enquote {\bibinfo {title} {{Fast Chain Contraction
  during Protein Folding: ``Foldability'' and Collapse Dynamics}},}\ }\href
  {\doibase 10.1103/PhysRevLett.90.168103} {\bibfield  {journal} {\bibinfo
  {journal} {Phys. Rev. Lett.}\ }\textbf {\bibinfo {volume} {90}},\ \bibinfo
  {eid} {168103} (\bibinfo {year} {2003})}\BibitemShut {NoStop}%
\bibitem [{\citenamefont {Chen}\ \emph {et~al.}(1998)\citenamefont {Chen},
  \citenamefont {Wood}, \citenamefont {Fink},\ and\ \citenamefont
  {Kliger}}]{Chen1998}%
  \BibitemOpen
  \bibfield  {author} {\bibinfo {author} {\bibfnamefont {E.}~\bibnamefont
  {Chen}}, \bibinfo {author} {\bibfnamefont {M.~J.}\ \bibnamefont {Wood}},
  \bibinfo {author} {\bibfnamefont {A.~L.}\ \bibnamefont {Fink}}, \ and\
  \bibinfo {author} {\bibfnamefont {D.~S.}\ \bibnamefont {Kliger}},\ }\bibfield
   {title} {\enquote {\bibinfo {title} {{Time-resolved circular dichroism
  studies of protein folding intermediates of cytochrome c.}}}\ }\href
  {\doibase 10.1021/bi972369f} {\bibfield  {journal} {\bibinfo  {journal}
  {Biochemistry}\ }\textbf {\bibinfo {volume} {37}},\ \bibinfo {pages}
  {5589--5598} (\bibinfo {year} {1998})}\BibitemShut {NoStop}%
\bibitem [{\citenamefont {{Niezborala}}\ and\ \citenamefont
  {{Hache}}(2006)}]{Niezborala2007}%
  \BibitemOpen
  \bibfield  {author} {\bibinfo {author} {\bibfnamefont {C.}~\bibnamefont
  {{Niezborala}}}\ and\ \bibinfo {author} {\bibfnamefont {F.}~\bibnamefont
  {{Hache}}},\ }\bibfield  {title} {\enquote {\bibinfo {title} {{Measuring the
  dynamics of circular dichroism in a pump-probe experiment with a
  Babinet-Soleil Compensator}},}\ }\href {\doibase 10.1364/JOSAB.23.002418}
  {\bibfield  {journal} {\bibinfo  {journal} {J. Opt. Soc. Am. B}\ }\textbf
  {\bibinfo {volume} {23}},\ \bibinfo {pages} {2418--2424} (\bibinfo {year}
  {2006})}\BibitemShut {NoStop}%
\bibitem [{\citenamefont {{Fidler}}\ \emph {et~al.}(2014)\citenamefont
  {{Fidler}}, \citenamefont {{Singh}}, \citenamefont {{Long}}, \citenamefont
  {{Dahlberg}},\ and\ \citenamefont {{Engel}}}]{FidlerEngel2014}%
  \BibitemOpen
  \bibfield  {author} {\bibinfo {author} {\bibfnamefont {A.~F.}\ \bibnamefont
  {{Fidler}}}, \bibinfo {author} {\bibfnamefont {V.~P.}\ \bibnamefont
  {{Singh}}}, \bibinfo {author} {\bibfnamefont {P.~D.}\ \bibnamefont {{Long}}},
  \bibinfo {author} {\bibfnamefont {P.~D.}\ \bibnamefont {{Dahlberg}}}, \ and\
  \bibinfo {author} {\bibfnamefont {G.~S.}\ \bibnamefont {{Engel}}},\
  }\bibfield  {title} {\enquote {\bibinfo {title} {{Dynamic localization of
  electronic excitation in photosynthetic complexes revealed with chiral
  two-dimensional spectroscopy}},}\ }\href {\doibase 10.1038/ncomms4286}
  {\bibfield  {journal} {\bibinfo  {journal} {Nature Comm.}\ }\textbf {\bibinfo
  {volume} {5}},\ \bibinfo {eid} {3286} (\bibinfo {year} {2014})}\BibitemShut
  {NoStop}%
\bibitem [{\citenamefont {{Parson}}(2006)}]{ParsonMOS}%
  \BibitemOpen
  \bibfield  {author} {\bibinfo {author} {\bibfnamefont {W.}~\bibnamefont
  {{Parson}}},\ }\href@noop {} {\emph {\bibinfo {title} {{Modern Optical
  Spectroscopy}}}}\ (\bibinfo  {publisher} {Springer},\ \bibinfo {address} {New
  York City},\ \bibinfo {year} {2006})\BibitemShut {NoStop}%
\bibitem [{\citenamefont {{Mukamel}}(1999)}]{MukamelPoNLS}%
  \BibitemOpen
  \bibfield  {author} {\bibinfo {author} {\bibfnamefont {S.}~\bibnamefont
  {{Mukamel}}},\ }\href@noop {} {\emph {\bibinfo {title} {{Principles of
  Nonlinear Optical Spectroscopy}}}}\ (\bibinfo  {publisher} {Oxford University
  Press},\ \bibinfo {address} {Oxford},\ \bibinfo {year} {1999})\BibitemShut
  {NoStop}%
\bibitem [{\citenamefont {Nobuyuki~Harada}(1983)}]{NobuyukiCDS}%
  \BibitemOpen
  \bibfield  {author} {\bibinfo {author} {\bibfnamefont {K.~N.}\ \bibnamefont
  {Nobuyuki~Harada}},\ }\href@noop {} {\emph {\bibinfo {title} {{Circular
  dichroic spectroscopy: exciton coupling in organic stereochemistry}}}}\
  (\bibinfo  {publisher} {University Science Books},\ \bibinfo {address}
  {Herndon, VA},\ \bibinfo {year} {1983})\BibitemShut {NoStop}%
\bibitem [{\citenamefont {Abramavicius}\ and\ \citenamefont
  {Mukamel}(2006)}]{Abramavicius2006-2}%
  \BibitemOpen
  \bibfield  {author} {\bibinfo {author} {\bibfnamefont {D.}~\bibnamefont
  {Abramavicius}}\ and\ \bibinfo {author} {\bibfnamefont {S.}~\bibnamefont
  {Mukamel}},\ }\bibfield  {title} {\enquote {\bibinfo {title}
  {{Chirality-induced signals in coherent multidimensional spectroscopy of
  excitons.}}}\ }\href {\doibase 10.1063/1.2104527} {\bibfield  {journal}
  {\bibinfo  {journal} {J. Chem. Phys.}\ }\textbf {\bibinfo {volume} {124}},\
  \bibinfo {pages} {034113} (\bibinfo {year} {2006})}\BibitemShut {NoStop}%
\bibitem [{\citenamefont {{Egorova}}(2015)}]{Egorova2015}%
  \BibitemOpen
  \bibfield  {author} {\bibinfo {author} {\bibfnamefont {D.}~\bibnamefont
  {{Egorova}}},\ }\bibfield  {title} {\enquote {\bibinfo {title} {{Detection of
  dark states in two-dimensional electronic photon-echo signals via
  ground-state coherence}},}\ }\href {\doibase 10.1063/1.4921636} {\bibfield
  {journal} {\bibinfo  {journal} {J. Chem. Phys}\ }\textbf {\bibinfo {volume}
  {142}},\ \bibinfo {eid} {212452} (\bibinfo {year} {2015})}\BibitemShut
  {NoStop}%
\bibitem [{\citenamefont {Fassioli}\ \emph {et~al.}(2013)\citenamefont
  {Fassioli}, \citenamefont {Dinshaw}, \citenamefont {Arpin},\ and\
  \citenamefont {Scholes}}]{Fassioli2013}%
  \BibitemOpen
  \bibfield  {author} {\bibinfo {author} {\bibfnamefont {F.}~\bibnamefont
  {Fassioli}}, \bibinfo {author} {\bibfnamefont {R.}~\bibnamefont {Dinshaw}},
  \bibinfo {author} {\bibfnamefont {P.~C.}\ \bibnamefont {Arpin}}, \ and\
  \bibinfo {author} {\bibfnamefont {G.~D.}\ \bibnamefont {Scholes}},\
  }\bibfield  {title} {\enquote {\bibinfo {title} {Photosynthetic light
  harvesting: excitons and coherence},}\ }\href {\doibase
  10.1098/rsif.2013.0901} {\bibfield  {journal} {\bibinfo  {journal} {J. R.
  Soc. Interface}\ }\textbf {\bibinfo {volume} {11}} (\bibinfo {year} {2013}),\
  10.1098/rsif.2013.0901},\
  \BibitemShut {NoStop}%
\bibitem [{\citenamefont {Engel}\ \emph {et~al.}(2007)\citenamefont {Engel},
  \citenamefont {Calhoun}, \citenamefont {Read}, \citenamefont {Ahn},
  \citenamefont {Mancal}, \citenamefont {Cheng}, \citenamefont {Blankenship},\
  and\ \citenamefont {Fleming}}]{Engel2007}%
  \BibitemOpen
  \bibfield  {author} {\bibinfo {author} {\bibfnamefont {G.~S.}\ \bibnamefont
  {Engel}}, \bibinfo {author} {\bibfnamefont {T.~R.}\ \bibnamefont {Calhoun}},
  \bibinfo {author} {\bibfnamefont {E.~L.}\ \bibnamefont {Read}}, \bibinfo
  {author} {\bibfnamefont {T.-K.}\ \bibnamefont {Ahn}}, \bibinfo {author}
  {\bibfnamefont {T.}~\bibnamefont {Mancal}}, \bibinfo {author} {\bibfnamefont
  {Y.-C.}\ \bibnamefont {Cheng}}, \bibinfo {author} {\bibfnamefont {R.~E.}\
  \bibnamefont {Blankenship}}, \ and\ \bibinfo {author} {\bibfnamefont {G.~R.}\
  \bibnamefont {Fleming}},\ }\bibfield  {title} {\enquote {\bibinfo {title}
  {{Evidence for wavelike energy transfer through quantum coherence in
  photosynthetic systems.}}}\ }\href {\doibase 10.1038/nature05678} {\bibfield
  {journal} {\bibinfo  {journal} {Nature}\ }\textbf {\bibinfo {volume} {446}},\
  \bibinfo {pages} {782--786} (\bibinfo {year} {2007})}\BibitemShut {NoStop}%
\bibitem [{\citenamefont {Calhoun}\ \emph {et~al.}(2009)\citenamefont
  {Calhoun}, \citenamefont {Ginsberg}, \citenamefont {Schlau-Cohen},
  \citenamefont {Cheng}, \citenamefont {Ballottari}, \citenamefont {Bassi},\
  and\ \citenamefont {Fleming}}]{Calhoun2009}%
  \BibitemOpen
  \bibfield  {author} {\bibinfo {author} {\bibfnamefont {T.~R.}\ \bibnamefont
  {Calhoun}}, \bibinfo {author} {\bibfnamefont {N.~S.}\ \bibnamefont
  {Ginsberg}}, \bibinfo {author} {\bibfnamefont {G.~S.}\ \bibnamefont
  {Schlau-Cohen}}, \bibinfo {author} {\bibfnamefont {Y.-C.}\ \bibnamefont
  {Cheng}}, \bibinfo {author} {\bibfnamefont {M.}~\bibnamefont {Ballottari}},
  \bibinfo {author} {\bibfnamefont {R.}~\bibnamefont {Bassi}}, \ and\ \bibinfo
  {author} {\bibfnamefont {G.~R.}\ \bibnamefont {Fleming}},\ }\bibfield
  {title} {\enquote {\bibinfo {title} {Quantum coherence enabled determination
  of the energy landscape in light-harvesting complex {II}},}\ }\href {\doibase
  10.1021/jp908300c} {\bibfield  {journal} {\bibinfo  {journal} {J. Phys. Chem.
  B}\ }\textbf {\bibinfo {volume} {113}},\ \bibinfo {pages} {16291--16295}
  (\bibinfo {year} {2009})},\  \BibitemShut {NoStop}%
\bibitem [{\citenamefont {Collini}\ \emph {et~al.}(2010)\citenamefont
  {Collini}, \citenamefont {Wong}, \citenamefont {Wilk}, \citenamefont {Curmi},
  \citenamefont {Brumer},\ and\ \citenamefont {Scholes}}]{Collini2010}%
  \BibitemOpen
  \bibfield  {author} {\bibinfo {author} {\bibfnamefont {E.}~\bibnamefont
  {Collini}}, \bibinfo {author} {\bibfnamefont {C.~Y.}\ \bibnamefont {Wong}},
  \bibinfo {author} {\bibfnamefont {K.~E.}\ \bibnamefont {Wilk}}, \bibinfo
  {author} {\bibfnamefont {P.~M.~G.}\ \bibnamefont {Curmi}}, \bibinfo {author}
  {\bibfnamefont {P.}~\bibnamefont {Brumer}}, \ and\ \bibinfo {author}
  {\bibfnamefont {G.~D.}\ \bibnamefont {Scholes}},\ }\bibfield  {title}
  {\enquote {\bibinfo {title} {Coherently wired light-harvesting in
  photosynthetic marine algae at ambient temperature},}\ }\href {\doibase
  10.1038/nature08811} {\bibfield  {journal} {\bibinfo  {journal} {Nature}\
  }\textbf {\bibinfo {volume} {463}},\ \bibinfo {pages} {644--647} (\bibinfo
  {year} {2010})}\BibitemShut {NoStop}%
\bibitem [{\citenamefont {Panitchayangkoon}\ \emph {et~al.}(2010)\citenamefont
  {Panitchayangkoon}, \citenamefont {Hayes}, \citenamefont {Fransted},
  \citenamefont {Caram}, \citenamefont {Harel}, \citenamefont {Wen},
  \citenamefont {Blankenship},\ and\ \citenamefont
  {Engel}}]{Panitchayangkoon2010}%
  \BibitemOpen
  \bibfield  {author} {\bibinfo {author} {\bibfnamefont {G.}~\bibnamefont
  {Panitchayangkoon}}, \bibinfo {author} {\bibfnamefont {D.}~\bibnamefont
  {Hayes}}, \bibinfo {author} {\bibfnamefont {K.~A.}\ \bibnamefont {Fransted}},
  \bibinfo {author} {\bibfnamefont {J.~R.}\ \bibnamefont {Caram}}, \bibinfo
  {author} {\bibfnamefont {E.}~\bibnamefont {Harel}}, \bibinfo {author}
  {\bibfnamefont {J.}~\bibnamefont {Wen}}, \bibinfo {author} {\bibfnamefont
  {R.~E.}\ \bibnamefont {Blankenship}}, \ and\ \bibinfo {author} {\bibfnamefont
  {G.~S.}\ \bibnamefont {Engel}},\ }\bibfield  {title} {\enquote {\bibinfo
  {title} {Long-lived quantum coherence in photosynthetic complexes at
  physiological temperature},}\ }\href {\doibase 10.1073/pnas.1005484107}
  {\bibfield  {journal} {\bibinfo  {journal} {Proc. Natl. Acad. Sci. USA}\
  }\textbf {\bibinfo {volume} {107}},\ \bibinfo {pages} {12766--12770}
  (\bibinfo {year} {2010})},\  \BibitemShut
  {NoStop}%
\bibitem [{\citenamefont {Collini}\ and\ \citenamefont
  {Scholes}(2009)}]{Collini2009}%
  \BibitemOpen
  \bibfield  {author} {\bibinfo {author} {\bibfnamefont {E.}~\bibnamefont
  {Collini}}\ and\ \bibinfo {author} {\bibfnamefont {G.~D.}\ \bibnamefont
  {Scholes}},\ }\bibfield  {title} {\enquote {\bibinfo {title} {{Electronic and
  vibrational coherences in resonance energy transfer along MEH-PPV chains at
  room temperature'}},}\ }\href {\doibase 10.1021/jp810757x} {\bibfield
  {journal} {\bibinfo  {journal} {J. Phys. Chem. A}\ }\textbf {\bibinfo
  {volume} {113}},\ \bibinfo {pages} {4223--4241} (\bibinfo {year}
  {2009})}\BibitemShut {NoStop}%
\bibitem [{\citenamefont {Halpin}\ \emph {et~al.}(2014)\citenamefont {Halpin},
  \citenamefont {Johnson}, \citenamefont {Tempelaar}, \citenamefont {Murphy},
  \citenamefont {Knoester}, \citenamefont {Jansen},\ and\ \citenamefont
  {Miller}}]{Halpin2014}%
  \BibitemOpen
  \bibfield  {author} {\bibinfo {author} {\bibfnamefont {A.}~\bibnamefont
  {Halpin}}, \bibinfo {author} {\bibfnamefont {P.~J.~M.}\ \bibnamefont
  {Johnson}}, \bibinfo {author} {\bibfnamefont {R.}~\bibnamefont {Tempelaar}},
  \bibinfo {author} {\bibfnamefont {R.~S.}\ \bibnamefont {Murphy}}, \bibinfo
  {author} {\bibfnamefont {J.}~\bibnamefont {Knoester}}, \bibinfo {author}
  {\bibfnamefont {T.~L.~C.}\ \bibnamefont {Jansen}}, \ and\ \bibinfo {author}
  {\bibfnamefont {R.~J.~D.}\ \bibnamefont {Miller}},\ }\bibfield  {title}
  {\enquote {\bibinfo {title} {{Two-dimensional spectroscopy of a molecular
  dimer unveils the effects of vibronic coupling on exciton coherences.}}}\
  }\href {\doibase 10.1038/nchem.1834} {\bibfield  {journal} {\bibinfo
  {journal} {Nature chem.}\ }\textbf {\bibinfo {volume} {6}},\ \bibinfo {pages}
  {196--201} (\bibinfo {year} {2014})}\BibitemShut {NoStop}%
\bibitem [{\citenamefont {Kolli}\ \emph {et~al.}(2012)\citenamefont {Kolli},
  \citenamefont {O'Reilly}, \citenamefont {Scholes},\ and\ \citenamefont
  {Olaya-Castro}}]{Kolli2012}%
  \BibitemOpen
  \bibfield  {author} {\bibinfo {author} {\bibfnamefont {A.}~\bibnamefont
  {Kolli}}, \bibinfo {author} {\bibfnamefont {E.~J.}\ \bibnamefont {O'Reilly}},
  \bibinfo {author} {\bibfnamefont {G.~D.}\ \bibnamefont {Scholes}}, \ and\
  \bibinfo {author} {\bibfnamefont {A.}~\bibnamefont {Olaya-Castro}},\
  }\bibfield  {title} {\enquote {\bibinfo {title} {{The fundamental role of
  quantized vibrations in coherent light harvesting by cryptophyte algae}},}\
  }\href@noop {} {\bibfield  {journal} {\bibinfo  {journal} {J. Chem. Phys.}\
  }\textbf {\bibinfo {volume} {{137}}} (\bibinfo {year} {{2012}})}\BibitemShut
  {NoStop}%
\bibitem [{\citenamefont {Christensson}\ \emph {et~al.}(2012)\citenamefont
  {Christensson}, \citenamefont {Kauffmann}, \citenamefont {Pullerits},\ and\
  \citenamefont {Manc{\v a}l}}]{Christensson2012}%
  \BibitemOpen
  \bibfield  {author} {\bibinfo {author} {\bibfnamefont {N.}~\bibnamefont
  {Christensson}}, \bibinfo {author} {\bibfnamefont {H.~F.}\ \bibnamefont
  {Kauffmann}}, \bibinfo {author} {\bibfnamefont {T.}~\bibnamefont
  {Pullerits}}, \ and\ \bibinfo {author} {\bibfnamefont {T.}~\bibnamefont
  {Manc{\v a}l}},\ }\bibfield  {title} {\enquote {\bibinfo {title} {Origin of
  long-lived coherences in light-harvesting complexes},}\ }\href {\doibase
  10.1021/jp304649c} {\bibfield  {journal} {\bibinfo  {journal} {J. Phys. Chem.
  B}\ }\textbf {\bibinfo {volume} {116}},\ \bibinfo {pages} {7449--7454}
  (\bibinfo {year} {2012})},\  \BibitemShut {NoStop}%
\bibitem [{\citenamefont {{Chin}}\ \emph {et~al.}(2013)\citenamefont {{Chin}},
  \citenamefont {{Prior}}, \citenamefont {{Rosenbach}}, \citenamefont
  {{Caycedo-Soler}}, \citenamefont {{Huelga}},\ and\ \citenamefont
  {{Plenio}}}]{Chin2013}%
  \BibitemOpen
  \bibfield  {author} {\bibinfo {author} {\bibfnamefont {A.~W.}\ \bibnamefont
  {{Chin}}}, \bibinfo {author} {\bibfnamefont {J.}~\bibnamefont {{Prior}}},
  \bibinfo {author} {\bibfnamefont {R.}~\bibnamefont {{Rosenbach}}}, \bibinfo
  {author} {\bibfnamefont {F.}~\bibnamefont {{Caycedo-Soler}}}, \bibinfo
  {author} {\bibfnamefont {S.~F.}\ \bibnamefont {{Huelga}}}, \ and\ \bibinfo
  {author} {\bibfnamefont {M.~B.}\ \bibnamefont {{Plenio}}},\ }\bibfield
  {title} {\enquote {\bibinfo {title} {{The role of non-equilibrium vibrational
  structures in electronic coherence and recoherence in pigment-protein
  complexes}},}\ }\href {\doibase 10.1038/nphys2515} {\bibfield  {journal}
  {\bibinfo  {journal} {Nat. Phys.}\ }\textbf {\bibinfo {volume} {9}},\
  \bibinfo {pages} {113--118} (\bibinfo {year} {{2013}})}\BibitemShut {NoStop}%
\bibitem [{\citenamefont {O'Reilly}\ and\ \citenamefont
  {Olaya-Castro}(2014)}]{OReilly2014}%
  \BibitemOpen
  \bibfield  {author} {\bibinfo {author} {\bibfnamefont {E.~J.}\ \bibnamefont
  {O'Reilly}}\ and\ \bibinfo {author} {\bibfnamefont {A.}~\bibnamefont
  {Olaya-Castro}},\ }\bibfield  {title} {\enquote {\bibinfo {title}
  {{Non-classicality of the molecular vibrations assisting exciton energy
  transfer at room temperature.}}}\ }\href {\doibase 10.1038/ncomms4012}
  {\bibfield  {journal} {\bibinfo  {journal} {Nature comm.}\ }\textbf {\bibinfo
  {volume} {5}},\ \bibinfo {pages} {3012} (\bibinfo {year} {2014})},\  \BibitemShut {NoStop}%
\bibitem [{\citenamefont {Womick}\ and\ \citenamefont
  {Moran}(2011)}]{Womick2011}%
  \BibitemOpen
  \bibfield  {author} {\bibinfo {author} {\bibfnamefont {J.~M.}\ \bibnamefont
  {Womick}}\ and\ \bibinfo {author} {\bibfnamefont {A.~M.}\ \bibnamefont
  {Moran}},\ }\bibfield  {title} {\enquote {\bibinfo {title} {Vibronic
  enhancement of exciton sizes and energy transport in photosynthetic
  complexes},}\ }\href {\doibase 10.1021/jp106713q} {\bibfield  {journal}
  {\bibinfo  {journal} {J. Phys. Chem. B}\ }\textbf {\bibinfo {volume} {115}},\
  \bibinfo {pages} {1347--1356} (\bibinfo {year} {2011})},\  \BibitemShut {NoStop}%
\bibitem [{\citenamefont {Richards}\ \emph {et~al.}(2012)\citenamefont
  {Richards}, \citenamefont {Wilk}, \citenamefont {Curmi}, \citenamefont
  {Quiney},\ and\ \citenamefont {Davis}}]{Richards2012}%
  \BibitemOpen
  \bibfield  {author} {\bibinfo {author} {\bibfnamefont {G.~H.}\ \bibnamefont
  {Richards}}, \bibinfo {author} {\bibfnamefont {K.~E.}\ \bibnamefont {Wilk}},
  \bibinfo {author} {\bibfnamefont {P.~M.~G.}\ \bibnamefont {Curmi}}, \bibinfo
  {author} {\bibfnamefont {H.~M.}\ \bibnamefont {Quiney}}, \ and\ \bibinfo
  {author} {\bibfnamefont {J.~A.}\ \bibnamefont {Davis}},\ }\bibfield  {title}
  {\enquote {\bibinfo {title} {Coherent vibronic coupling in light-harvesting
  complexes from photosynthetic marine algae},}\ }\href {\doibase
  10.1021/jz201600f} {\bibfield  {journal} {\bibinfo  {journal} {J. Phys. Chem.
  Lett.}\ }\textbf {\bibinfo {volume} {3}},\ \bibinfo {pages} {272--277}
  (\bibinfo {year} {2012})},\  \BibitemShut {NoStop}%
\bibitem [{\citenamefont {Novelli}\ \emph {et~al.}(2015)\citenamefont
  {Novelli}, \citenamefont {Nazir}, \citenamefont {Richards}, \citenamefont
  {Roozbeh}, \citenamefont {Wilk}, \citenamefont {Curmi},\ and\ \citenamefont
  {Davis}}]{Novelli2015}%
  \BibitemOpen
  \bibfield  {author} {\bibinfo {author} {\bibfnamefont {F.}~\bibnamefont
  {Novelli}}, \bibinfo {author} {\bibfnamefont {A.}~\bibnamefont {Nazir}},
  \bibinfo {author} {\bibfnamefont {G.~H.}\ \bibnamefont {Richards}}, \bibinfo
  {author} {\bibfnamefont {A.}~\bibnamefont {Roozbeh}}, \bibinfo {author}
  {\bibfnamefont {K.~E.}\ \bibnamefont {Wilk}}, \bibinfo {author}
  {\bibfnamefont {P.~M.~G.}\ \bibnamefont {Curmi}}, \ and\ \bibinfo {author}
  {\bibfnamefont {J.~A.}\ \bibnamefont {Davis}},\ }\bibfield  {title} {\enquote
  {\bibinfo {title} {Vibronic resonances facilitate excited-state coherence in
  light-harvesting proteins at room temperature},}\ }\href {\doibase
  10.1021/acs.jpclett.5b02058} {\bibfield  {journal} {\bibinfo  {journal} {J.
  Phys. Chem. Lett.}\ }\textbf {\bibinfo {volume} {6}},\ \bibinfo {pages}
  {4573--4580} (\bibinfo {year} {2015})}\BibitemShut {NoStop}%
\bibitem [{\citenamefont {{Singh}}\ \emph {et~al.}(2015)\citenamefont
  {{Singh}}, \citenamefont {{Westberg}}, \citenamefont {{Wang}}, \citenamefont
  {{Dahlberg}}, \citenamefont {{Gellen}}, \citenamefont {{Gardiner}},
  \citenamefont {{Cogdell}},\ and\ \citenamefont {{Engel}}}]{Singh2015}%
  \BibitemOpen
  \bibfield  {author} {\bibinfo {author} {\bibfnamefont {V.~P.}\ \bibnamefont
  {{Singh}}}, \bibinfo {author} {\bibfnamefont {M.}~\bibnamefont {{Westberg}}},
  \bibinfo {author} {\bibfnamefont {C.}~\bibnamefont {{Wang}}}, \bibinfo
  {author} {\bibfnamefont {P.~D.}\ \bibnamefont {{Dahlberg}}}, \bibinfo
  {author} {\bibfnamefont {T.}~\bibnamefont {{Gellen}}}, \bibinfo {author}
  {\bibfnamefont {A.~T.}\ \bibnamefont {{Gardiner}}}, \bibinfo {author}
  {\bibfnamefont {R.~J.}\ \bibnamefont {{Cogdell}}}, \ and\ \bibinfo {author}
  {\bibfnamefont {G.~S.}\ \bibnamefont {{Engel}}},\ }\bibfield  {title}
  {\enquote {\bibinfo {title} {{Towards quantification of vibronic coupling in
  photosynthetic antenna complexes}},}\ }\href {\doibase 10.1063/1.4921324}
  {\bibfield  {journal} {\bibinfo  {journal} {J. Chem. Phys.}\ }\textbf
  {\bibinfo {volume} {142}},\ \bibinfo {eid} {212446} (\bibinfo {year}
  {2015})}\BibitemShut {NoStop}%
\bibitem [{\citenamefont {Tiwari}, \citenamefont {Peters},\ and\ \citenamefont
  {Jonas}(2012)}]{Tiwari2012}%
  \BibitemOpen
  \bibfield  {author} {\bibinfo {author} {\bibfnamefont {V.}~\bibnamefont
  {Tiwari}}, \bibinfo {author} {\bibfnamefont {W.~K.}\ \bibnamefont {Peters}},
  \ and\ \bibinfo {author} {\bibfnamefont {D.~M.}\ \bibnamefont {Jonas}},\
  }\bibfield  {title} {\enquote {\bibinfo {title} {Electronic resonance with
  anticorrelated pigment vibrations drives photosynthetic energy transfer
  outside the adiabatic framework},}\ }\href {\doibase 10.1073/pnas.1211157110}
  {\bibfield  {journal} {\bibinfo  {journal} {Proc. Natl. Acad. Sci. USA}\ }
  (\bibinfo {year} {2012}),\ 10.1073/pnas.1211157110},\ 
  \BibitemShut {NoStop}%
\bibitem [{\citenamefont {Plenio}, \citenamefont {Almeida},\ and\ \citenamefont
  {Huelga}(2013)}]{Plenio2013}%
  \BibitemOpen
  \bibfield  {author} {\bibinfo {author} {\bibfnamefont {M.~B.}\ \bibnamefont
  {Plenio}}, \bibinfo {author} {\bibfnamefont {J.}~\bibnamefont {Almeida}}, \
  and\ \bibinfo {author} {\bibfnamefont {S.~F.}\ \bibnamefont {Huelga}},\
  }\bibfield  {title} {\enquote {\bibinfo {title} {{Origin of long-lived
  oscillations in 2D-spectra of a quantum vibronic model: electronic versus
  vibrational coherence.}}}\ }\href {\doibase 10.1063/1.4846275} {\bibfield
  {journal} {\bibinfo  {journal} {J. chem. phys.}\ }\textbf {\bibinfo {volume}
  {139}},\ \bibinfo {pages} {235102} (\bibinfo {year} {2013})},\  \BibitemShut {NoStop}%
\bibitem [{\citenamefont {Liebel}\ \emph {et~al.}(2015)\citenamefont {Liebel},
  \citenamefont {Schnedermann}, \citenamefont {Wende},\ and\ \citenamefont
  {Kukura}}]{Liebel2015}%
  \BibitemOpen
  \bibfield  {author} {\bibinfo {author} {\bibfnamefont {M.}~\bibnamefont
  {Liebel}}, \bibinfo {author} {\bibfnamefont {C.}~\bibnamefont
  {Schnedermann}}, \bibinfo {author} {\bibfnamefont {T.}~\bibnamefont {Wende}},
  \ and\ \bibinfo {author} {\bibfnamefont {P.}~\bibnamefont {Kukura}},\
  }\bibfield  {title} {\enquote {\bibinfo {title} {Principles and applications
  of broadband impulsive vibrational spectroscopy},}\ }\href {\doibase
  10.1021/acs.jpca.5b05948} {\bibfield  {journal} {\bibinfo  {journal} {The J.
  Phys. Chem. A}\ }\textbf {\bibinfo {volume} {119}},\ \bibinfo {pages}
  {9506--9517} (\bibinfo {year} {2015})},\ \bibinfo {note} {pMID:
  26262557}\BibitemShut {NoStop}%
\bibitem [{\citenamefont {{Lim}}\ \emph {et~al.}(2015)\citenamefont {{Lim}},
  \citenamefont {{Pale{\v c}ek}}, \citenamefont {{Caycedo-Soler}},
  \citenamefont {{Lincoln}}, \citenamefont {{Prior}}, \citenamefont {{von
  Berlepsch}}, \citenamefont {{Huelga}}, \citenamefont {{Plenio}},
  \citenamefont {{Zigmantas}},\ and\ \citenamefont {{Hauer}}}]{Lim2015}%
  \BibitemOpen
  \bibfield  {author} {\bibinfo {author} {\bibfnamefont {J.}~\bibnamefont
  {{Lim}}}, \bibinfo {author} {\bibfnamefont {D.}~\bibnamefont {{Pale{\v
  c}ek}}}, \bibinfo {author} {\bibfnamefont {F.}~\bibnamefont
  {{Caycedo-Soler}}}, \bibinfo {author} {\bibfnamefont {C.~N.}\ \bibnamefont
  {{Lincoln}}}, \bibinfo {author} {\bibfnamefont {J.}~\bibnamefont {{Prior}}},
  \bibinfo {author} {\bibfnamefont {H.}~\bibnamefont {{von Berlepsch}}},
  \bibinfo {author} {\bibfnamefont {S.~F.}\ \bibnamefont {{Huelga}}}, \bibinfo
  {author} {\bibfnamefont {M.~B.}\ \bibnamefont {{Plenio}}}, \bibinfo {author}
  {\bibfnamefont {D.}~\bibnamefont {{Zigmantas}}}, \ and\ \bibinfo {author}
  {\bibfnamefont {J.}~\bibnamefont {{Hauer}}},\ }\bibfield  {title} {\enquote
  {\bibinfo {title} {{Vibronic origin of long-lived coherence in an artificial
  molecular light harvester}},}\ }\href {\doibase 10.1038/ncomms8755}
  {\bibfield  {journal} {\bibinfo  {journal} {Nature Comm.}\ }\textbf {\bibinfo
  {volume} {6}},\ \bibinfo {eid} {7755} (\bibinfo {year} {2015})},\ 
  \BibitemShut {NoStop}%
\bibitem [{\citenamefont {{Valkunas}}, \citenamefont {Abramavicius},\ and\
  \citenamefont {Mancal}(2013)}]{ValkunasMEDaR}%
  \BibitemOpen
  \bibfield  {author} {\bibinfo {author} {\bibfnamefont {L.}~\bibnamefont
  {{Valkunas}}}, \bibinfo {author} {\bibfnamefont {D.}~\bibnamefont
  {Abramavicius}}, \ and\ \bibinfo {author} {\bibfnamefont {T.}~\bibnamefont
  {Mancal}},\ }\href@noop {} {\emph {\bibinfo {title} {{Molecular excitation
  dynamics and relaxation}}}}\ (\bibinfo  {publisher} {Wiley-VCH},\ \bibinfo
  {address} {Weinheim},\ \bibinfo {year} {2013})\BibitemShut {NoStop}%
\bibitem [{\citenamefont {{Chen}}\ \emph {et~al.}(2010)\citenamefont {{Chen}},
  \citenamefont {{Zheng}}, \citenamefont {{Shi}},\ and\ \citenamefont
  {{Yan}}}]{Chen2010}%
  \BibitemOpen
  \bibfield  {author} {\bibinfo {author} {\bibfnamefont {L.}~\bibnamefont
  {{Chen}}}, \bibinfo {author} {\bibfnamefont {R.}~\bibnamefont {{Zheng}}},
  \bibinfo {author} {\bibfnamefont {Q.}~\bibnamefont {{Shi}}}, \ and\ \bibinfo
  {author} {\bibfnamefont {Y.}~\bibnamefont {{Yan}}},\ }\bibfield  {title}
  {\enquote {\bibinfo {title} {{Two-dimensional electronic spectra from the
  hierarchical equations of motion method: Application to model dimers}},}\
  }\href {\doibase 10.1063/1.3293039} {\bibfield  {journal} {\bibinfo
  {journal} {J. Chem. Phys.}\ }\textbf {\bibinfo {volume} {132}},\ \bibinfo
  {pages} {024505} (\bibinfo {year} {2010})}\BibitemShut {NoStop}%
\bibitem [{\citenamefont {Breuer}\ and\ \citenamefont {F.}(2002)}]{BreuerTOQS}%
  \BibitemOpen
  \bibfield  {author} {\bibinfo {author} {\bibfnamefont {H.~P.}\ \bibnamefont
  {Breuer}}\ and\ \bibinfo {author} {\bibfnamefont {P.}~\bibnamefont {F.}},\
  }\href@noop {} {\emph {\bibinfo {title} {{The theory of open quantum
  systems}}}}\ (\bibinfo  {publisher} {Oxford University Press},\ \bibinfo
  {address} {Oxford},\ \bibinfo {year} {2002})\BibitemShut {NoStop}%
\bibitem [{\citenamefont {Condon}(1926)}]{Condon1926}%
  \BibitemOpen
  \bibfield  {author} {\bibinfo {author} {\bibfnamefont {E.}~\bibnamefont
  {Condon}},\ }\bibfield  {title} {\enquote {\bibinfo {title} {A theory of
  intensity distribution in band systems},}\ }\href {\doibase
  10.1103/PhysRev.28.1182} {\bibfield  {journal} {\bibinfo  {journal} {Phys.
  Rev.}\ }\textbf {\bibinfo {volume} {28}},\ \bibinfo {pages} {1182--1201}
  (\bibinfo {year} {1926})}\BibitemShut {NoStop}%
\bibitem [{\citenamefont {Rosenfeld}(1928)}]{Rosenfeld1928}%
  \BibitemOpen
  \bibfield  {author} {\bibinfo {author} {\bibfnamefont {L.}~\bibnamefont
  {Rosenfeld}},\ }\bibfield  {title} {\enquote {\bibinfo {title}
  {{Quantenmechanische Theorie der natuerlichen optischen Aktivitaet von
  Fluessigkeiten und Gasen}},}\ }\href {\doibase 10.1007/BF01342393} {\bibfield
   {journal} {\bibinfo  {journal} {Z. Phys.}\ }\textbf {\bibinfo {volume}
  {52}},\ \bibinfo {pages} {161--174} (\bibinfo {year} {1928})}\BibitemShut
  {NoStop}%
\bibitem [{\citenamefont {Deflores}, \citenamefont {Nicodemus},\ and\
  \citenamefont {Tokmakoff}(2007)}]{Deflores2007}%
  \BibitemOpen
  \bibfield  {author} {\bibinfo {author} {\bibfnamefont {L.~P.}\ \bibnamefont
  {Deflores}}, \bibinfo {author} {\bibfnamefont {R.~a.}\ \bibnamefont
  {Nicodemus}}, \ and\ \bibinfo {author} {\bibfnamefont {A.}~\bibnamefont
  {Tokmakoff}},\ }\bibfield  {title} {\enquote {\bibinfo {title}
  {{Two-dimensional Fourier transform spectroscopy in the pump-probe
  geometry.}}}\ }\href {\doibase 10.1364/OL.32.002966} {\bibfield  {journal}
  {\bibinfo  {journal} {Opt. lett.}\ }\textbf {\bibinfo {volume} {32}},\
  \bibinfo {pages} {2966--2968} (\bibinfo {year} {2007})}\BibitemShut {NoStop}%
\bibitem [{\citenamefont {{Cho}}(2003)}]{Cho2003}%
  \BibitemOpen
  \bibfield  {author} {\bibinfo {author} {\bibfnamefont {M.}~\bibnamefont
  {{Cho}}},\ }\bibfield  {title} {\enquote {\bibinfo {title} {{Two-dimensional
  circularly polarized pump-probe spectroscopy}},}\ }\href {\doibase
  10.1063/1.1599344} {\bibfield  {journal} {\bibinfo  {journal} {J. Chem.
  Phys.}\ }\textbf {\bibinfo {volume} {119}},\ \bibinfo {pages} {7003--7016}
  (\bibinfo {year} {2003})}\BibitemShut {NoStop}%
\bibitem [{\citenamefont {{Abramavicius}}, \citenamefont {{Zhuang}},\ and\
  \citenamefont {{Mukamel}}(2006)}]{Abramavicius2006}%
  \BibitemOpen
  \bibfield  {author} {\bibinfo {author} {\bibfnamefont {D.}~\bibnamefont
  {{Abramavicius}}}, \bibinfo {author} {\bibfnamefont {W.}~\bibnamefont
  {{Zhuang}}}, \ and\ \bibinfo {author} {\bibfnamefont {S.}~\bibnamefont
  {{Mukamel}}},\ }\bibfield  {title} {\enquote {\bibinfo {title} {{Probing
  molecular chirality via excitonic nonlinear response}},}\ }\href {\doibase
  10.1088/0953-4075/39/24/003} {\bibfield  {journal} {\bibinfo  {journal} {J.
  Phys. B}\ }\textbf {\bibinfo {volume} {39}},\ \bibinfo {pages} {5051--5066}
  (\bibinfo {year} {2006})}\BibitemShut {NoStop}%
\bibitem [{\citenamefont {{Wagni{\`e}re}}(1982)}]{Wagniere1982}%
  \BibitemOpen
  \bibfield  {author} {\bibinfo {author} {\bibfnamefont {G.}~\bibnamefont
  {{Wagni{\`e}re}}},\ }\bibfield  {title} {\enquote {\bibinfo {title} {{The
  evaluation of three-dimensional rotational averages}},}\ }\href {\doibase
  10.1063/1.442747} {\bibfield  {journal} {\bibinfo  {journal} {J. Chem.
  Phys.}\ }\textbf {\bibinfo {volume} {76}},\ \bibinfo {pages} {473--480}
  (\bibinfo {year} {1982})}\BibitemShut {NoStop}%
\bibitem [{\citenamefont {Tempelaar}, \citenamefont {Jansen},\ and\
  \citenamefont {Knoester}(2014)}]{Tempelaar2014}%
  \BibitemOpen
  \bibfield  {author} {\bibinfo {author} {\bibfnamefont {R.}~\bibnamefont
  {Tempelaar}}, \bibinfo {author} {\bibfnamefont {T.~L.~C.}\ \bibnamefont
  {Jansen}}, \ and\ \bibinfo {author} {\bibfnamefont {J.}~\bibnamefont
  {Knoester}},\ }\bibfield  {title} {\enquote {\bibinfo {title} {{Vibrational
  Beatings Conceal Evidence of Electronic Coherence in the FMO Light-Harvesting
  Complex.}}}\ }\href {\doibase 10.1021/jp510074q} {\bibfield  {journal}
  {\bibinfo  {journal} {J. phys. chem. B}\ }\textbf {\bibinfo {volume} {118}},\
  \bibinfo {pages} {12865--72} (\bibinfo {year} {2014})}\BibitemShut {NoStop}%
\bibitem [{\citenamefont {Hauer}(1991)}]{Hauer1991}%
  \BibitemOpen
  \bibfield  {author} {\bibinfo {author} {\bibfnamefont {J.~F.}\ \bibnamefont
  {Hauer}},\ }\bibfield  {title} {\enquote {\bibinfo {title} {{Application of
  Prony analysis to the determination of modal content and equivalent models
  for measured power system response}},}\ }\href {\doibase 10.1109/59.119247}
  {\bibfield  {journal} {\bibinfo  {journal} {IEEE Trans. Power Syst.}\
  }\textbf {\bibinfo {volume} {6}},\ \bibinfo {pages} {1062--1068} (\bibinfo
  {year} {1991})}\BibitemShut {NoStop}%
\bibitem [{\citenamefont {{Schlau-Cohen}}, \citenamefont {{Ishizaki}},\ and\
  \citenamefont {{Fleming}}(2011)}]{SchlauCohen2011}%
  \BibitemOpen
  \bibfield  {author} {\bibinfo {author} {\bibfnamefont {G.~S.}\ \bibnamefont
  {{Schlau-Cohen}}}, \bibinfo {author} {\bibfnamefont {A.}~\bibnamefont
  {{Ishizaki}}}, \ and\ \bibinfo {author} {\bibfnamefont {G.~R.}\ \bibnamefont
  {{Fleming}}},\ }\bibfield  {title} {\enquote {\bibinfo {title}
  {{Two-dimensional electronic spectroscopy and photosynthesis: Fundamentals
  and applications to photosynthetic light-harvesting}},}\ }\href@noop {}
  {\bibfield  {journal} {\bibinfo  {journal} {Chem. Phys.}\ }\textbf {\bibinfo
  {volume} {387}},\ \bibinfo {pages} {1--22} (\bibinfo {year}
  {2011})}\BibitemShut {NoStop}%
\bibitem [{\citenamefont {Savikhin}, \citenamefont {Buck},\ and\ \citenamefont
  {Struve}(1997)}]{Savikhin1997}%
  \BibitemOpen
  \bibfield  {author} {\bibinfo {author} {\bibfnamefont {S.}~\bibnamefont
  {Savikhin}}, \bibinfo {author} {\bibfnamefont {D.~R.}\ \bibnamefont {Buck}},
  \ and\ \bibinfo {author} {\bibfnamefont {W.~S.}\ \bibnamefont {Struve}},\
  }\bibfield  {title} {\enquote {\bibinfo {title} {{Oscillating anisotropies in
  a bacteriochlorophyll protein: Evidence for quantum beating between exciton
  levels}},}\ }\href {\doibase 10.1016/S0301-0104(97)00223-1} {\bibfield
  {journal} {\bibinfo  {journal} {Chem. Phys}\ }\textbf {\bibinfo {volume}
  {223}},\ \bibinfo {pages} {303--312} (\bibinfo {year} {1997})}\BibitemShut
  {NoStop}%
\bibitem [{\citenamefont {Edington}, \citenamefont {Riter},\ and\ \citenamefont
  {Beck}(1995)}]{Edington1995}%
  \BibitemOpen
  \bibfield  {author} {\bibinfo {author} {\bibfnamefont {M.~D.}\ \bibnamefont
  {Edington}}, \bibinfo {author} {\bibfnamefont {R.~E.}\ \bibnamefont {Riter}},
  \ and\ \bibinfo {author} {\bibfnamefont {W.~F.}\ \bibnamefont {Beck}},\
  }\bibfield  {title} {\enquote {\bibinfo {title} {{Evidence for Coherent
  Energy Transfer in Allophycocyanin Trimers}},}\ }\href {\doibase
  10.1021/j100043a001} {\bibfield  {journal} {\bibinfo  {journal} {J. Phys.
  Chem.}\ }\textbf {\bibinfo {volume} {99}},\ \bibinfo {pages} {15699--15704}
  (\bibinfo {year} {1995})}\BibitemShut {NoStop}%
\bibitem [{\citenamefont {Smith}\ and\ \citenamefont
  {Jonas}(2011)}]{Smith2011}%
  \BibitemOpen
  \bibfield  {author} {\bibinfo {author} {\bibfnamefont {E.~R.}\ \bibnamefont
  {Smith}}\ and\ \bibinfo {author} {\bibfnamefont {D.~M.}\ \bibnamefont
  {Jonas}},\ }\bibfield  {title} {\enquote {\bibinfo {title} {{Alignment,
  vibronic level splitting, and coherent coupling effects on the pump - Probe
  polarization anisotropy}},}\ }\href {\doibase 10.1021/jp201928s} {\bibfield
  {journal} {\bibinfo  {journal} {J. Phys. Chem. A}\ }\textbf {\bibinfo
  {volume} {115}},\ \bibinfo {pages} {4101--4113} (\bibinfo {year}
  {2011})}\BibitemShut {NoStop}%
\bibitem [{\citenamefont {{Shore}}\ and\ \citenamefont
  {{Knight}}(1993)}]{Shore1993}%
  \BibitemOpen
  \bibfield  {author} {\bibinfo {author} {\bibfnamefont {B.~W.}\ \bibnamefont
  {{Shore}}}\ and\ \bibinfo {author} {\bibfnamefont {P.~L.}\ \bibnamefont
  {{Knight}}},\ }\bibfield  {title} {\enquote {\bibinfo {title} {{The
  Jaynes-Cummings Model}},}\ }\href {\doibase 10.1080/09500349314551321}
  {\bibfield  {journal} {\bibinfo  {journal} {J. Mod. Opt.}\ }\textbf {\bibinfo
  {volume} {40}},\ \bibinfo {pages} {1195--1238} (\bibinfo {year}
  {1993})}\BibitemShut {NoStop}%
\bibitem [{\citenamefont {Braak}(2011)}]{Braak2011}%
  \BibitemOpen
  \bibfield  {author} {\bibinfo {author} {\bibfnamefont {D.}~\bibnamefont
  {Braak}},\ }\bibfield  {title} {\enquote {\bibinfo {title} {Integrability of
  the rabi model},}\ }\href {\doibase 10.1103/PhysRevLett.107.100401}
  {\bibfield  {journal} {\bibinfo  {journal} {Phys. Rev. Lett.}\ }\textbf
  {\bibinfo {volume} {107}},\ \bibinfo {pages} {100401} (\bibinfo {year}
  {2011})}\BibitemShut {NoStop}%
\bibitem [{\citenamefont {Dhar}, \citenamefont {Rogers},\ and\ \citenamefont
  {Nelson}(1994)}]{Dhar1994}%
  \BibitemOpen
  \bibfield  {author} {\bibinfo {author} {\bibfnamefont {L.}~\bibnamefont
  {Dhar}}, \bibinfo {author} {\bibfnamefont {J.~A.}\ \bibnamefont {Rogers}}, \
  and\ \bibinfo {author} {\bibfnamefont {K.~A.}\ \bibnamefont {Nelson}},\
  }\bibfield  {title} {\enquote {\bibinfo {title} {Time-resolved vibrational
  spectroscopy in the impulsive limit},}\ }\href {\doibase 10.1021/cr00025a006}
  {\bibfield  {journal} {\bibinfo  {journal} {Chem. Rev.}\ }\textbf {\bibinfo
  {volume} {94}},\ \bibinfo {pages} {157--193} (\bibinfo {year}
  {1994})}\BibitemShut {NoStop}%
\bibitem [{\citenamefont {Tanimura}\ and\ \citenamefont
  {Kubo}(1989)}]{Tanimura1989}%
  \BibitemOpen
  \bibfield  {author} {\bibinfo {author} {\bibfnamefont {Y.}~\bibnamefont
  {Tanimura}}\ and\ \bibinfo {author} {\bibfnamefont {R.}~\bibnamefont
  {Kubo}},\ }\href {\doibase 10.1143/JPSJ.58.1850} {\enquote {\bibinfo {title}
  {{Time Evolution of a Quantum System in Contact with a Nearly
  Gaussian-Markoffian Noise Bath}},}\ } (\bibinfo {year} {1989})\BibitemShut
  {NoStop}%
\bibitem [{\citenamefont {Ishizaki}\ and\ \citenamefont
  {Tanimura}(2005)}]{Ishizaki2005}%
  \BibitemOpen
  \bibfield  {author} {\bibinfo {author} {\bibfnamefont {A.}~\bibnamefont
  {Ishizaki}}\ and\ \bibinfo {author} {\bibfnamefont {Y.}~\bibnamefont
  {Tanimura}},\ }\bibfield  {title} {\enquote {\bibinfo {title} {{Quantum
  dynamics of system strongly coupled to low-temperature colored noise bath:
  Reduced hierarchy equations approach}},}\ }\href {\doibase
  10.1143/JPSJ.74.3131} {\bibfield  {journal} {\bibinfo  {journal} {J. Phys.
  Soc. Jpn.}\ }\textbf {\bibinfo {volume} {74}},\ \bibinfo {pages} {3131--3134}
  (\bibinfo {year} {2005})}\BibitemShut {NoStop}%
\bibitem [{\citenamefont {Zhu}\ \emph {et~al.}()\citenamefont {Zhu},
  \citenamefont {Liu}, \citenamefont {Xie},\ and\ \citenamefont
  {Shi}}]{Zhu2012}%
  \BibitemOpen
  \bibfield  {author} {\bibinfo {author} {\bibfnamefont {L.}~\bibnamefont
  {Zhu}}, \bibinfo {author} {\bibfnamefont {H.}~\bibnamefont {Liu}}, \bibinfo
  {author} {\bibfnamefont {W.}~\bibnamefont {Xie}}, \ and\ \bibinfo {author}
  {\bibfnamefont {Q.}~\bibnamefont {Shi}},\ }\bibfield  {title} {\enquote
  {\bibinfo {title} {Explicit system-bath correlation calculated using the
  hierarchical equations of motion method},}\ }\href@noop {} {\ }\BibitemShut
  {NoStop}%
\bibitem [{\citenamefont {Viti}, \citenamefont {Petrucci},\ and\ \citenamefont
  {Barone}(1997)}]{Piero1997}%
  \BibitemOpen
  \bibfield  {author} {\bibinfo {author} {\bibfnamefont {V.}~\bibnamefont
  {Viti}}, \bibinfo {author} {\bibfnamefont {C.}~\bibnamefont {Petrucci}}, \
  and\ \bibinfo {author} {\bibfnamefont {P.}~\bibnamefont {Barone}},\
  }\bibfield  {title} {\enquote {\bibinfo {title} {Prony methods in nmr
  spectroscopy},}\ }\href {\doibase
  10.1002/(SICI)1098-1098(1997)8:6<565::AID-IMA9>3.0.CO;2-8} {\bibfield
  {journal} {\bibinfo  {journal} {Int. J. Imaging Syst. Technol.}\ }\textbf
  {\bibinfo {volume} {8}},\ \bibinfo {pages} {565--571} (\bibinfo {year}
  {1997})}\BibitemShut {NoStop}%
\bibitem [{\citenamefont {Luthon}\ \emph {et~al.}(1989)\citenamefont {Luthon},
  \citenamefont {Blanpain}, \citenamefont {Decorps},\ and\ \citenamefont
  {Albrand}}]{Luthon1989}%
  \BibitemOpen
  \bibfield  {author} {\bibinfo {author} {\bibfnamefont {F.}~\bibnamefont
  {Luthon}}, \bibinfo {author} {\bibfnamefont {R.}~\bibnamefont {Blanpain}},
  \bibinfo {author} {\bibfnamefont {M.}~\bibnamefont {Decorps}}, \ and\
  \bibinfo {author} {\bibfnamefont {J.}~\bibnamefont {Albrand}},\ }\bibfield
  {title} {\enquote {\bibinfo {title} {Parametric spectrum analysis of 2d
  \{NMR\} signals. application to in vivo j spectroscopy},}\ }\href {\doibase
  http://dx.doi.org/10.1016/0022-2364(89)90090-5} {\bibfield  {journal}
  {\bibinfo  {journal} {. Mag. Res. (1969)}\ }\textbf {\bibinfo {volume}
  {81}},\ \bibinfo {pages} {538 -- 551} (\bibinfo {year} {1989})}\BibitemShut
  {NoStop}%
\bibitem [{\citenamefont {Meunier}\ and\ \citenamefont
  {Brouaye}(1998)}]{Meunier1998}%
  \BibitemOpen
  \bibfield  {author} {\bibinfo {author} {\bibfnamefont {M.}~\bibnamefont
  {Meunier}}\ and\ \bibinfo {author} {\bibfnamefont {F.}~\bibnamefont
  {Brouaye}},\ }\bibfield  {title} {\enquote {\bibinfo {title} {{Fourier
  transform, wavelets, Prony analysis: tools for harmonics and quality of
  power}},}\ }in\ \href {\doibase 10.1109/ICHQP.1998.759842} {\emph {\bibinfo
  {booktitle} {Proc. of the Int. Conf. on Harmonics and Quality of Power}}},\
  Vol.~\bibinfo {volume} {1}\ (\bibinfo {year} {1998})\ pp.\ \bibinfo {pages}
  {71--76}\BibitemShut {NoStop}%
\bibitem [{\citenamefont {Volpato}\ and\ \citenamefont
  {Collini}(2015)}]{Volpato2015}%
  \BibitemOpen
  \bibfield  {author} {\bibinfo {author} {\bibfnamefont {A.}~\bibnamefont
  {Volpato}}\ and\ \bibinfo {author} {\bibfnamefont {E.}~\bibnamefont
  {Collini}},\ }\bibfield  {title} {\enquote {\bibinfo {title} {Time-frequency
  methods for coherent spectroscopy},}\ }\href {\doibase 10.1364/OE.23.020040}
  {\bibfield  {journal} {\bibinfo  {journal} {Opt. Express}\ }\textbf {\bibinfo
  {volume} {23}},\ \bibinfo {pages} {20040--20050} (\bibinfo {year}
  {2015})}\BibitemShut {NoStop}%
\end{thebibliography}
\end{document}